\documentclass[structabstract]{aa}  
\usepackage{graphicx}
\usepackage{txfonts}
\begin{document}

\title{PMAS Optical Integral Field Spectroscopy of Luminous Infrared
  Galaxies.}
\subtitle{
 II.-- Spatially resolved stellar populations and excitation
 conditions.\thanks{Based on observations collected at
    the German-Spanish Astronomical Center, Calar Alto, jointly
    operated by the Max-Planck-Institut f\"ur Astronomie Heidelberg and
    the  Instituto de Astrof\'{\i}sica de Andaluc\'{\i}a (CSIC).}}

\author{Almudena Alonso-Herrero\inst{1}, 
Macarena Garc\'{\i}a-Mar\'{\i}n\inst{2}, Javier Rodr\'{\i}guez
Zaur\'{\i}n\inst{1}, 
Ana Monreal-Ibero\inst{3},
  Luis Colina\inst{1}, Santiago Arribas\inst{1}} 

\institute{Departamento de Astrof\'{\i}sica Molecular 
e Infrarroja, Instituto de Estructura de la Materia, CSIC, 
E-28006 Madrid, Spain\\
\email{aalonso@damir.iem.csic.es}
\and
I. Physikalisches Institut, Universit\"at zu K\"oln, 50937
 K\"oln, Germany
\and
European Organisation for Astronomical Research in the Southern
Hemisphere, Karl-Schwarzschild-Strasse 2 D-85748 
Garching bei M\"unchen, Germany}

\date{Preprint}

\abstract
{The general properties (e.g., activity class, star formation rates,
metallicities, extinctions, average ages, etc) of luminous and
ultraluminous infrared galaxies (LIRGs and ULIRGs, respectively) in
the local universe
are well known since large samples of these objects have been the subject
of numerous spectroscopic works over the last three decades. There
are, however, relatively few 
  studies of the spatially-resolved spectroscopic 
properties of large samples of LIRGs and ULIRGs using integral field
spectroscopy (IFS).} 
{We are carrying out an IFS survey of local
  ($z<0.26$) samples of LIRGs and ULIRGs 
to characterize their two-dimensional spectroscopic properties. 
The main goal of this paper is to
 study the spatially resolved properties of the
stellar populations and the excitation 
conditions in a sample of LIRGs.} 
{We analyze optical (3800-7200\AA) IFS data 
taken with the Potsdam Multi-Aperture 
Spectrophotometer (PMAS) of the central 
few kiloparsecs of eleven LIRGs. To study the stellar populations we
fit the optical stellar continuum and the hydrogen recombination lines
of selected regions in the galaxies. We analyze the excitation
conditions of the gas using the spatially resolved
 properties of the brightest optical emission
lines. We complemented the PMAS observations with existing {\it
  HST}/NICMOS near-infrared continuum and Pa$\alpha$ imaging.}
{The optical continua of selected regions in our LIRGs
  are well fitted with a
  combination of an evolved ($\sim 0.7-10\,$Gyr) stellar population with 
 an ionizing stellar population ($1-20\,$Myr). The latter population 
is more obscured than the evolved population, and
has visual extinctions in good agreement
with those obtained from the Balmer decrement. Except for NGC~7771, we
found no clear evidence for an important contribution to the optical
light from an intermediate-aged stellar population ($\sim 100-500\,$Myr). Even after correcting for the presence of stellar absorption,
a large fraction of spaxels with low observed equivalent widths of
H$\alpha$ in emission  still show enhanced [N\,{\sc
  ii}]$\lambda$6584/H$\alpha$ 
and [S\,{\sc ii}]$\lambda\lambda$6717,6731/H$\alpha$ ratios.
These ratios are likely to be produced by a combination of photoionization in H\,{\sc ii}
regions and diffuse emission. These regions of enhanced line ratios
are generally coincident with low surface brightness H\,{\sc ii} regions and
diffuse emission 
detected in the H$\alpha$ and Pa$\alpha$ images.
Using the PMAS spatially resolved  line
ratios and the NICMOS Pa$\alpha$ photometry of H\,{\sc ii} regions we find that
the fraction of diffuse emission in LIRGs varies from galaxy to
galaxy, and it is generally 
less than 60\% as found in other starburst galaxies. }
{}

\keywords{galaxies: evolution  --- galaxies: nuclei --- galaxies: star
  formation ---
  galaxies: structure --- infrared: galaxies}

\authorrunning{Alonso-Herrero et al.}
%\titlerunning{PMAS Optical Integral Field Spectroscopy of Luminous Infrared
%  Galaxies}

\maketitle

\section{Introduction}

Luminous and Ultraluminous Infrared Galaxies (LIRGs and ULIRGs,
respectively) with infrared $8-1000\,\mu$m luminosities 
$L_{\rm IR}= 10^{11}-10^{12}\,{\rm
  L}_\odot$ and $L_{\rm IR}= 10^{12}-10^{13}\,{\rm
  L}_\odot$, respectively (see Sanders \& Mirabel 1996)
are among the most luminous objects in the local universe, and are
believed to be powered by strong star formation and/or AGN activity. 
The main properties of local LIRGs and ULIRGs are well known since these
two classes of galaxies have been extensively studied using    
imaging and spectroscopy over the last three decades. These
properties include, among others, morphologies (e.g., 
Veilleux, Kim, \& Sanders 2002; Sanders \& Ishida 2004; Scoville et
al. 2000; Bushouse et al. 2002; Alonso-Herrero et al. 2006),  
the nuclear activity class, 
star formation rates, and extinctions 
(e.g., Heckman,
  Armus, \& Miley 1987; Armus, Heckman, \& Miley 1989;  Kim et al. 1995,
  1998; Veilleux et al. 1995, 1999; Goldader et al. 1995; 
Wu et al. 1998; Rigopoulou et al. 1999;  Heckman et  al. 2000; Yuan,
Kewley, \& Sanders 2010), 
stellar populations (e.g., Poggianti \& Wu 2000; Chen et al. 2009;
Rodr\'{\i}guez Zaur\'{\i}n, Tadhunter, \& Gonz\'alez Delgado 2009,
2010a), metallicities (e.g., Rupke, Veilleux \&   Baker 2008),
and molecular gas content (e.g., Sanders et al. 1991; Gao \& Solomon
2004; Graci\'a-Carpio et al. 2006, 2008).

\begin{table*}
\scriptsize
\caption{Statistical properties of the spatially resolved of line
  emission} 
\begin{tabular}{lcccccccccccccccccccc}

\hline
\hline
           & \multicolumn{4}{c}{[OIII]$\lambda$5007/H$\beta$} &
           \multicolumn{4}{c}{[NII]$\lambda$6584/H$\alpha$} & 
\multicolumn{4}{c}{[SII]$\lambda\lambda$6717,6731/H$\alpha$}
           & \multicolumn{4}{c}{[OI]$\lambda$6300/H$\alpha$} &
           \multicolumn{4}{c}{EW(H$\alpha)_{\rm em}$}\\
& med & ave & $\sigma$ & $N$ & med &ave & $\sigma$ & $N$ & med &ave & $\sigma$ & $N$
           & med  &ave & $\sigma$ & $N$ & med & ave & $\sigma$ & $N$ \\
\hline
NGC~23          &0.65& 0.74& 0.44&132& 0.67& 0.77& 0.32&214& 0.64& 0.71& 0.38&207& 0.09& 0.15& 0.15&172&11.&20.&20.&214\\
MCG~+12-02-001  &0.76& 0.87& 0.47&104& 0.38& 0.38& 0.13&233& 0.54& 0.62& 0.32&171& 0.04& 0.07& 0.07& 81&46.&57.&36.&233\\
UGC~1845        &1.16& 1.24& 0.75& 20& 1.20& 1.39& 0.84&171& 0.39& 0.50& 0.37& 55& 0.07& 0.15& 0.15& 44& 7.&11.&11.&178\\
NGC~2388        &0.52& 0.64& 0.61& 79& 0.58& 0.68& 0.37&171& 0.42& 0.51& 0.44&150& 0.06& 0.13& 0.16& 79&16.&22.&17.&171\\
MCG~+02-20-003  &0.66& 0.82& 0.47&118& 0.44& 0.51& 0.37&220& 0.69& 0.79& 0.44&191& 0.07& 0.12& 0.12& 93&22.&30.&23.&220\\
NGC~5936        &0.39& 0.45& 0.26& 71& 0.42& 0.44& 0.09&256& 0.30& 0.32& 0.08&238& 0.05& 0.06& 0.04& 26&30.&34.&16.&256\\
NGC~6701        &0.58& 0.57& 0.20& 28& 0.65& 0.80& 0.46&250& 0.45& 0.61& 0.57&177& 0.08& 0.15& 0.19& 46& 9.&12.&10.&250\\
NGC~7469$^*$    &1.17& 1.76& 1.59& 77& 0.64& 0.72& 0.37&235& 0.51& 0.57& 0.24&196& 0.07& 0.09& 0.09& 75&20.&27.&19.&234\\
NGC~7591        &0.58& 0.92& 0.98& 29& 0.54& 0.70& 0.45&241& 1.03& 1.29& 0.86&208& 0.19& 0.26& 0.21& 70& 7.&11.&11.&242\\
NGC~7771        &0.59& 0.73& 0.50& 81& 0.70& 0.90& 0.65&270& 0.74& 0.89& 0.67&254& 0.13& 0.19& 0.19&125& 7.&13.&17.&272\\
\hline
All             & 0.66 & 0.85 & 0.77 & 739 & 0.56 & 0.71 & 0.51 & 2261
& 0.52   & 0.71 & 0.64 & 1852 & 0.08 & 0.14 & 0.15 & 811 & 17 & 24 &
24 & 2261\\ 

\hline
\end{tabular}

Notes.--- For each line ratio and the EW(H$\alpha)_{\rm em}$ we list the values
of the median, average, standard deviation and number of spaxels with
measurements. The statistics is done using the values of the
individual spaxels and thus are not light weighted. That is, the average
value of a given line ratio is not necessarily equal to that measured
from the integrated spectra of the galaxy. The line ratios are not
corrected for extinction or stellar absorption.\\
$^*$The statistics is only for the narrow component of
the hydrogen lines.

\end{table*}

There are, however, relatively few
  studies of the spatially-resolved spectroscopic 
properties of LIRGs and ULIRGs
  using integral field spectroscopy (IFS). 
Most works have focused on individual famous objects or small samples (e.g.,
Colina, Arribas, \& Borne 1999; Arribas, Colina, \& Clements 2001; 
Murphy et al. 2001; 
L\'{\i}pari et al. 2004a,b; 
Colina, Arribas, \& Monreal-Ibero 2005; Monreal-Ibero, Arribas, \&
Colina 2006; Garc\'{\i}a-Mar\'{\i}n et al. 2006; Reunanen,
Tacconi-Garman, \& Ivanov 2007; Bedregal et al. 2009). 
IFS instruments working in
the optical and infrared spectral ranges on 4 and 8\,m-class
telescopes  are now ubiquitous, and thus 
this situation is changing rapidly. Our group is involved
in an ambitious project intended to characterize the detailed optical
and infrared spectroscopic 
properties of local samples of LIRGs and ULIRGs using a
variety of IFS facilities.

This is the second paper in a series presenting observations with 
the  Potsdam Multi-Aperture Spectrophotometer (PMAS, Roth et al. 2005) 
of the northern portion of a sample of local LIRGs defined by Alonso-Herrero et
al. (2006). The PMAS sample is in turn  
part of the larger IFS survey of nearby ($z<0.26$) LIRGs and ULIRGs 
in the northern and the southern
hemispheres using different IFS instruments. The optical ones include, 
apart from the PMAS instrument, 
the VIMOS instrument (Le F\`evre et al. 2003) on the VLT, and 
the INTEGRAL+WYFFOS system  (Arribas et al. 1998; Bingham et al. 1994) 
on the William Herschel  Telescope (WHT). 
The first results of this project can be
found in Alonso-Herrero et al. (2009, hereafter Paper~I) 
for the PMAS atlas of LIRGs,
Garc\'{\i}a-Mar\'{\i}n et al. (2009a,b) for the INTEGRAL results of
ULIRGs, and Arribas et al. (2008), Monreal-Ibero et al. (2010a) and Rodr\'{\i}guez Zaur\'{\i}n
et al. (2010b) for the VIMOS results. Additionally, Pereira-Santaella
et al.  (2010) studied the spatially resolved mid-infrared properties of LIRGs
using the spectral mapping capability of IRS on {\it Spitzer}, and
Bedregal et al. (2009) presented a detailed near-IR IFS study of a
local LIRG using SINFONI on the VLT.

In this paper we study in detail the stellar populations,  
excitation conditions, and diffuse emission 
in the central regions (a few kiloparsecs) 
of a sample of 11 LIRGs using the PMAS data. The paper is organized as
follows. We present the observations and data analysis in
Section~2. In Section~3 we describe the modelling of the stellar
populations. The results regarding the morphology, stellar
populations, excitation conditions, and diffuse emission are 
presented in Sections~4, 5, and 6. We  give our conclusions in
Section~7.

\section{Observations and Data Analysis}

\subsection{Sample, observations and Data Reduction}
The observations and data reduction of the PMAS data are described 
in detail in Paper~I. Briefly, we used PMAS 
on the 3.5\,m telescope at the German-Spanish
Observatory of Calar Alto (Spain) to observe a sample of 11 local
LIRGs. These comprise the majority of the northern hemisphere portion
of the volume-limited ($v=2750-5200\,{\rm km\,s}^{-1}$) sample of
LIRGs of Alonso-Herrero et 
al. (2006). This volume-limited sample was originally drawn from the
{\it IRAS} Revised Bright Galaxy Sample (RBGS, Sanders et al. 2003). 
The range of IR luminosities of the PMAS sample is $\log (L_{\rm
  IR}/{\rm L}_\odot) =11.05-11.59$, and the galaxies are at an average
distance of 61\,Mpc (for the assumed cosmology $H_0=75\,{\rm km \,
  s}^{-1} \, {\rm  Mpc}^{-1}$). As the sample is flux-limited, it is 
composed mostly by moderate IR luminosity systems, with an average $ \log
(L_{\rm IR}/{\rm L}_\odot) =11.32$ for the full sample of
Alonso-Herrero et al. (2006).

The  PMAS observations were taken with  
the Lens Array Mode configuration, which is
made of  a $16 \times 16$ array of microlenses coupled with fibers 
called hereafter spaxels. We used the 1\arcsec \, magnification that
provides a field
of view (FoV) of $16\arcsec \times 16\arcsec$.  All the galaxies were
observed with a single pointing, except  NGC~7771 for which we
obtained two pointings to construct a mosaic with a $28\arcsec \times
16\arcsec$ FoV. We covered the
wavelength 
range of $3800-7200\,$\AA, using the V300 grating with 
a spectral resolution of 
6.8\,\AA  \ full width half maximum (FWHM). The full 
description of the data reduction procedure can be found in
Paper~I.

In addition to the PMAS data, in this paper we
make use of the {\it HST}/NICMOS F110W ($\lambda_{\rm c}=1.1\,\mu$m)
and F160W  ($\lambda_{\rm c}=1.6\,\mu$m) continuum
observations to construct continuum near-IR color maps, as discussed
by Alonso-Herrero et al. (2006). The only additional step needed for
the reduced NICMOS images was to rotate and trim them 
to match the orientation and FoV, respectively, of the PMAS
images.

\begin{table}[h]
\caption{Definition of indices of the stellar absorption features.}
\begin{tabular}{lccccccc}
\hline
\hline
Name   & \multicolumn{2}{c}{Blue Cont.} & 
\multicolumn{2}{c}{Red Cont.} &
\multicolumn{2}{c}{Line} & Ref\\   
   & $\lambda_{\rm c}$& $\Delta \lambda$ & $\lambda_{\rm c}$& $\Delta
   \lambda$ & $\lambda_{\rm c}$& $\Delta \lambda$ \\
\hline
$C_{6500}/C_{4800}$ & 6500 & 50 & 4800 & 50 & -- & -- & 1\\
D$_n(4000)$ & 3900 & 100 & 4050 & 100 & -- & -- & 2\\
H$\delta_{\rm A}$  & 4060 & 38.5 & 4145 & 32.5 & 4102 & 38.5 & 3\\
H$\beta_{\rm A}$ & 4776 & 12 & 4948 & 12 &4861 & 30 &4\\
H$\alpha_{\rm A}$ & 6510 & 8 & 6616 & 8 & 6563 & 30  & 4\\
\hline
\end{tabular}

Notes.--- All the wavelengths are in \AA. \\
References: 1. Kim et al. (1995) and Veilleux et
al. (1995). 2. Balogh et al. (1999). 3. Worthey \& Ottaviani
(1997). 4. Gonz\'alez Delgado et al. (2005).
\end{table}

\subsection{PMAS spatially resolved emission 
line ratios and equivalent widths}

We constructed spectral maps of the brightest emission lines 
in an automated fashion using our own IDL routines as
  well as the IDL-based MPFITEXPR 
algorithm\footnote{http://www.purl.com/net/mpfit
developed by Markwardt (2008).  
We fitted } the lines to Gaussian functions and the adjacent 
continuum to a straight line, on a spaxel-by-spaxel basis, as described in 
more detail Paper~I.  These fits provided the flux, equivalent width (EW), and
full width half maximum of the emission lines.  
In this paper, we generated spectral maps of the 
following optical line ratios:  [O\,{\sc iii}]$\lambda$5007/H$\beta$, 
[O\,{\sc i}]$\lambda$6300/H$\alpha$, [N\,{\sc  ii}]$\lambda$6584/H$\alpha$,  
and [S\,{\sc ii}]$\lambda\lambda$6717,6731/H$\alpha$.  We also
  generated maps of the EW of
the H$\alpha$ line in emission (EW(H$\alpha)_{\rm em})$, which are thus
measured as positive numbers.
The maps of the line ratios and EW are not corrected for the presence
of H$\alpha$ and H$\beta$ in absorption. 
The spectral maps of the observed line ratios together with those of
the observed flux of H$\alpha$ and  EW(H$\alpha)_{\rm em}$ are shown
in Figure~1, except for IC~860, for which the line emission is compact
(see Paper~I).  In Table~1 we list for each of the emission
line ratios the number of
spaxels where it was possible to obtain a measurement, the median, the
average value and the standard deviation, for each galaxy and the full
PMAS sample.

\begin{table*}
\begin{center}
\caption{Observed properties of selected star-forming regions.}
\begin{tabular}{cccccccc}
\hline
\hline
Region & Pos x & Pos y & Size & EW(H$\alpha)_{\rm em}$ & D$_n(4000)$ &
H$\delta_{\rm A}$ & $C_{6500}/C_{4800}$\\
& arcsec & arcsec & ${\rm pc}\times {\rm pc}$ & \AA\\
\hline
\multicolumn{8}{c}{NGC~23}\\
\hline
HII-1 & $-3.0$ & $-3.5$ & $289\times578$ & 93.9 & 1.37 & --   & 1.14\\
HII-2 & $-3.5$ & $+2.5$ & $578\times578$ & 50.8 & 1.31 & $1.95^\dag$ & 1.12\\
HII-3 & $+0.5$ & $+3.5$ & $578\times578$ & 47.2 & 1.20 & $2.68^\dag$ & 0.98\\
HII-4 & $+1.5$ & $-2.5$ & $578\times578$ & 26.9 & 1.22 & 5.01 & 0.80\\
\hline
\multicolumn{8}{c}{NGC~2388}\\
\hline
HII-1 & $+4.5$ & $-0.5$ & $560\times560$ & 66.3 & 1.27 & 3.87 & 1.41\\
HII-2 & $-2.5$ & $+0.5$ & $560\times560$ & 64.0 & 1.35 & --   & 1.69\\
\hline
\multicolumn{8}{c}{NGC~7771}\\
\hline
HII-1 & $-2.5$ & $-2.0$ & $277\times554$ & 89.0 & 1.18 & --  & 0.99\\
HII-2 & $-5.5$ & $-0.5$ & $554\times554$ & 39.8 & 1.46 & --  & 1.57\\
HII-3 & $-3.5$ & $+1.5$ & $554\times554$ & 26.8 & 1.44 & --  & 1.45\\
Nucleus2 & $-2.0$ & $+1.0$ & $277\times277$ & 34.0 & 1.21 & 2.00 & 1.48\\
\hline

\end{tabular}

Notes. --- The positions of the extracted regions are given relative
to that of the nucleus of the galaxy. 
All the reported values of EW(H$\alpha)_{\rm em}$ are in
emission. \\
The values of H$\delta_{\rm A}$ are not corrected for
H$\delta$ nebular emission. $^\dag$The nebular
H$\delta$ line is seen in emission within the stellar absorption
feature. 

\end{center}
\end{table*}

We also produced  standard optical diagnostic diagrams  using the
  brightest emission lines (Baldwin, Phillips, \& Terlevich 1981;
  Veilleux \& Osterbrock 1987) on a spaxel-by-spaxel basis for 
the galaxies in our
sample. These diagrams provide useful information on the excitation
conditions of different regions in galaxies, such as,  photoionization
by young stars, shocks, and AGN photoionization. In Paper~I we
presented such 
diagrams for the nuclear 
and integrated measurements of our sample. The spatially-resolved 
  diagnostic
diagrams for each of the galaxies in our sample are shown in
Figure~2. We also produced [N\,{\sc
  ii}]$\lambda$6584/H$\alpha$  vs. [S\,{\sc
  ii}]$\lambda\lambda$6717,6731/H$\alpha$ diagrams (Figure~3) for each of the
galaxies in our 
sample. These diagrams have an advantage over the  Veilleux \&
  Osterbrock (1987) diagrams in that
they contain more data points, as the [O\,{\sc iii}]$\lambda$5007/H$\beta$ line
ratio can be strongly affected by both the presence of an underlying stellar
absorption and extinction. As we shall see in Sections~5 and 6, 
the effects of the Balmer absorption stellar features on the observed
line ratios are not negligible in
regions with low values of EW(H$\alpha)_{\rm em}$ or EW(H$\beta)_{\rm
  em}$. For this reason, the individual measurements in
  diagrams of Figures~2 and 3 are color-coded  according to
  arbitrarily chosen ranges of EW(H$\alpha)_{\rm 
  em}$.
All these diagrams and the effects of the correction for stellar
  absorption features will be discussed in Section~6. 

The errors of the line ratios depend on the observed values of 
EW(H$\alpha)_{\rm em}$ and the S/N of the spectra. We estimated 
the typical uncertainties by comparing the line ratios measured
automatically with our IDL routines 
with those fitted manually 
with the {\it splot} routine within {\sc IRAF} for 
selected spaxels in each galaxy. For each galaxy, the comparison was made for 
spaxels within the smallest observed range of EW(H$\alpha)_{\rm em}$
where the uncertainties are the highest. By choosing spaxels with low
values of EW(H$\alpha)_{\rm em}$, we basically estimated an upper
limit to the uncertainties of the observed line ratios. 
As can be seen from Figure~3 the largest uncertainties in the 
[N\,{\sc  ii}]$\lambda$6584/H$\alpha$  and 
[S\,{\sc  ii}]$\lambda\lambda$6717,6731/H$\alpha$ 
ratios are $10-25$\%, and $15-40$\%, respectively, 
depending on the range of EW(H$\alpha)_{\rm em}$ and the galaxy.

\subsection{Extraction and Analysis of the 1D spectra of selected regions}

For each galaxy we extracted  the nuclear and integrated spectra as
done in Paper~I. Briefly, we identified the position of the
optical nucleus as the peak of the 6200\,\AA \, continuum emission,
and  extracted the nuclear spectrum using the corresponding spaxel. The
physical size covered by the nuclear spectrum for each galaxy
was given in Paper~I, and it is typically the approximate central 300\,pc. 
The integrated spectrum of each galaxy was extracted by defining
$\sim 6200\,$\AA \, continuum 
isophotes and then summing up all the spaxels contained within the
chosen external continuum isophote to include as much as possible of 
the PMAS FoV. The area covered by the integrated spectrum can be seen
in figure~1 in Paper~I, and it is generally the central $5-8\,$kpc,
depending on the galaxy.

\begin{figure*}

\setcounter{figure}{3}
\hspace{0.5cm}
\includegraphics[width=8.5cm,angle=-90]{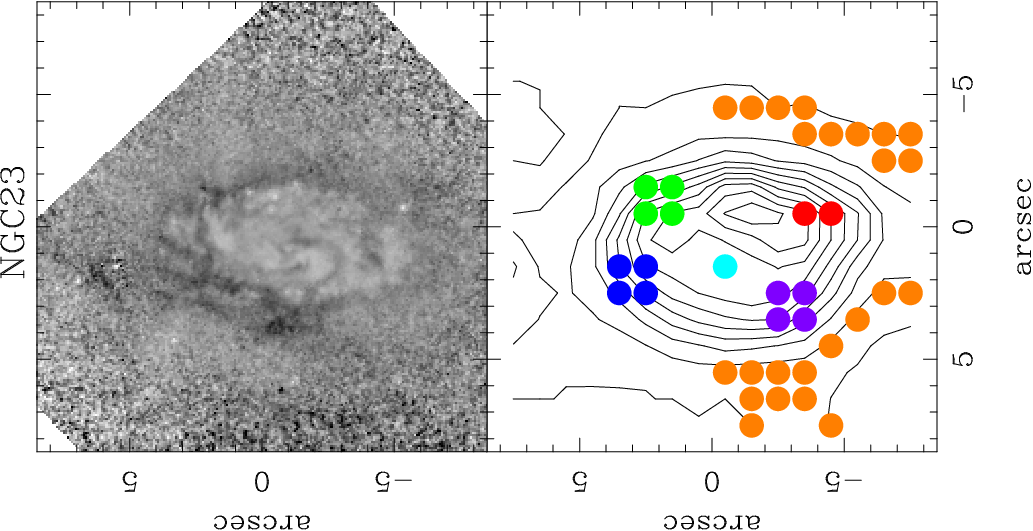}
\hspace{0.5cm}
\includegraphics[width=8.5cm,angle=-90]{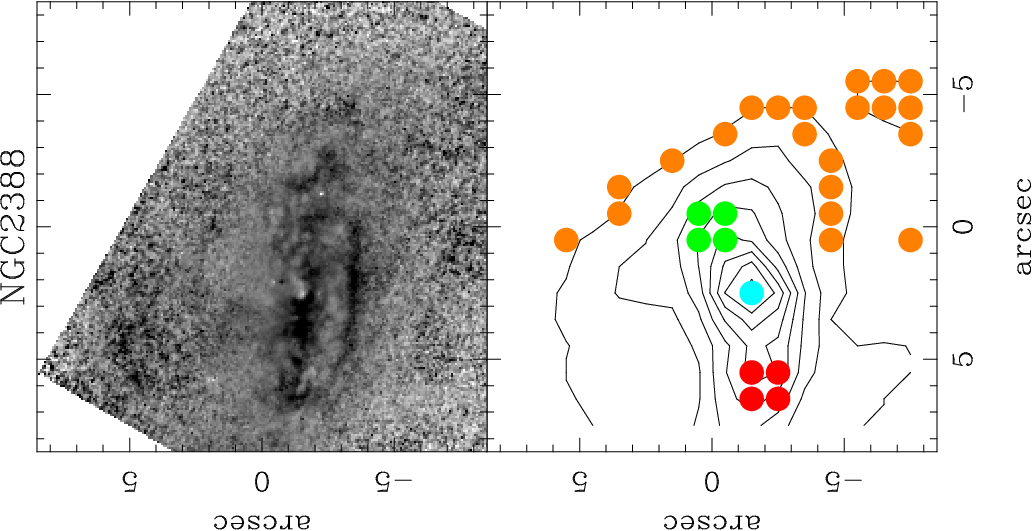}
\hspace{0.5cm}
\includegraphics[width=8.5cm,angle=-90]{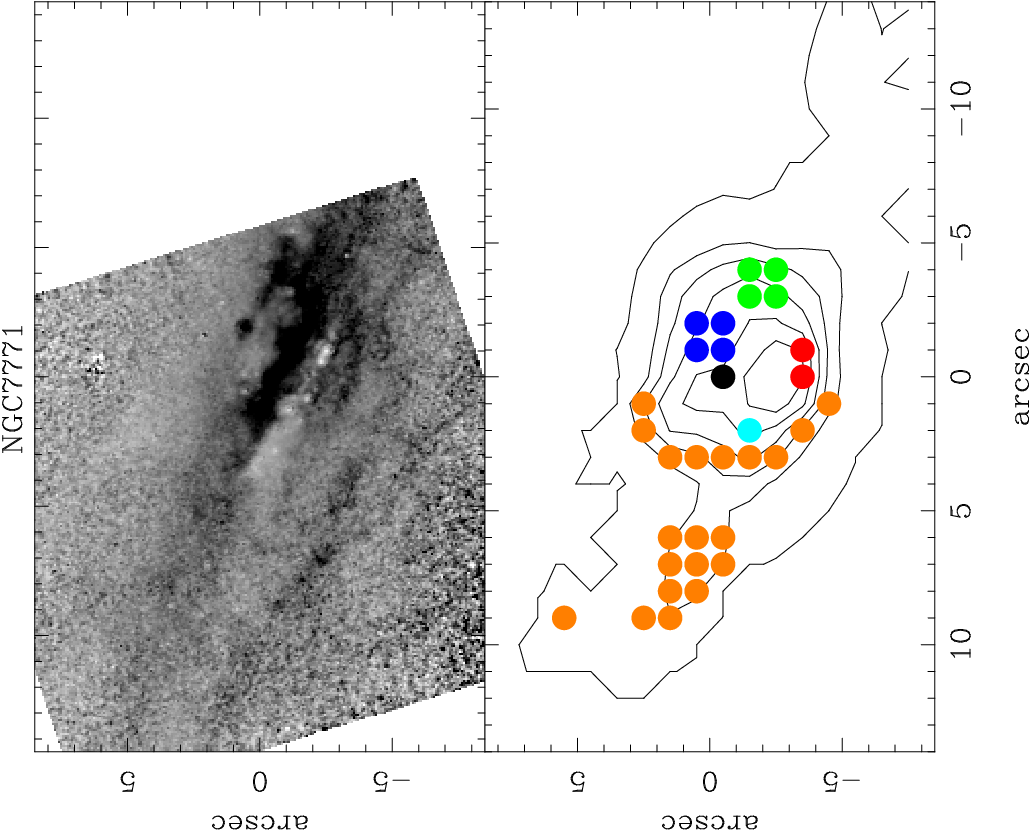}

\caption{{\it Upper panels:} {\it HST}/NICMOS
  $1.6\,\mu$m to $1.1\,\mu$m flux ratio maps. Dark regions show a
  deficit of $1.1\,\mu$m flux, that is, a redder ${\rm F160W}-{\rm
    F110W}$ color. All three galaxies are shown on the same flux ratio
scale. {\it Lower panels:}  The contours are the observed H$\alpha$
  fluxes of the three galaxies for which extracted spectra of
  circumnuclear H\,{\sc ii} regions. For each galaxy
  we show the regions of interest: the nucleus (defined as the peak of the
  6200\,\AA \, continuum emission, light blue dots), bright H\,{\sc ii}
  regions (red, blue, purple, and green dots). The orange dots are the 
spaxels used to produce  the average spectrum of regions with
relatively low EW of H$\alpha$ (see Section~2.3). 
For  NGC~7771 we additionally extracted a spectrum of the possible
  location of the true nucleus (shown as a black dot), 
based on the H$\alpha$ velocity field
  (see Paper~I for details). } 

\end{figure*}

For the three galaxies observed under photometric conditions: NGC~23,
NGC~2388, and NGC~7771, we  extracted spectra of a number of bright H\,{\sc ii}
regions. The locations of the selected regions are shown in the lower panels of
Figure~4. These regions were chosen to probe, for each galaxy, a
range of EW(H$\alpha$)$_{\rm em}$ and continuum slopes as can be seen
from Figure~5. 
Veilleux et al. (1995) found that the
extinction in (U)LIRGs powered by star formation is  correlated with 
the shape of the optical continuum. 
We estimated the shape of the optical continuum by measuring the ratio
($C_{6500}/C_{4800}$) between the continuum fluxes (see Figure~4) near
H$\beta$ and H$\alpha$ at 4800 and 6500\,\AA, respectively. The
definition of this ratio is given in Table~2. 
Similarly the gas extinctions are
related to the {\it HST}/NICMOS ${\rm F160W} -
{\rm F110W}$ colors (see Alonso-Herrero et al. 2006). Figure~4 shows
how the selected H\,{\sc ii} regions in these galaxies span a range in 
${\rm F160W}$ to ${\rm F110W}$ flux ratios, and there is a good
correspondence between red near-IR colors and steep optical continua.
Table~3 gives details of the observed properties 
of these regions. Given
the good quality of the spectra for these three galaxies, we
will use them to do a detailed modelling of the stellar populations in
these galaxies (Section~3.1).

%Additionally for NGC~7771
%we extracted the spectrum of the possible location of the ¨true¨
%nucleus, as opposed to the optical peak, of the galaxy. The location
%of the ¨true¨ nucleus is based on the physical point that makes the
%H$\alpha$ velocity field of this galaxy symmetric (see Paper I). As
%can be seen from the near-IR color map (Figure~3), 
%the ¨true¨ nucleus of NGC~7771 appears to
%be located behind a large amount of obscuring material. 

As we will explain in Section~5.1, we can use the 4000\AA-break
and the Balmer lines in absorption, and in particular the H$\delta$
line, to infer an estimate of the average age of the stellar
populations. Although the H$\delta$ nebular emission line is much weaker
than H$\alpha$ and H$\beta$, it can also be observed in emission in
bright/young H\,{\sc ii} regions (see for instance the HII-1 region in NGC~23,
Figures~4 and 5). To minimize the contamination from the
Balmer lines in emission, for each galaxy we selected regions with low
values of EW(H$\alpha)_{\rm em}$ so we could attempt to
 measure the H$\delta$ feature in
absorption. We produced the average spectrum of this low EW emission by 
summing up the spaxels with the specified range of 
EW(H$\alpha$)$_{\rm em}$ for each galaxy. 
The individual spaxels had typically 
$20\,{\rm \AA} < {\rm EW(H}\alpha)_{\rm em} < 6\,{\rm \AA}$, but the
specific range depended on the galaxy and the S/N of the data. From the 
comparison between the maps of the observed EW(H$\alpha$)$_{\rm em}$ in Figure~1
and the {\it HST}/NICMOS Pa$\alpha$ maps shown in Paper~I, it is
clear that the low EW(H$\alpha$)$_{\rm em}$ values
are associated with regions of diffuse emission or low 
surface brightness H\,{\sc ii} regions. 
%In some cases, we  only included spaxels
%from one side 
%of the velocity field (see Figure~3 for a few examples). 
Table~4 lists the values of EW(H$\alpha$)$_{\rm em}$ as measured from the
average spectrum of regions of low EW of H$\alpha$ of each galaxy. 
All the extracted spectra were shifted to rest-frame wavelengths
prior to the analysis and fitting the stellar populations.

\begin{table*}
\begin{center}
\caption{Observed properties for the nuclei, average of regions with
  low EW(H$\alpha)_{\rm em}$ and
  integrated emission.}
\begin{tabular}{lccccccccc}
\hline
\hline
Galaxy  &   \multicolumn{3}{c}{Nuclear} &
\multicolumn{3}{c}{Low-EW(H$\alpha)_{\rm em}$ Region} 
& \multicolumn{3}{c}{Integrated$^*$}\\
& EW(H$\alpha)_{\rm em}$ & D$_n(4000)$ &
H$\delta_{\rm A}$  & EW(H$\alpha)_{\rm em}$ & D$_n(4000)$ &
H$\delta_{\rm A}$  & EW(H$\alpha)_{\rm em}$ & D$_n(4000)$ &
H$\delta_{\rm A}$  \\   
 & \AA & & & \AA & & & \AA\\
\hline
\hline
NGC~23   & 12.8 & 1.40 & 3.80
 & 7.3 & 1.34 & 3.54 & 32.8 & 1.29 & $3.13^\dag$\\ 
MCG~+12-02-001 & 145.6 & -- & -- & -- & -- & -- & 92.5 & -- & --\\ 
UGC~1845       & 20.4 & 1.60 & 6.66 & 10.5 & -- & -- & 25.1 & -- &  --\\ 
NGC~2388       & 53.1 & 1.30 & $2.58^\dag$ & 10.0 & 1.40 & 4.22 & 28.8 &
1.35 & 3.21\\ 
MCG~+02-20-003 & 143.6 & 1.42 & $2.01^\dag$ &-- & -- &-- & 39.9 & 1.36 & 3.85 \\ 
IC~860        & 1.5 & 1.42 & 5.67 &-- &-- &-- & -- & 1.36 & 4.40\\ 
NGC~5936      & 66.1  & 1.33 & 2.29 & 16.0 & 1.25 &2.92 & 38.6 & 1.12
& $2.40^\dag$\\  
NGC~6701      & 34.8 & 1.23 & $2.20^\dag$ & 14.0 & 1.24 & 2.58 & 17.2 &
1.15 & 2.57 \\  
%NGC~7469      &  \\ 
NGC~7591      &  31.4 & 1.55 & -- &-- &-- &-- & 9.2 & 1.39 & 2.15\\ 
NGC~7771      & 23.0  & 1.18 & 5.27& 12.0 & 1.28 &4.33 & 15.5 & 1.36 &
$2.22^\dag$\\  
\hline
\end{tabular} 
\end{center}

Notes.--- $^*$Integrated spectra refers to the central $\sim 5\,$kpc to $\sim
8\,$kpc regions, depending on the galaxy (see Paper~I for details).\\ 
All the reported values of EW(H$\alpha)_{\rm em}$ are for the
  line observed in emission.\\ 
The values of H$\delta_{\rm A}$ are not corrected for
H$\delta$ nebular emission. $^\dag$The nebular
H$\delta$ line is clearly seen in emission within the stellar absorption
feature. 
\end{table*}

\begin{figure}

\setcounter{figure}{4}
\includegraphics[width=8.5cm,angle=-90]{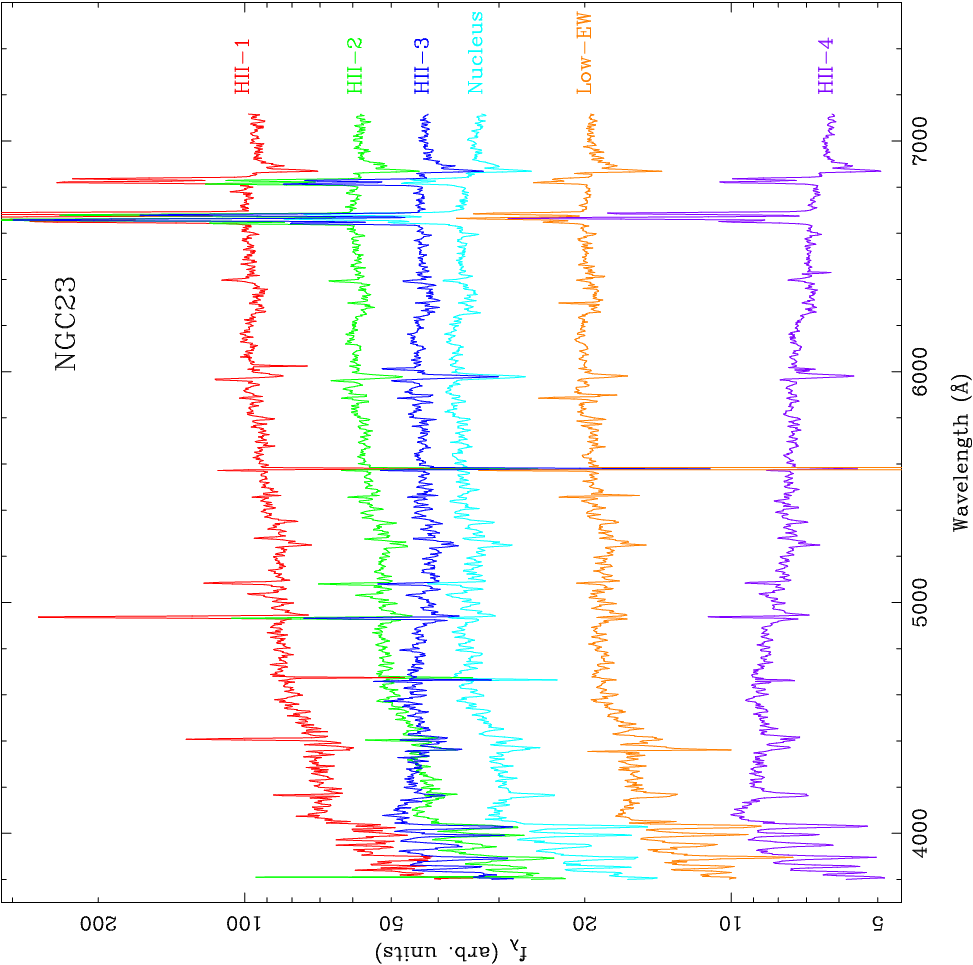}
\includegraphics[width=8.5cm,angle=-90]{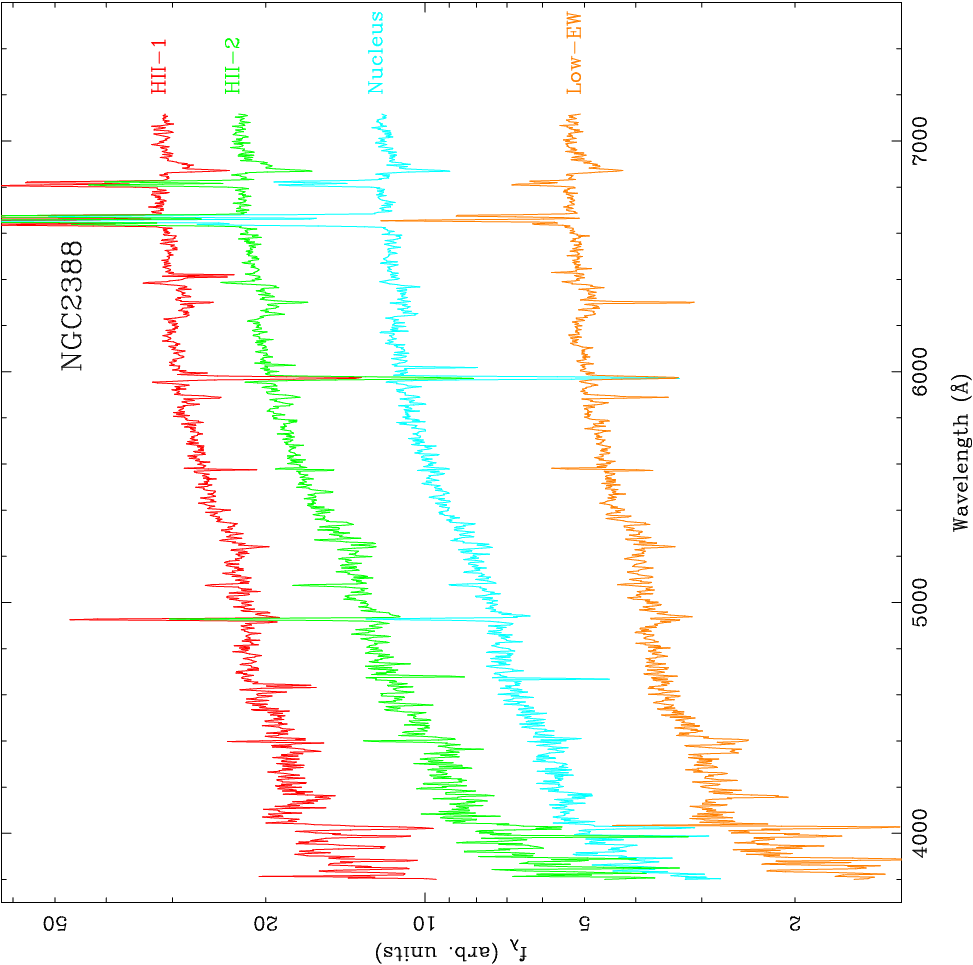}
\includegraphics[width=8.5cm,angle=-90]{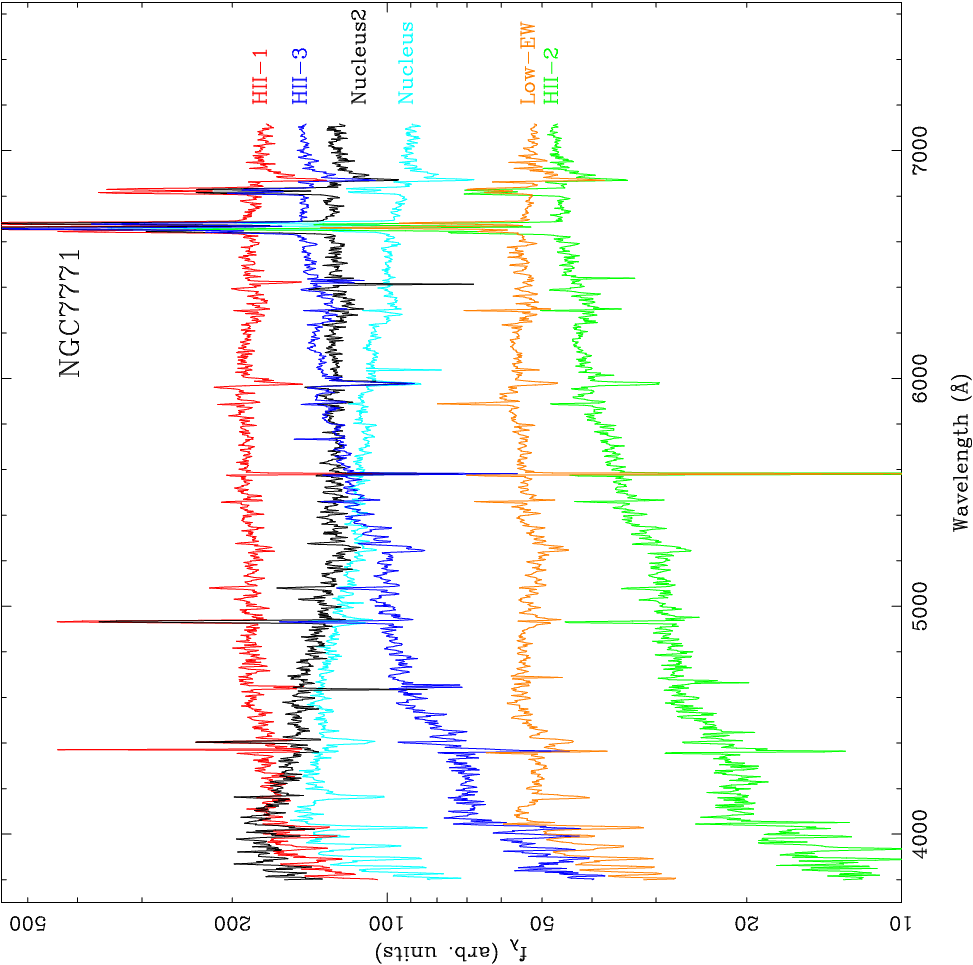}
\caption{PMAS spectra of selected regions (see Figure~3 and Table~3),
  arbitrarily scaled, of the three galaxies observed under photometric
  conditions: NGC~23, NGC~2388, and NGC~7771. The color used for
  plotting the spectra corresponds to those of Figure~4 (lower panel).} 

\end{figure}

In Section~5 we will study the stellar populations of LIRGs using
  the 4000\,\AA-break and the H$\delta$ stellar feature.
We adopted the definition of  Balogh et al. (1999)
for the ${\rm D}_n(4000)$ index and that of Worthey \& Ottaviani (1997)  
for the  H$\delta_{\rm A}$ index.\footnote{We note that the
value of this index is positive for absorption, and throughout this
paper we will use this symbol to indicate the feature  in absorption.}
We measured these indices 
for the nuclear and integrated spectra, as well as the average spectra
of regions of low EW(H$\alpha)_{\rm em}$.
The pseudo-continuum bands of these two
indices, as well as the line window for the 
H$\delta_{\rm A}$ index are listed in Table~2. 
For the majority of the selected
H\,{\sc ii} regions (see Table~3), as well as the  nuclei and
integrated emission (Table~4), the
measured values H$\delta_{\rm A}$ are only lower limits as the
observed values of EW(H$\alpha)_{\rm em}$ in emission imply a contribution
from nebular H$\delta$ emission to the index. In Tables~3 and 4 we
marked those regions where we clearly detected the nebular H$\delta$ line 
in emission within
the stellar absorption. We also measured the  H$\alpha$ flux and the 
EW(H$\alpha$)$_{\rm em}$ of the line in emission in all the extracted
spectra. 

%To estimate the correction to H$\delta_{\rm A}$ for H$\delta$
%nebular emission, we assumed case B recombination and used the
%measured H$\alpha$ fluxes.  **** need to make this correction ****

\section{Modelling of the Stellar Populations}

As discussed in Paper~I the nuclear and integrated spectra of our
sample of LIRGs show evidence for the presence of an ionizing stellar
population, plus a more evolved stellar population as indicated by the
presence of strong absorption stellar features in the blue part of the
spectrum.  This is also apparent for the H\,{\sc ii} regions in our
sample of LIRGs, including those with the largest ${\rm
EW(H}\alpha)_{\rm em}$ 
(see Figure~5). These strong absorption stellar features
appear to be a general property of local LIRGs and ULIRGs (see e.g.,
Armus et al. 1989; Veilleux et al. 1995; Kim et al. 1995, 1998;
Marcillac et al. 2006; Chen et al. 2009; Rodr\'{\i}guez Zaur\'{\i}n et
al. 2009), as well as intermediate-redshift LIRGs (see e.g., Hammer et
al. 2005; Marcillac et al. 2006; Caputi et al. 2008).  Moreover, it
has been suggested that LIRGs represent phases in the life of galaxies
with episodic and extremely efficient star formation (Hammer et
al. 2005).  This suggests that using one single stellar
population  (SSP) may not be appropriate for modelling individual
regions and the integrated emission of LIRGs.

\subsection{Modelling of the stellar continuum}

For the modelling of the stellar continuum we used the Bruzual \&
Charlot (2003, BC03 hereafter) models with solar metallicity,
instantaneous star formation, and a Salpeter (Salpeter 1955) initial
mass function (IMF) with lower and upper mass cutoffs of $m_{\rm l}
=0.1\,{\rm M}_\odot$ and $m_{\rm u} =100\,{\rm M}_\odot$,
respectively. These models have a spectral resolution of 3\,\AA \,
across the whole wavelength range from 3200 to 9500\,\AA.  Using these
assumptions we generated template spectra covering ages of between
1\,Myr and 10\,Gyr. The outputs of these models are normalized
to a total mass of $1\,{\rm M}_\odot$ formed in the burst of star
formation. The dust attenuation was modelled using the Calzetti et
al. (2000) extinction law, which is appropriate for starburst
galaxies.

To fit the optical spectra of the selected regions 
 (see Section~2.3) we  used the CONFIT code
(Robinson et al. 2000), which assumes two stellar populations plus a
power law in some cases. Briefly, CONFIT fits the continuum shape of
the extracted spectra using a minimum $\chi^2$ technique.  For each
spectrum, CONFIT measures the flux in $\sim 50-70$ wavelength bins,
chosen to be as evenly distributed in wavelength as possible, and to
avoid strong emission lines and atmospheric absorption features (see
Rodr\'{\i}guez Zaur\'{\i}n et al. 2009 for details). The relative flux
calibration error of 10\% measured was assumed during the modelling.

For this  work we  used a large number of
combinations of two stellar populations to determine which stellar
populations dominate the optical emission of our galaxies. 
We divided the two stellar populations 
into  an {\it ionizing stellar population} ($<20\,$Myr, in
intervals of 1, 2, ... 10, 20\,Myr) with a varying reddening (${\rm E(B-
V)}_{\rm young}
\leq$ 2.0 in increasing steps of 0.1) and an {\it evolved stellar
population} with ages between $100\,{\rm Myr}$ and $10\,{\rm Gyr}$
(100, 300, 500, 700\,Myr; 1, 2, ... 5\,Gyr and 10 Gyr) with moderate 
reddening (${\rm E(B- V)}_{\rm evolved} = 0, \, 0.2, \, 0.4$). This
preferential dust extinction is based on the scenario where the
youngest stellar populations are still partially embedded in their
dusty birth places, whereas the more evolved stellar populations have
already moved away from their natal clouds (Calzetti et al. 1994).
Poggianti \& Wu (2000) demonstrated that this scenario was compatible
with the observed optical spectra of infrared bright galaxies. 

The choice of an ionizing population is driven by the fact that all
the spectra modelled in this paper show H$\alpha$ in emission with
${\rm EW(H}\alpha)_{\rm em}>7\,$\AA \,(except for the nuclear region
of IC~860, see Tables~3 and 4), which sets an upper limit to the age
of approximately $10-20\,$Myr for an instantaneous burst (see e.g.,
Leitherer et al. 1999 and below). 
We did not include very
old stellar populations ($>10\,$Gyr) because these do not appear to
have a strong contribution to the optical light of 
local LIRGs and ULIRGs (see e.g., Chen
et al. 2009; Rodr\'{\i}guez Zaur\'{\i}n et al. 2009).
We assumed that both stellar
populations were formed in instantaneous bursts, which seems an
appropriate assumption since the physical scales of the selected
regions are a few hundred parsecs (see Table~3) and contain, at most,
a few H\,{\sc ii} regions, as shown in Paper~I. 

%The continuum spectrum of the combined two stellar populations is
%computed as: $\eta*f_\lambda{\rm (young)} + (1-\eta)*f_\lambda{\rm
%(evolved)}$, and thus $\eta$ represents the fraction in mass of young
%stars.

A priori, fits with $\chi^2_{\rm red} \leq 1.0$ should be considered
acceptable fits to the overall shape of the continuum (see discussion
in Tadhunter et al. 2005). However, the absolute value of $\chi^2_{\rm
red}$ is strongly dependent on the estimated errors. 
We found  that combinations 
with $\chi^2_{\rm red} > 0.2$ produced poor fits to the overall shape of the
continuum.  Moreover, for most regions 
we found acceptable solutions for $\chi^2_{\rm red} < 2 \times
\chi^2_{\rm min}$, where $\chi^2_{\rm min}$ was the minimum value of
$\chi^2_{\rm red}$ for a given region. Out of these solutions,  
we selected the best fitting models based on
a visual inspection of the fits to those absorption features with
relatively little emission line contamination. These include high
order Balmer lines, the CaII K $\lambda 3934$ line, the G-band
$\lambda 4305$, and the MgIb $\lambda 5173$ band. Finally, we rejected
solutions with ages of the young ionizing stellar populations older
than the upper limits (that is, before subtracting the stellar
continuum produced by the evolved stars) set by the EW(H$\alpha)_{\rm
em}$ values (see Section~3.2). 

\subsection{Modelling of the hydrogen recombination emission lines}

To model the properties of the hydrogen recombination emission lines,
we used the Starburst99 model (Leitherer et al.  1999) with the same
IMF and metallicity assumptions as above to generate the time
evolution of EW(H$\alpha)_{\rm em}$ (and also for H$\beta$) for an
instantaneous burst of star formation.  This model is better qualified
for the modelling of populations containing hot massive stars (see
V\'azquez \& Leitherer 2005 for details).  The EW of the H$\alpha$
emission line resulting from subtracting the modelled continuum
arising from the non-ionizing stellar population from the observed
spectra can be used to put further constraints on the age ($t_{\rm
  neb}$) of the
ionizing stellar populations. Additionally, we used the H$\alpha$/H$\beta$
emission line ratio measured after subtracting the stellar continuum
(from both the ionizing and non-ionizing stellar populations) to 
provide an independent estimate of the extinction to the gas
(E(B-V)$_{\rm neb}$). This gas
is ionized by the young stellar population assumed in the previous section.

\section{The maps of the EW(H$\alpha)_{\rm em}$  and optical line ratios}

The maps of EW(H$\alpha)_{\rm em}$ for our sample of LIRGs, covering
on average the central $\sim 4.7\,$kpc are shown in Figure~1. The map
of NGC~7771 covers the central $\sim 7.7\,{\rm kpc} \times 4.4\,{\rm
  kpc}$. To first order
the EW of the nebular Balmer emission lines can be used as indicators of the
age of the ionizing stellar populations. The values of 
EW(H$\alpha)_{\rm em}$ in our sample of LIRGs (see Figure~1), 
except for the nuclear region of IC~860 (Table~4),  
indicate ages of the young stellar populations
of between 5 and $\sim 10-20\,$Myr (see e.g., Leitherer et
al. 1999, and Section~3.1). 
For the majority of the LIRGs in this sample, as well as
for our VLT/VIMOS LIRGs of Rodr\'{\i}guez Zaur\'{\i}n et al. (2010b), 
the largest values of the EW of H$\alpha$ are not coincident with the
peak of the optical continuum  (the nucleus). However, it is important
to note that the  EW of the
Balmer nebular emission lines are also sensitive to 
the mass of the underlying non-ionizing population. In this sense, the EW of the
Balmer emission lines also provide an estimate of the ratio of the current star
formation rate compared with the averaged past star formation, that
is, the burst strength (see e.g., Kennicutt et al. 1987; 
Alonso-Herrero et al. 1996), whether
this refers to the integrated emission of a galaxy or to individual
regions within galaxies. In cases of small burst strengths, the observed
values of the EW only provide upper limits to the age of the current
star formation burst. Thus, the most
likely explanation for the smaller nuclear EW(H$\alpha)_{\rm em}$,
when compared to 
those of circumnuclear H\,{\sc ii} regions observed in some galaxies,
is a larger contribution 
from the underlying (more evolved) stellar population 
(see Kennicutt et al. 1989) and/or a slightly more evolved stellar
population.

The regions with the largest EW(H$\alpha)_{\rm em}$  show
values of the [N\,{\sc  ii}]$\lambda$6584/H$\alpha$ 
and [S\,{\sc ii}]$\lambda\lambda$6717,6731/H$\alpha$ line ratios (see
Figure~1) typical of H\,{\sc
  ii} regions in normal star-forming galaxies. In contrast, 
most nuclear regions in our LIRGs 
tend to show slightly larger lines
ratios than the H\,{\sc ii} regions of the same galaxy. This is
clearly seen in the  diagnostic
diagrams of Figure~2 where for 
each galaxy we plot the spatially resolved (on a spaxel by spaxel basis) line
ratios as a function of the observed value of EW(H$\alpha)_{\rm em}$ in
emission. This is in line with findings for the nuclei and H\,{\sc ii} regions
in normal star forming galaxies (see e.g., Kennicutt et
al. 1989). We note, however, that the line ratios of regions with low 
EW(H$\alpha)_{\rm em}$ will have the largest corrections for the
presence of stellar Balmer absorption lines as we shall see in Section~6.2.

The presence of extra-nuclear regions with enhanced values of the
[N\,{\sc  ii}]$\lambda$6584/H$\alpha$ 
and [S\,{\sc ii}]$\lambda\lambda$6717,6731/H$\alpha$ ratios  
relative to those of H\,{\sc ii } regions is again a
common property not only of the LIRGs studied here, but also of our
VLT/VIMOS sample of LIRGs (see Monreal-Ibero et al. 2010a) and our
WHT/INTEGRAL sample of ULIRGs (see
Garc\'{\i}a-Mar\'{\i}n et al. 2010).  In our sample
of LIRGs these regions tend to be associated with diffuse emission
rather than with high surface brightness H\,{\sc ii } regions, as
is also the case for  normal and starburst galaxies (Wang, Heckman, \&
Lehnert 1998). Moreover, for our VLT/VIMOS sample of LIRGs 
Monreal-Ibero et al. (2010a) found a correlation between the enhanced
optical ratios and increasing gas velocity dispersion 
in interacting and merger LIRGs, and this correlation is attributed to
the presence of shocks associated with the interaction processes. Our
PMAS sample is mostly composed of isolated galaxies and weakly
interacting galaxies (see Paper~I), and thus it is unlikely these
processes are responsible for the enhanced line ratios. However, since
the majority of the regions
with enhanced   [N\,{\sc  ii}]$\lambda$6584/H$\alpha$ 
and [S\,{\sc ii}]$\lambda\lambda$6717,6731/H$\alpha$ ratios in our sample are
observed in regions of relatively low EW(H$\alpha)_{\rm em}$, 
we will postpone the
discussion of this issue after the  line ratios are corrected for 
the presence of underlying Balmer stellar absorption features (see
Section~6.2).

\begin{figure}

\setcounter{figure}{5}
\includegraphics[width=6.5cm,angle=-90]{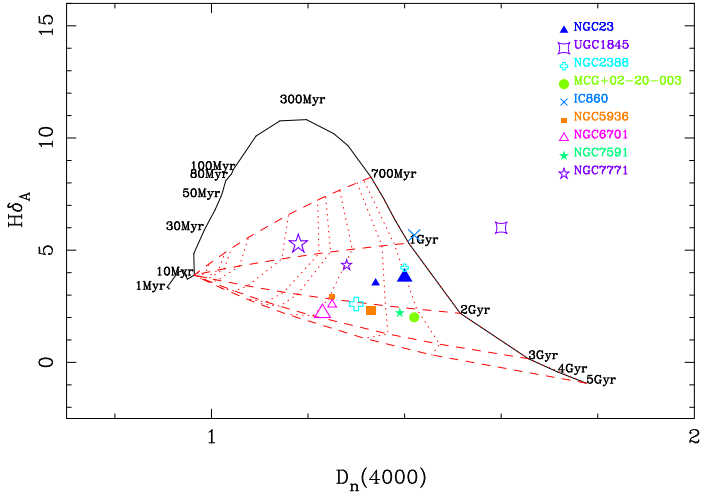}
\caption{${\rm D}_n(4000)$ vs. H$\delta_{\rm A}$ diagram. The small
  symbols are the measurements from the average spectrum of the
  regions with low EW(H$\alpha)_{\rm em}$ 
of each LIRG, whereas the large symbols are the
  measurements corresponding to the nuclear regions (see Table~4). 
For NGC~7591 we plot the values corresponding to the integrated
spectrum. 
Note that in the
  majority of nuclei the H$\delta_{\rm A}$ indices are lower limits due to
  possible contamination from the nebular H$\delta$ emission line. 
The black solid line is the time evolution (from 1\,Myr to 5\,Gyr) 
as predicted by the BC03
models using solar metallicity, a Salpeter IMF, and an instantaneous
burst of star formation. The dashed lines are combinations of
different evolved populations (ages 700\,Myr, 1\,Gyr, 2\,Gyr, 3\,Gyr, and
5\,Gyr) and a  young stellar population of 10\,Myr. The dotted
lines represent the fraction in mass of young stars with values of 
$0.001$, 0.002, 0.005, 0.008, 0.01, 0.02,
     0.05, 0.08, 0.1, 0.2, 0.5, and 0.8, from right to left.} 

\end{figure}

\begin{figure}

\setcounter{figure}{6}
\includegraphics[width=6.5cm,angle=-90]{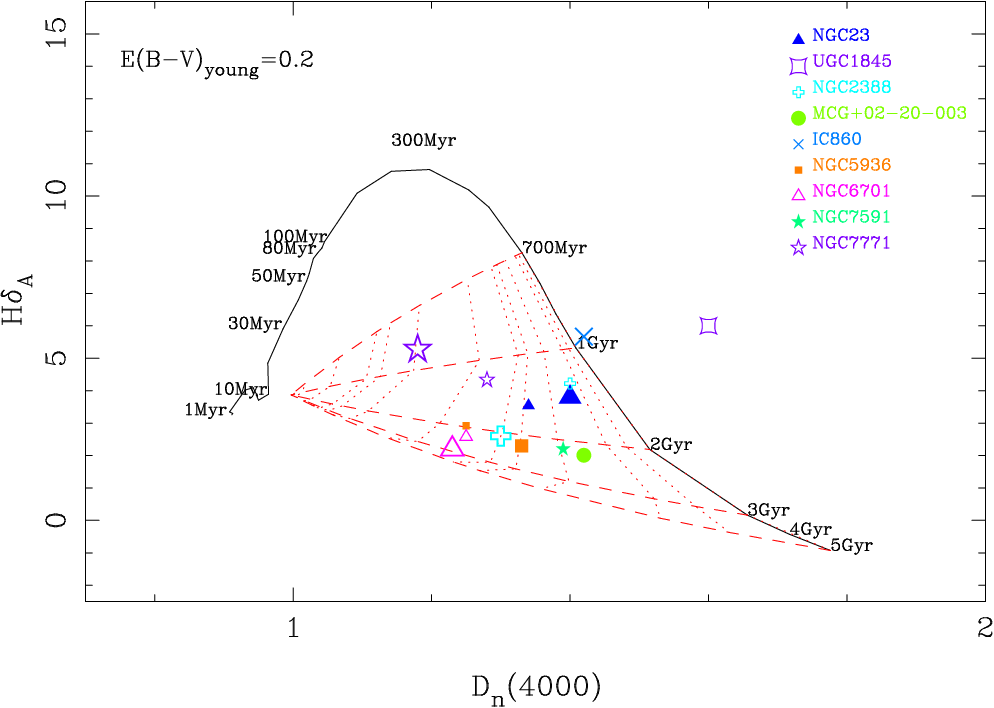}
\includegraphics[width=6.5cm,angle=-90]{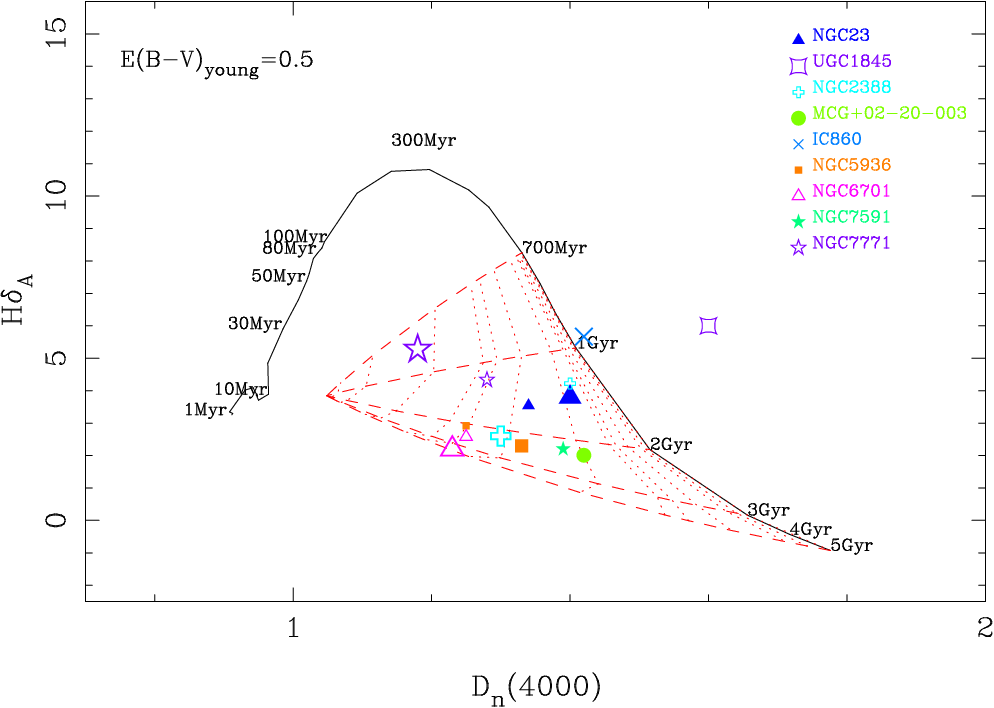}
\includegraphics[width=6.5cm,angle=-90]{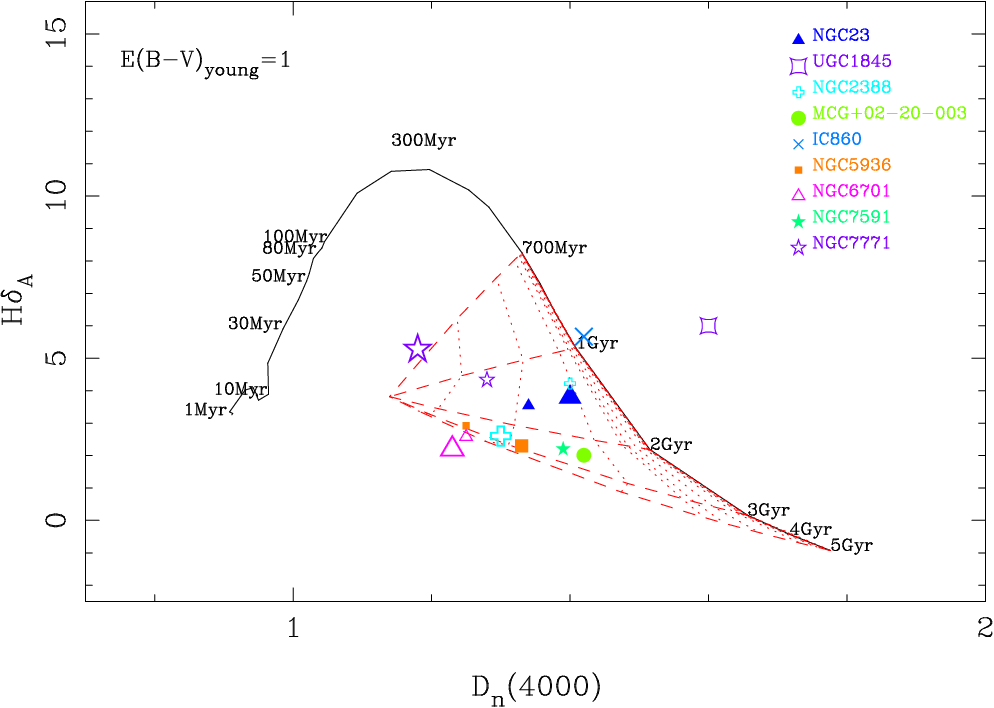}
\caption{Same as Figure~6, but showing the effects of reddening the
  young stellar population with ${\rm E(B-V)_{\rm young}}=0.2$ (upper
  panel), ${\rm
    E(B-V)}_{\rm young}=0.5$ (middle panel), and 
${\rm E(B-V)}_{\rm young}=1$ (bottom panel),
 and using the Calzetti et al. (2000)
  extinction law.} 

\end{figure}

\section{The Stellar Populations of local LIRGs}

\subsection{Results using the 4000\AA \, break and Balmer line
  absorption features} 

Kauffmann et al. (2003a) used a method 
based on the 4000\AA-break and the Balmer H$\delta$ feature in absorption
to constrain the mean (light-weighted) age
of the stellar population of a galaxy and the mass fraction formed in
recent bursts of star formation. Therefore, this method 
provides  information about the light-averaged properties of the stellar
populations.  From the stellar continuum spectra generated with BC03, 
we measured  the ${\rm D}_n(4000)$  and the
H$\delta_{\rm A}$ indices (see Section~2.3, and Table~2 for
definitions of the indices). For an
instantaneous burst  of star formation 
the ${\rm D}_n(4000)$ index increases monotonically with the age of
the stellar population, while the depth of the H$\delta$ line absorption
feature increases until about $300-400\,$Myr after the burst, and then decreases
again (black line in Figure~6). 
Thus the presence of high order Balmer lines in absorption is 
usually interpreted as the signature of an intermediate age ($100\,{\rm
  Myr}-1\,{\rm Gyr}$) stellar population. 
The evolution of these two indices is shown in Figure~6 using the BC03
models and is similar to the results of  Kauffmann et
al. (2003a) and Gonz\'alez-Delgado et al. (2005).

\begin{table*}
\begin{center}
\caption{Ranges of ages, extinctions and light fractions from young
  stars as derived from the stellar continuum modelling.}
\begin{tabular}{lccccccc}
\hline
\hline
Galaxy  &  Region & $t_{\rm evolved}$ & E(B-V)$_{\rm evolved}$ & $t_{\rm young}$ & E(B-V)$_{\rm young}$ 
& $f_{\rm NB}$  \\   
& & Gyr &   &  Myr & & \% \\
\hline
\hline
NGC~23   & Nucleus & $0.7-5$ & $0-0.4$ & $\le 10$ & $0.4-1.2$ & $15-56$  \\
         & Low-EW(H$\alpha)_{\rm em}$ & $0.7-1$ & $0-0.4$ & $\le 20$
         & $0.4-1.0$ & $20-55$\\
         & HII-1    & $2-10$ & $0-0.4$ & $\le6$  & $0.4-0.9$ & $30-60$  \\
         & HII-2    & $ 0.7-1$ & $0.2-0.4$ & $4-7$  & $0.4-1.0$ & $20-42$ \\
         & HII-3    & $ 0.7-1$ & 0.4 & $<7$  & $0.4$ & $44-57$ \\
         & HII-4    &$0.5-1$ & $0-0.4$ & $<8$ & $0.2-0.6$ &$35-55$ \\
\\
NGC~2388 & Nucleus & $2-5$ & $0.4$ & $6-7$ & $0.9- 1.0$ & $53-65$ \\
         & Low-EW(H$\alpha)_{\rm em}$ & 1 & $0.4$ & $7-20$ &
         $0.7-0.9$ &
         $26-40$ \\
         & HII-1    & $0.7-1$ & $0-0.4$ & $6-7$ & $0.8-1.0$ & $63-74$\\
         & HII-2    & $1$ & $0.4$ & $6$  & 1.3 & $40$ \\
\\
MCG~+02-20-003 & Nucleus & $0.7- 2$ & $0.2-0.4$ & $\le 6$ & $1.5-1.8$ & $43-73$ \\
\\
IC~860   & Nucleus & $0.5-1$ & $0.2-0.4$ &  $\le 20$ & $0.3-1.4$ &$18-53$\\
\\
NGC~5936 & Nucleus & $1-5$ & $0-0.2$& $5-7$ & $1.0-1.1$ & $76-84$\\
         & Low-EW(H$\alpha)_{\rm em}$ & $0.5-10$ & $0-0.4$ & $7-10$ & $0.3-0.8$ & $52-71$\\
\\
NGC~6701 & Nucleus & $0.7-5$ & $0-0.4$ & $\le 7$ & $0.4-0.5$ & $43-68$ \\
         & Low-EW(H$\alpha)_{\rm em}$ & $0.7-2$ & $0-0.2$ & $7-10$ & $0.2-0.4$ & $52-71$\\
\\
NGC~7591 & Nucleus & $2-10$ & $0-0.4$ & $\le 7$ & $0.4-1.5$ & $20-63$ \\ 
\\
NGC~7771 & Nucleus &  $0.3-0.5$ & 0.2 &$4-7$ & $0.3-0.6$ & $37-53$ \\
         & Nucleus2& $2-5$ & 0.4& $8-20$ & $0.4-0.6$ & $48-63$  \\
         & Low-EW(H$\alpha)_{\rm em}$ & $0.5-0.7$   & $0.4$ & $6-10$ & $0.4-0.6$ & $45-57$\\
         & HII-1    & $0.7-10$ & $0-0.4$ & $\le 6$ & $0.4-0.9$ &
         $50-75$ \\
         & HII-2    & $2-10$ & $0-0.4$ & $\le 7$ & $0.4-1.1$ &
         $30-70$ \\
         & HII-3    & $2-10$ & $0-0.4$ & $\le 8$ & $0.3-1.1$ &
         $25-58$ \\ 

\hline
\end{tabular} 
\end{center}
\end{table*}

\begin{table*}
\begin{center}
\caption{Examples of stellar continuum and nebular modelling.}
\begin{tabular}{lccccccccc}
\hline
\hline
Galaxy  &  Region & \multicolumn{5}{c}{Stellar continuum parameters} &
 \multicolumn{2}{c}{Nebular fit}\\
& & $t_{\rm evolved}$ & E(B-V)$_{\rm evolved}$ & $t_{\rm young}$ & E(B-V)$_{\rm young}$ 
& $f_{\rm NB}$  & $t_{\rm neb}$ & E(B-V)$_{\rm neb}$\\   
& & Gyr &   &  Myr & & \% &  Myr\\
\hline
\hline
NGC~23   & Nucleus & 2 & 0 & 7 & 0.7 & 31 & 6.9 &
0.6  \\
NGC~23         & HII-4    & 1 & 0.2 & 5 & 0.2 & 44  & 5.7 & 0.5\\
NGC~2388 & HII-1    & 1 & 0 & 6 &1.0 &  
      80   & 6.0 & 1.0\\
IC~860   & Nucleus & 1  & 0.4 &  9 & 0.3 & 27  & 8.3 &$-$\\
NGC~6701         & Low-EW(H$\alpha)_{\rm em}$ & 2 & 0 & 8 & 0.2 & 58 &
7.6 & 0.3\\
NGC~7771 & Nucleus &  0.5 & 0.2 & 7 & 0.3 & 40  & 6.1
& 0.5 \\
\hline
\end{tabular} 
\end{center}

\end{table*}

As discussed in Section~3, we need a combination of (at least)
two stellar populations to reproduce the observed properties of LIRGs,
including the 
${\rm D}_n(4000)$ and H$\delta_{\rm A}$ indices. This is clear from
Figure~6, where a single stellar population formed in an instantaneous
burst (black line) does not
reproduce the observed indices of the nuclei and regions of 
low-EW(H$\alpha)_{\rm em}$ in
our sample of LIRGs. A model with a constant
star formation rate predicts   ${\rm D}_n(4000)<1.2$ for all ages
(see figure~21 of Caputi
et al. 2008), and thus it is not appropriate for our galaxies either.

Figure~6 shows the result of
combining two stellar populations. In this diagram, the choice of the age
of the ionizing stellar population is not critical
because ${\rm D}_n(4000)$ and H$\delta_{\rm A}$ do not
vary much during 
approximately the first $\sim 10-20\,$Myr of the evolution of a single 
stellar population formed in an instantaneous burst (see e.g.,
Gonz\'alez-Delgado et al. 2005). 
Thus with this kind of  diagrams we cannot constrain 
the age of the youngest stellar populations in LIRGs. As we 
shall see in Section~5.2, the combined modelling of the stellar continuum and
the nebular emission lines puts strong constraints on the properties
of the ionizing stellar population. In Figure~6 we then combined a
10\,Myr population 
with evolved stellar populations with ages of 700\,Myr, 1\,Gyr,
2\,Gyr, 3\,Gyr, and 5\,Gyr. The ages of the evolved stellar population
are based on the location in this diagram of the
observed values for the nuclei and regions of low-EW(H$\alpha)_{\rm
  em}$ in our sample of
LIRGs. A scenario where the evolved stellar
population was formed in an instantaneous burst and the current star
formation is taking place at a constant rate (see Sarzi et al. 2007
for ${\rm D}_n(4000)$ vs. H$\delta_{\rm A}$ diagrams generated under
this assumption) would underpredict the strength of
the H$\delta$ absorption feature for most of the selected regions in our LIRGs, 
as was  the case for the star-forming regions in nuclear rings in the
sample of Sarzi et al. (2007).

From Figure~6 it is clear that the main effect of combining a young
population and
an evolved $1-2\,$Gyr stellar population  is the change in the observed
value of the ${\rm D}_n(4000)$ index, which becomes smaller as the
fraction in mass of the young stellar population increases. For a
combination with a ¨younger¨
evolved stellar population ($\sim 700\,$Myr) the effect is observed in both
indices, as is the case for older evolved populations
($>2\,$Gyr). Based on this figure and taking into account that the
measured H$\delta_{\rm A}$ are lower limits, the age of the evolved
stellar population is between 1 and 
3\,Gyr for most of the galaxies in our
sample. The only exception is  NGC~7771 that
appears to show the presence of an intermediate age (700\,Myr-1\,Gyr)
stellar population. We also show in Figure~6 a range of mass fractions
for the young stellar population, which in general are relatively
small for the nuclei and regions of low-EW(H$\alpha)_{\rm em}$.

In Figure~7 we show the effects on the ${\rm D}_n(4000)$ vs. H$\delta_{\rm A}$
diagram 
from the combination of a reddened young stellar population and an
unreddened evolved population. Figure~7 clearly demonstrates that not
accounting for the extinction 
to the young stellar population would make us underestimate its mass
fraction as well as underestimate the age of the
evolved stellar population.  However, it is also apparent from
Figures~6 and 7, that it is not possible to 
disentangle the effects of extinction, ages of the stellar
populations, and mass contributions from this kind of diagrams
alone.

\begin{figure*}
\setcounter{figure}{7}

\hspace{1cm}
\includegraphics[width=6.7cm,angle=-90]{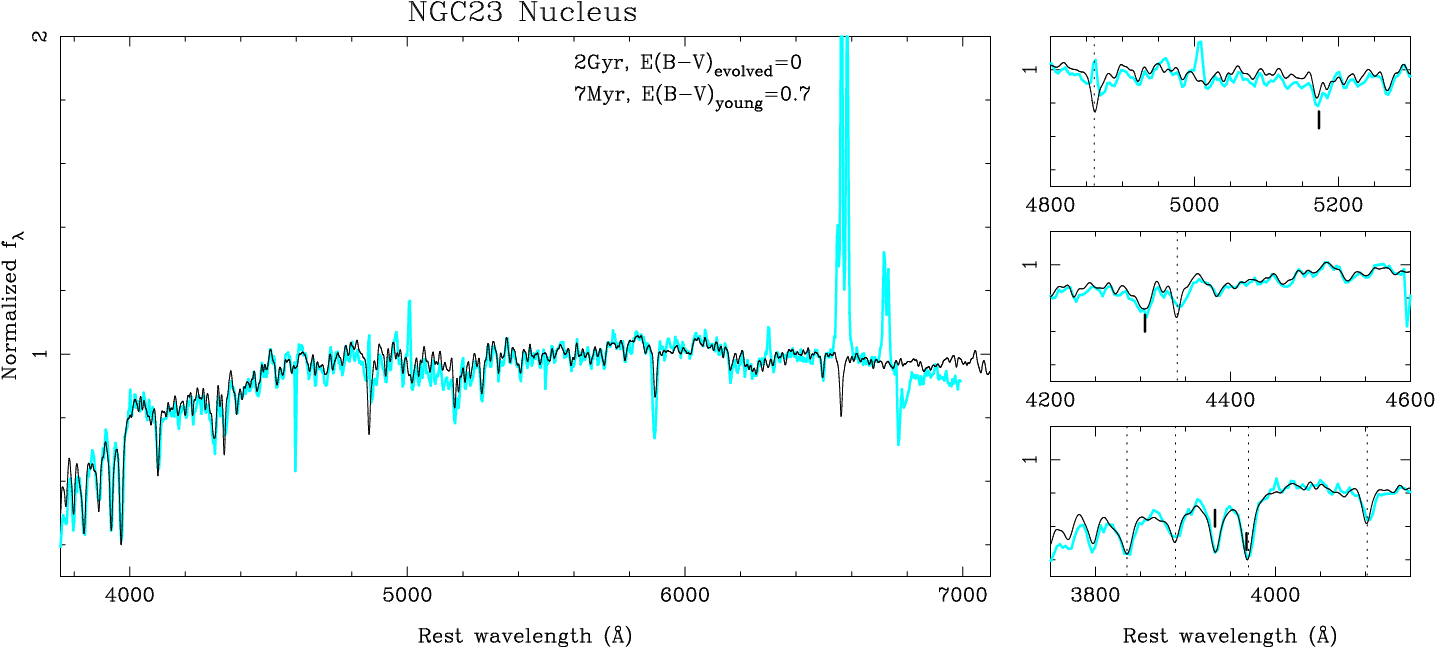}

\vspace{0.5cm}
\hspace{1cm}
\includegraphics[width=6.7cm,angle=-90]{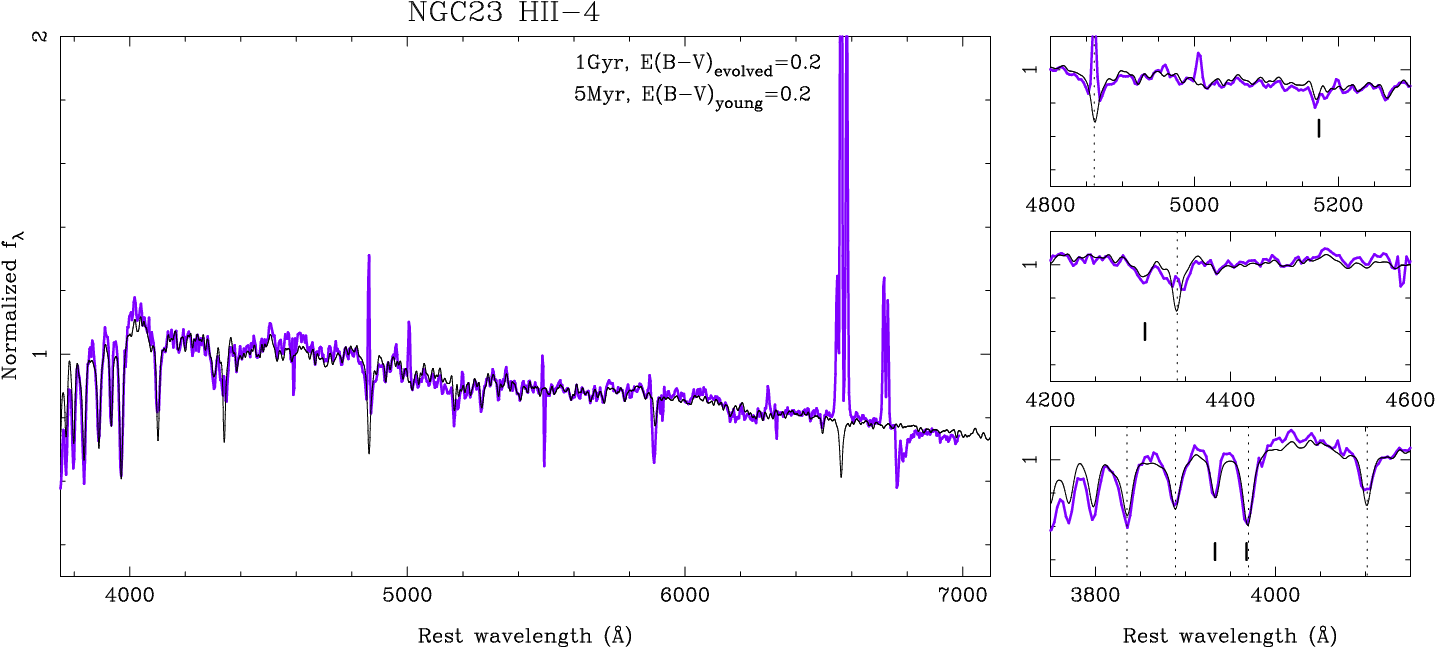}

\vspace{0.5cm}
\hspace{1cm}
\includegraphics[width=6.7cm,angle=-90]{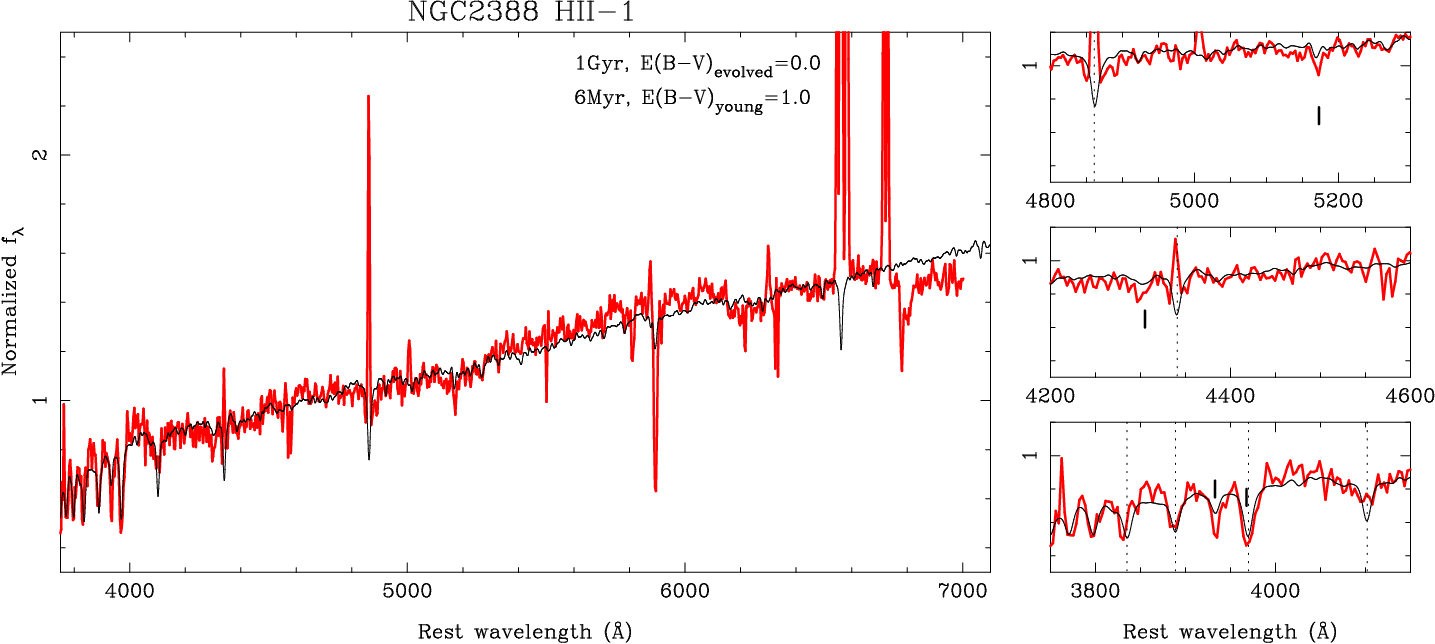}

\caption{(a) Examples of fits to the stellar continuum of regions in
  our sample of galaxies. The bluest rest-frame wavelength used for the
  modelling is 3800\,\AA. For each region, the left panel shows the
  observed spectrum (thick color line) and the model spectrum (thin
  black line) in arbitrary units. The spectra are normalized to unity
  at a wavelength within the normalizing bin ($4400-4800\,$\AA). 
The model parameters for each region
  are given at the top right of each plot.
The right panels are blow-ups of some spectral regions of interest.
We mark  the high order Balmer lines as well as  H$\delta$, H$\gamma$, and, 
H$\beta$ as dotted lines. The Ca\,{\sc
    ii} H and Ca\,{\sc ii} K lines (lower
right panel), the 
G-band (middle right panel), and the Mg\,{\sc i}b band (upper
right panel) are marked as thick solid lines. The spectral resolution
of the BC03 models has been slightly degraded to
match approximately that of our PMAS spectra.} 

\end{figure*}

\begin{figure*}
\setcounter{figure}{7}

\hspace{1cm}
\includegraphics[width=6.7cm,angle=-90]{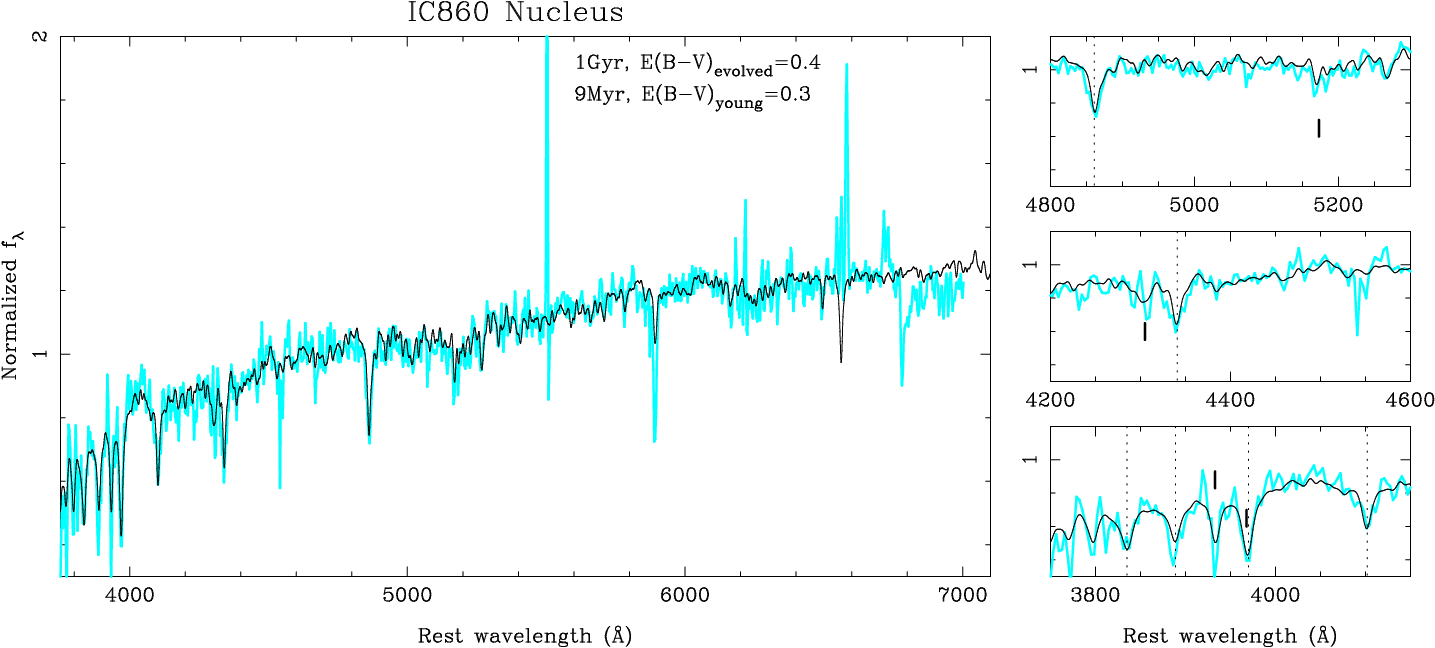}

\vspace{0.5cm}
\hspace{1cm}
\includegraphics[width=6.7cm,angle=-90]{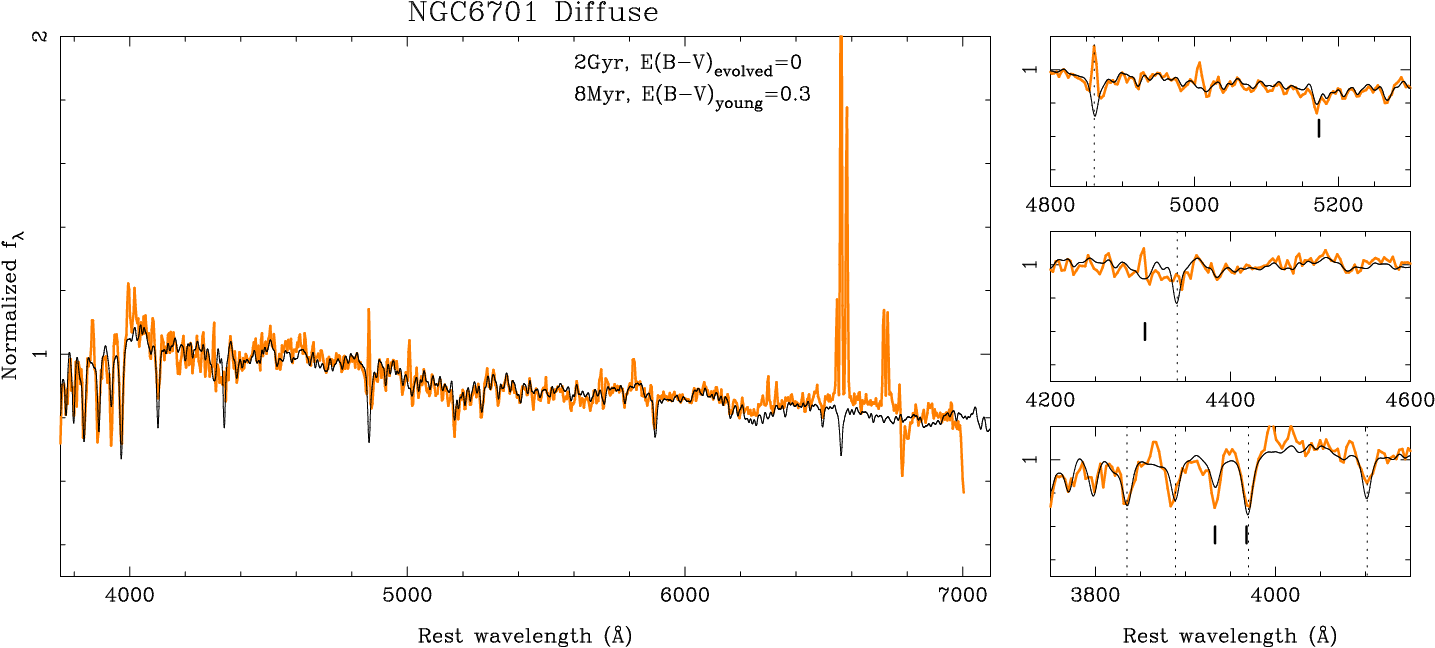}

\vspace{0.5cm}
\hspace{1cm}
\includegraphics[width=6.7cm,angle=-90]{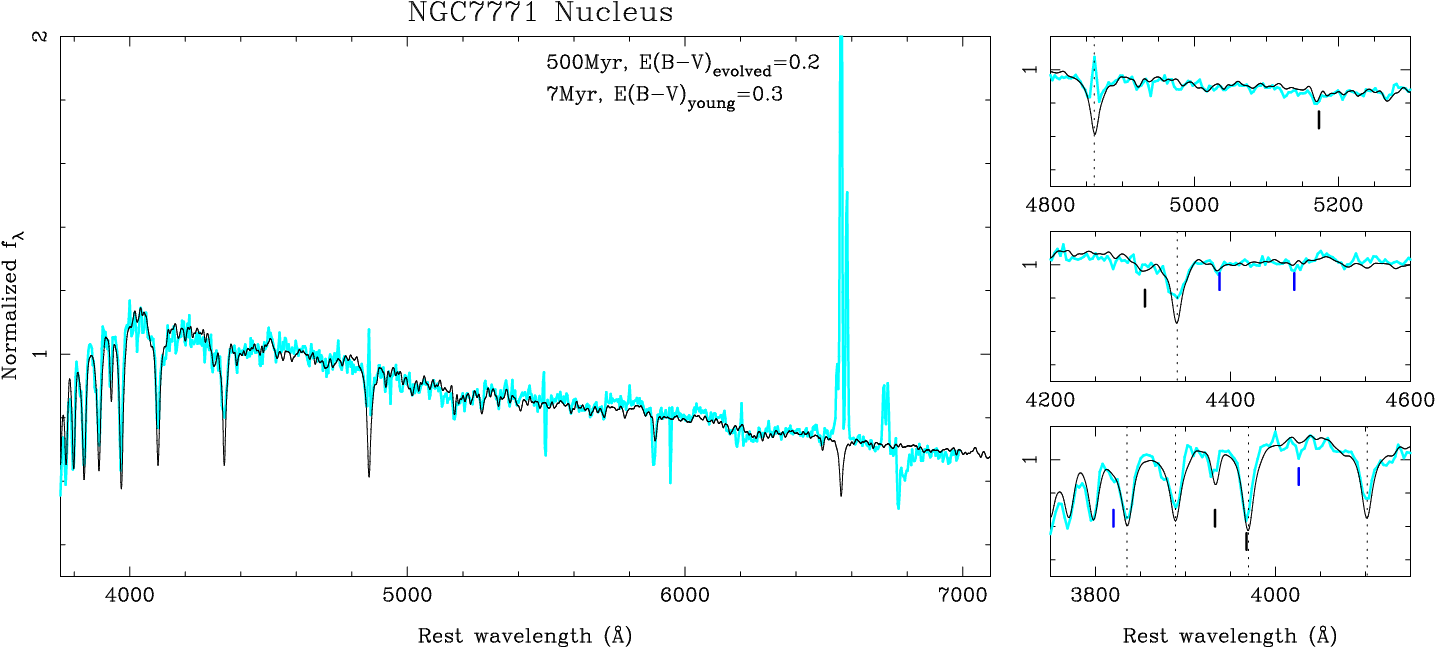}

\caption{(b). Same as Figure~8a. For the nuclear region of NGC~7771 
we also show the positions of the
  He\,{\sc i} features at 3820, 4026, 4388, and 4471\,\AA.} 

\end{figure*}

\subsection{Results of the modelling of the stellar continuum and
  nebular emission}

As explained in Section~2.3, we selected a number of regions in our
sample of LIRGs for the study of their stellar populations. These
include the nuclear regions, the average spectra of regions 
of low-EW(H$\alpha)_{\rm em}$,
as well as a number of bright H\,{\sc ii} regions in NGC~23, NGC~2388,
and NGC~7771. We excluded from this analysis three galaxies for
various reasons. The nuclear optical spectra of MCG~+12-02-001 appears
to be completely dominated by a young ionizing stellar population
(that is, we do not see any evidence for the presence of absorption
features), and thus it was not possible to constrain the evolved
stellar population. The signal-to-noise ratio of the nuclear spectrum
of UGC~1845 did not allow us to constrain the properties of the
stellar populations. Finally we excluded NGC~7469 because of the
possible contamination of the optical spectra from the AGN non-stellar
continuum. We refer the reader to D\'{\i}az-Santos et al. (2007) for a
detailed study of the stellar populations of the ring of star
formation of NGC~7469 using high angular resolution {\it HST}
photometric data.

Table~5 summarizes for each galaxy and region the acceptable ranges of
ages and extinctions for the evolved ($t_{\rm evolved}$ and
E(B-V)$_{\rm evolved}$) and ionizing ($t_{\rm young}$ and
E(B-V)$_{\rm young}$) stellar populations
derived from the stellar continuum fit  as explained in Section~3.1.
We also list in this table the fraction of the light in the
normalising bin emitted by the ionizing (young) stellar component
($f_{\rm NB}$),  for the models that produce adequate fits.
Because of the different shapes of the extracted spectra it was not
always possible to select the same normalizing bin. However, the
normalizing bin was always selected to be located within the
wavelength range of $4400-4800\,$\AA\,and usually spanning $\sim$
100\AA.  
Figure~8 presents a few examples of fits to the stellar continuum for a
number of regions in our sample.

For the majority of the regions studied here except for
the optical nucleus of NGC~7771 (see Section~5.2.1), we find that stellar
populations with ages between 100\,Myr and 500\,Myr do not make a
strong contribution to the optical light. Ages for the
evolved stellar populations of between 0.7 and 5\,Gyr (or even 10\,Gyr
in some cases) provided reasonable fits to the optical continua.  
These ages are within the range of ages derived for 
massive spiral galaxies (Gallazzi et al. 2005 and references therein).
Another noteworthy result is that the 
regions with low-EW(H$\alpha)_{\rm em}$ 
in the galaxies tend to have slightly older ionizing stellar
populations and less extinction than other regions for the same
galaxy. The ages of the evolved stellar populations agree with our
findings in the previous section using the ${\rm D}_n(4000)$ and
H$\delta_{\rm A}$ diagram, keeping in mind that 
H$\delta_{\rm A}$ may only provide a lower limit to the age of the
evolved stars for some of the regions.  
We finally note that for models including evolved
  plus young stellar 
  components  it is particularly hard to distinguish between ages in
  the range $2-10\,$Gyr. Therefore, in some cases we are not able to put strong
  constraints on the age of the stellar populations for ages older
  than $\sim 1-2\,$Gyr.

As can be seen from Table~5 the  ionizing stellar
populations always contribute a minimum of $f_{\rm NB}
\geq 15\%$ to the optical emission in {\it all} the modelled
spectra, and they dominate the emission ($f_{\rm NB} \geq 50\%$) in, 
at least, 7 of the 23 spectra modeled. On the other hand, 
in the nuclear regions of NGC~23 and IC~860, and
NGC~7771, the evolved stellar populations may be the main contributors
to the optical light (see also Table~6). Our
spectra do not sample the near-UV spectral region, which is 
important for constraining the properties of the ionizing
stellar populations. Therefore, the ages of such stellar populations
are not well constrained when fitting  the stellar continuum
alone. 

We were able to put tighter constraints on the young stellar
populations  when we included the nebular fitting. 
Table~6  gives the parameters of the stellar continuum and nebular
fits for those regions shown in Figure~8. The nebular
ages and extinctions ($t_{\rm neb}$ and
E(B-V)$_{\rm neb}$) were derived as explained in Section~3.2, for an
acceptable combination of stellar populations close or at the minimum value of
$\chi^2_{\rm red}$. We find ages of the ionizing
stellar populations from the EW of H$\alpha$ 
of between 5.6 and 8.8\,Myr. The extinctions to the
ionizing stellar stars range  between ${\rm E(B-V)}=0.2$
and ${\rm E(B-V)}=1.8$ (Tables~5 and 6).  
The extinctions to
the ionizing stellar populations are always significantly greater than
those to the evolved stars, and generally consistent with 
those derived to the young stars from the stellar continuum
modelling.

Our results are, in a broad brush sense, consistent with
those of Rodr\'{\i}guez Zaur\'{\i}n et al. (2009) for their sample of
ULIRGs. That is, the optical spectra can be modeled using a combination of
an evolved plus a young stellar population. However, 
the extracted spectra of our LIRGs show, in most cases,
deeper G and Mg\,{\sc i}b bands than those of the ULIRGs of the Rodr\'{\i}guez
Zaur\'{\i}n et al. (2009) sample. This suggests that the evolved
stellar populations 
are somewhat older for the LIRGs in our sample. In fact, 
Rodr\'{\i}guez Zaur\'{\i}n et al. (2009) found adequate fits for their
ULIRGs which included evolved stellar populations of $0.3 - 0.5\,$Gyr, 
while this was rarely the case for our sample of LIRGs. 
The stellar populations dominating the optical light
of ULIRGs could be the result of enhanced
star formation coinciding with the first pass of the merging nuclei,
along with a further, more 
intense, episode  of star formation occurring as the nuclei finally merge
together (Rodr\'{\i}guez Zaur\'{\i}n et
al. 2010a). Our sample of LIRGs on the other hand, is
mostly composed of relatively isolated spiral galaxies and weakly interacting
galaxies (see Paper~I), with moderate IR luminosities (see Section~2.1). 
Given the relatively small fraction of
strongly interacting/merger systems in our sample compared to the 
Rodr\'{\i}guez Zaur\'{\i}n et al. (2009) ULIRG sample, it may not be  
unexpected that intermediate-age ($\sim 100-500\,$Myr)
stellar populations do 
not dominate the optical light of our LIRGs. The only exception in
our sample are the central regions of 
NGC~7771 (see discussion in Section~5.2.2). 
 Moderate luminosity spiral-like LIRGs may be 
 constantly forming stars and may have
not undergone a major burst of star formation in the last $1-2\,$Gyr,
as is the case of {\it normal} spiral galaxies
(Kauffmann et al. 2003b).  Our results are also in accord with the
observational findings of Poggianti \& Wu (2000) that most isolated
systems in their sample of IR-bright galaxies showed on average more
moderate Balmer absorption features than the interacting systems. The
effects of possible minor mergers are likely to be difficult to evaluate using
the stellar populations as models and observations suggest that it is
the satellite galaxy rather than the primary galaxy that is more
susceptible to enhanced star formation (Woods \& Geller 2007; Cox et
al. 2008).

\subsubsection{The central regions of NGC~7771}

There is dynamical evidence that NGC~7771 is weakly interacting with
NGC~7770 (Keel 1993) and is located in a group of galaxies. It is then
possible that the interaction process resulted in a strong burst of
star formation in the past. For instance,  
the ring in this galaxy, clearly detected in our Pa$\alpha$ images
(see Paper~I), 
appears to have a complex star formation history with
evidence for multiple generations of stars (Davies, Alonso-Herrero, \&
Ward 1997; Smith et al. 1999; Reunanen et al. 2000). 

The optical nucleus of NGC~7771 is the only region in our sample of
LIRGs for which adequate fits were obtained with a large
contribution to the optical light from an intermediate-age ($300-500\,$Myr)
stellar population (see Table~5 and Figure~8). The fit to the average spectrum of 
regions with low
 EW(H$\alpha)_{\rm em}$ also required an evolved
stellar population of $500-700\,$Myr. 
In the case of the nuclear spectrum there is evidence for an additional
stellar population based on the presence of He\,{\sc i} absorption
features at various wavelengths, which are indicated in
Figure~8.b. These absorption lines are strongest for stellar
populations with ages in the range $20-50 \,$Myr (Gonz\'alez-Delgado
et al. 1999; 2005), and are not observed in stellar populations older
than 100\,Myr (the lifetime of B stars). The presence of this
important population of non-ionizing stars dominated by B stars was
already infered by Davies et al. (1997). Given the clear evidence for
the presence of B stars, for  this region we also tried combinations
including stellar populations of $30-80\,$Myr for the young
component. We found 
that a combination with a dominant ($f_{\rm NB} =
70-80$\%), low reddening ($E(B-V)_{\rm young} \sim$ 0.2) stellar population of
$40-50\,$Myr plus an unreddened evolved stellar population of few
Gyr provided an acceptable fit. We note that with this combination the
stellar population responsible for ionizing the gas is not accounted
for. 
Therefore, in this particular case, it is likely that models
including a larger number of stellar components (at least three) would
be more adequate. We again emphasize that the nuclear region of
NGC~7771 is the only region where we find a clear evidence for these
He\,{\sc i} features. 
In this respect, it is not clear if the optical nucleus of this galaxy may be a special region in
this galaxy (see Davies et al. 1997 for a discussion on this
issue). It is important to recall at this point that  
the optical nucleus of NGC~7771 is probably not the true
nucleus of the galaxy, as it does not coincide with the center of the
bright ring of star formation, the peak of near-IR emission (see
Paper~I), or even with the region with the largest value of
EW(H$\alpha)_{\rm em}$ (see Figure~1.j).

%{\bf Finally, the modeling results obtained for this, and the other
%galaxies and regions modeled emphasize the ``degeneracy'' inherent to
%this type of studies and highlight the dangers of concentrating just
%on the best mathematical solution in some cases i.e., the one
%providing the best reduced $\chi^{2}$ value.}

\begin{figure}

\setcounter{figure}{8}
\includegraphics[width=6.5cm,angle=-90]{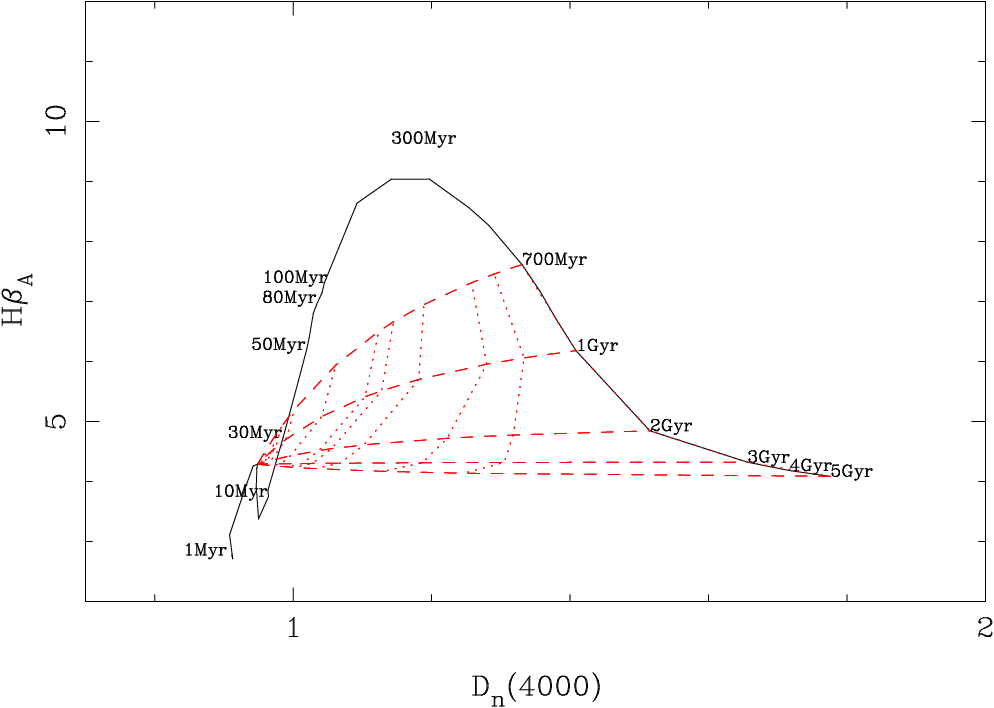}
\includegraphics[width=6.5cm,angle=-90]{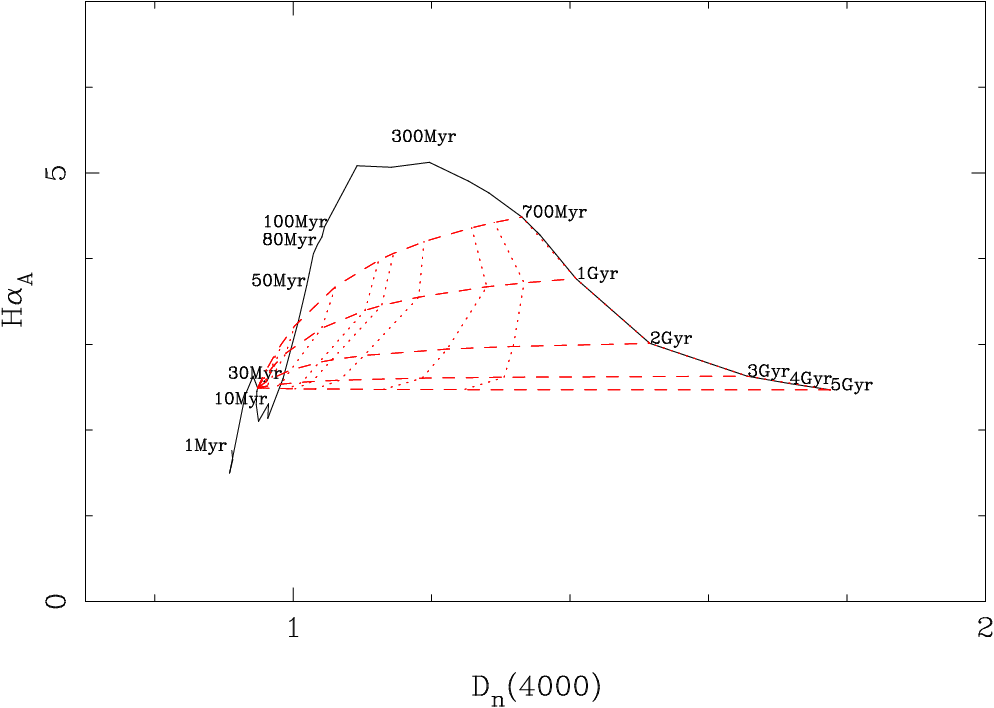}
\caption{Theoretical ${\rm D}_n(4000)$ vs. H$\beta_{\rm A}$  (upper panel)
  and ${\rm D}_n(4000)$ vs. H$\alpha_{\rm A}$ (lower panel) diagrams. 
The black solid line is
  the time evolution (from 1\,Myr to 5\,Gyr)  
as predicted by the BC03
models using solar metallicity, a Salpeter IMF, and an instantaneous
burst of star formation. The dashed lines are combinations of
different evolved populations (ages 700\,Myr, 1\,Gyr, 2\,Gyr, 3\,Gyr, and
5\,Gyr) and a  young ionizing stellar population of 6\,Myr. The dotted
lines represent the contribution in mass of the two stellar
populations, as in Figure~6.
} 

\end{figure} 

\subsection{Predictions for the H$\beta$ and H$\alpha$ stellar absorption 
features}

To study in detail the optical line ratios and in particular those of
the diffuse regions, we need to correct for the presence of 
H$\beta$ and H$\alpha$ stellar absorption features. 
However, given the strong star formation activity in our sample, it is
difficult to measure reliably these  absorption features because of 
the strong contamination produced by the Balmer recombination lines in
emission. An alternative approach is to  generate
theoretical diagrams of ${\rm D}_n(4000)$ vs. H$\alpha_{\rm A}$ and 
${\rm D}_n(4000)$ vs. H$\beta_{\rm A}$   measured from the spectra
  generated from the combination of the BC03 young and evolved stellar
  populations as explained in Section~3.1.  
 The H$\beta_{\rm A}$ and H$\alpha_{\rm A}$
  indices for the stellar absorption features 
are defined in a similar way to that used for the H$\delta$
  absorption feature (see Section~2.3), and thus have a positive
  value. The line windows,   
and the blue and red pseudo-continuum windows  
defined by Gonz\'alez Delgado et al. (2005) are given in Table~1.
 For the ages of the evolved stellar population we used those
  derived in Section~5.2. 
For the ionizing stellar
population we chose an age of 6\,Myr, which is representative of our
sample of LIRGs (Tables~5 and 6), 
although the results are not strongly dependent of
chosen age of this population. For the  mass
fractions we used the same values as those shown in Figure~6.

From the 
${\rm D}_n(4000)$ vs. H$\beta_{\rm A}$ diagram (Figure~9, upper
panel), we can see that for ages of the evolved stellar population of
between 1 and 5\,Gyr, the predicted average value of the
  H$\beta_{\rm A}$ index is
$\sim 5 \pm 1\,$\AA. This value is almost independent of the mass fraction
in young stars, except for cases where the stellar mass is dominated
by the contribution from ionizing stars. For NGC~7771, which is the
clearest case in our sample for the presence of an intermediate age
stellar population, the predicted value would be H$\beta_{\rm A} \sim
7\pm 1\,$\AA. In this case, the predicted value is more sensitive to the
mass in young stars and the age of the evolved stellar population. 
For comparison, the measurements of H$\beta_{\rm A}$
reported by Kim et al. (1995) for the central 2\,kpc of  
the galaxies in common with our
sample are 3 and 6\,\AA, although Kim et al. (1995) pointed out these
values were {\it admittedly subjective} because of the method they used
for fitting their data. 

Figure~9 (lower panel) shows the predictions for the 
${\rm D}_n(4000)$ vs. H$\alpha_{\rm
  A}$ diagram. In the case of  H$\alpha_{\rm A}$ there is a very small
dependence of the predicted  value  with the mass fraction in
young stars. For the majority of our LIRGs with ages of the stellar
population $1-5\,$Gyr the average 
predicted value for the index 
would be H$\alpha_{\rm A} \sim 3\pm 0.5 \,$\AA, whereas for
NGC~7771 we would predict H$\alpha_{\rm A} \sim
4\pm 0.5\,$\AA. For comparison,  
 Moustakas \& Kennicutt (2006b) found an 
average H$\alpha$ stellar absorption correction of $2.8 \pm 0.4\,$\AA \, 
for the integrated spectra of a sample of nearby star forming galaxies.

\section{Excitation Conditions in local LIRGs}

\subsection{Spatially resolved diagnostic diagrams}

Diagnostic diagrams using bright optical emission lines
(Balwin et al. 1981 and Veilleux \& Osterbrock 1987) are
useful for differentiating between the various sources of 
excitation  of the gas
in the nuclei of galaxies and their integrated
emission. With optical IFS data we can additionally study the  distribution of
the ionization structure of spatially resolved regions (that is, on a
spaxel-by-spaxel basis) of 
nearby galaxies (e.g., 
Garc\'{\i}a-Mar\'{\i}n et al. 2006; Garc\'{\i}a-Lorenzo et
al. 2008; Blanc et al. 2009; Stoklasov\'a et al. 2009; Monreal-Ibero
et al. 2010b; Garc\'{\i}a-Mar\'{\i}n et al. 2010), 
as well as of H\,{\sc ii} regions in
nearby galaxies (e.g., Rela\~no et al 2010). 
The diagnostic diagrams for the
spatially resolved measurements for each of the LIRGs in our sample
are shown in Figure~2. The measurements for the individual spaxels are
color coded according to the observed value of the EW(H$\alpha)_{\rm
  em}$ in emission. In 
these diagrams we plotted the  
empirical and theoretical boundaries derived by Kauffmann et
al. (2003c) and Kewley et al. (2001). These boundaries are shown for
reference as they may provide clues about the dominant excitation
mechanism: ionization by young stars, shocks, or AGN ionization.

\begin{figure*}

\setcounter{figure}{9}
\includegraphics[width=7cm,angle=-90]{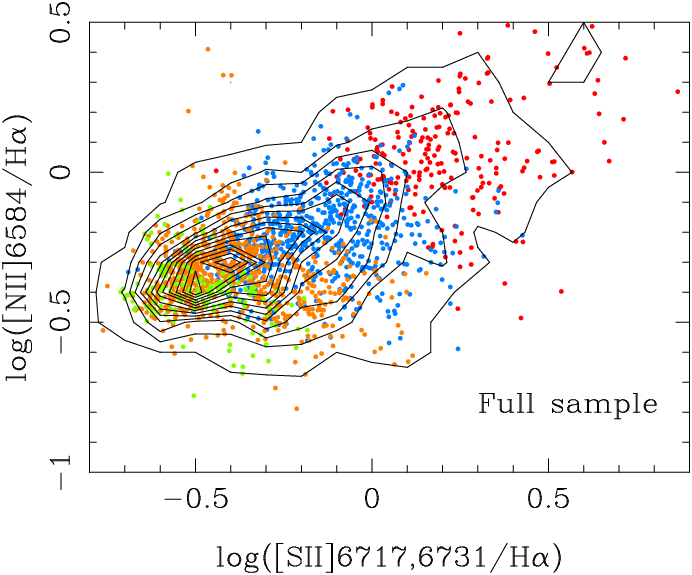}
\hspace{0.5cm}
\includegraphics[width=7cm,angle=-90]{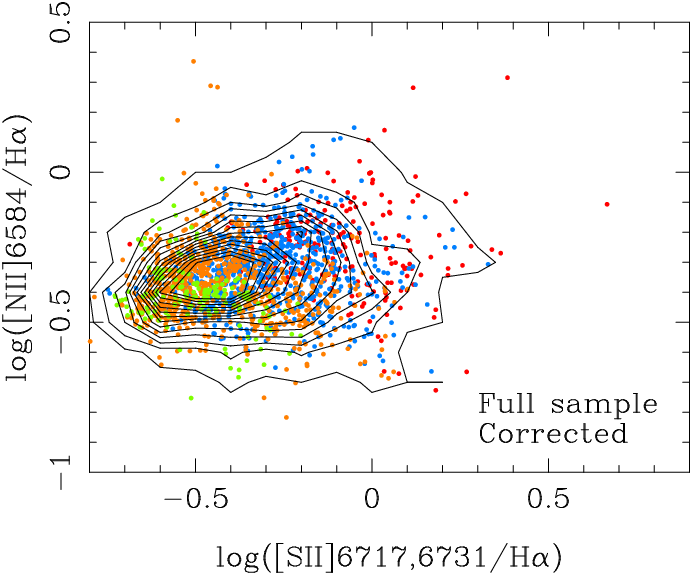}
\caption{Spatially resolved  
[N\,{\sc ii}]$\lambda$6584/H$\alpha$ vs. 
[S\,{\sc ii}]$\lambda\lambda$6717,6731/H$\alpha$ diagrams for the full PMAS
sample. The individual 
  measurements (a total of 1852 spaxels) in both panels are color
  coded according to the observed EW(H$\alpha)_{\rm em}$ in emission
  as in Figure~2.  
The left panel shows the ratios not corrected for underlying
H$\alpha$ stellar absorption, whereas the right panel are the line
ratios corrected for underlying
H$\alpha$ stellar absorption as described in Section~5.3. The contours
are plotted to help assess the data point density in bins of 0.1\,dex in
both emission line ratios.} 

\end{figure*} 

The first result worth noticing is that a large number of the
spatially resolved measurements in the   
[O\,{\sc iii}]$\lambda$5007/H$\beta$ vs. 
[N\,{\sc ii}]$\lambda$6584/H$\alpha$  
diagram (Fig.~2, left panels) fall  in the composite region, and in
particular a large fraction of those spaxels with EW(H$\alpha)_{\rm
  em}<20\,$\AA.  A similar result is found for ULIRGs
(Garc\'{\i}a-Mar\'{\i}n et al. 2010). 
On the other hand, most of the spaxels with EW(H$\alpha)_{\rm em} >60\,$\AA\,
are located in the H\,{\sc ii} region of the diagram, with the only
exception of some regions in NGC~7469. There are also differences from galaxy to galaxy. 
For instance the majority of the spaxels of MCG~+12-02-001 and
NGC~5936 are located
in the H\,{\sc ii} region of the diagram, while other cases such as
UGC~1845 most spaxels are in the composite region. 
The composite region on this diagram lies 
between the observational AGN/H\,{\sc ii} boundary  and the 
``maximum starburst line''  of Kewley et al. (2001), above which the
observed line ratios cannot be explained by star formation alone. 
Kewley et al. (2006) interpreted the observed ratios in the composite
region as produced by a metal-rich stellar population and an 
AGN (but see Cid Fernandes et al. 2010 and
references therein, for an opposing point of view).  While the Kewley
et al. 
argument is valid for nuclear and even integrated spectra of galaxies,
in our sample a large fraction of spaxels with enhanced line ratios
are detected in the extra-nuclear regions of galaxies without an AGN.

The number of  spatially resolved measurements located 
in the {\it LINER} region of the 
diagrams  of the other two diagnostic diagrams (Fig.~2, middle and right panels), 
is smaller than in the  diagram with the [N\,{\sc
  ii}]$\lambda$6584/H$\alpha$ ratio
(see also Veilleux et al. 1995). This is in part a sensitivity issue,
especially for the relatively faint [O\,{\sc i}]$\lambda$6300 line,
which is, on the other hand, a very good shock tracer. 
Still, a large 
fraction of spaxels  with EW(H$\alpha)_{\rm em}<20\,$\AA, mostly in
NGC~23, MCG+02-20-003, and NGC~7771, are in the {\it LINER} region of
these two diagnostic diagrams. 
The presence of extra-nuclear regions in LIRGs and ULIRGs  
with  LINER-like excitation has
often been interpreted as an indication for the presence of large
scale shocks where interactions are playing a
major role (Monreal-Ibero, Arribas, \& Colina 2006;
Monreal-Ibero et al. 2010a) or to shocks due to the presence of
outflowing nuclear gas (see e.g., Armus et al. 1989). Calzetti et
al. (2004) using a  [O\,{\sc iii}]$\lambda$5007/H$\beta$ vs. 
[S\,{\sc ii}]$\lambda\lambda$6717,6731/H$\alpha$  diagram with spatially resolved
measurements of four nearby starburst galaxies, found evidence for
the presence of non-photoionized gas. These authors also demonstrated
that shocks from supernovae and stellar winds
are able to provide sufficient mechanical energy to account for the
non-photoionized gas in their galaxies. In the case of
this sample of LIRGs, most galaxies do not show evidence
for strong interactions either from their morphologies or from the observed
H$\alpha$ velocity fields (see Paper~I). Our LIRGs, however present
very high central H$\alpha$ surface brightnesses and thus star
formation rates per surface area (see Alonso-Herrero et al. 2006),
which would imply a contribution from supernovae. 
One important caveat to keep in mind is 
that the observed line ratios in the diagnostic diagrams of Figure~2 
have not been corrected for the presence
of stellar absorption. 
These corrections will be most relevant for
those spaxels with the lowest equivalent widths of the hydrogen
recombination emission lines, as we shall see in the following section.

\subsection{Spatially resolved [N\,{\sc ii}]$\lambda$6584 /H$\alpha$ vs. 
[S\,{\sc ii}]$\lambda\lambda$6717,6731/H$\alpha$ diagrams}

In our own Galaxy and other galaxies  enhanced [N\,{\sc
ii}]$\lambda$6584/H$\alpha$ and [S\,{\sc
ii}]$\lambda\lambda$6717,6731/H$\alpha$ line ratios  appear to be
associated with the presence of diffuse ionized gas (DIG\footnote{This
  gas is also referred to as diffuse ionized medium or DIM, and warm
  ionized medium or WIM.}, see the recent
review by Haffner et al. 2009 and references therein). 
The presence of DIG emission has been detected spectroscopically 
in the extra-planar emission of
edge-on galaxies (Rand 1996; Miller \& Veilleux 2003). A similar
result was infered based on the 
enhanced low-ionization emission relative  to that of H\,{\sc ii}
regions of the integrated emission of galaxies
(see e.g., Lehnert \& Heckman 1994; Wang,
Heckman, \& Lehnert 1997; Moustakas \& Kennicutt 2006a) as well as on spatially
resolved measurements (see e.g., Calzetti et al. 2004). 
In Paper~I we already discussed the possibility of the presence of
this diffuse emission in our sample of LIRGs, 
as in general the integrated line ratios of the galaxies are greater than the 
typical values observed in disk H\,{\sc ii} regions (see Kennicutt et
al. 1989). Moreover, 
the regions of enhanced line ratios in our LIRGs are generally not
associated with regions of high Pa$\alpha$ surface brightness, that
is, bright H\,{\sc ii} regions. Clear
examples of this are NGC~23, NGC~7591 and NGC~7771 (see
Figure~1 and Paper~I). Regions of enhanced low ionization emission can
also be associated to the presence of an AGN in the form of an
ionization cone, as is the case of one of the nuclei of the 
interacting LIRG 
Arp~299 (Garc\'{\i}a-Mar\'{\i}n et al. 2006).

To study in more detail the excitation conditions in our sample of
LIRGs, we produced  spatially resolved [S\,{\sc
ii}]$\lambda\lambda$6717,6731/H$\alpha$ vs. [N\,{\sc
ii}]$\lambda$6584/H$\alpha$ diagrams for each of the galaxies (see Figure~3). The advantages
of these diagrams over the 
standard diagnostic diagrams are twofold. First they contain more data
points than the diagnostic diagrams (see  the statistics  in
Table~1 for the number of 
spaxels with measurements for each line ratio) because H$\beta$ in LIRGs is heavily affected
by extinction and stellar absorption. Second, as we showed in
Section~5.3, the corrections for the presence of stellar absorption in
H$\alpha$ are less
dependent on the results of the stellar population models than those
for H$\beta$.

A comparison between the observed  [S\,{\sc
ii}]$\lambda\lambda$6717,6731/H$\alpha$ vs. [N\,{\sc
ii}]$\lambda$6584/H$\alpha$ ratios and predictions from 
different models (ionization by young stars, shocks, AGN
photoionization)   can shed some light on the dominant excitation
conditions in our sample of LIRGs. 
In Figure~3 we show the Dopita et al. (2006)
models for evolving H\,{\sc ii} regions. In these models  
the ionization parameter is replaced by the $\mathcal R$ parameter,
which is defined as the ratio of the mass of the ionizing cluster to 
the pressure of the interstellar medium. We chose models with 
solar and twice solar metallicity based on the 
derived abundances of our galaxies from the
integrated line ratios over the central few kpc (Paper~I) corrected for stellar
absorption (see Section~5.3). To estimate the abundances, we 
used the Pettini \& Pagel (2004)
empirical calibration  
based on the [O\,{\sc iii}]$\lambda$5007/H$\beta$ and [N\,{\sc
  ii}]$\lambda$6584/H$\alpha$ ratios, also known as the O3N2 index
(see Alloin et al. 1979). All the LIRGs in our sample have
near solar or super-solar abundances (see Table~7), for a solar
abundance of $12+\log{\rm (O/H)} = 8.66$ (Asplund et al. 2004), and are
within the derived abundances of the large sample of LIRGs studied by
Rupke et al. (2008). 

Although we do not intend to
use the Dopita et al. models for dating the H\,{\sc ii} regions, it is
clear  that line ratios of  [S\,{\sc
ii}]$\lambda\lambda$6717,6731/H$\alpha \sim 0.6-1$ could  be produced by  
evolved and metal rich H\,{\sc ii} regions (see details in Dopita et al. 2006). 
As can be seen from Figure~3, the  [S\,{\sc
ii}]$\lambda\lambda$6717,6731/H$\alpha$ vs. [N\,{\sc
ii}]$\lambda$6584/H$\alpha$ diagrams of the central regions 
of some galaxies (e.g., MCG+12-02-001 and NGC~5936) could be mostly
explained as emission coming from H\,{\sc ii} regions, although it has
to be noted that there are no spaxels with ${\rm
  EW(H}\alpha)_{\rm em}<5\,$\AA \, in the central regions of these
galaxies. On the other hand, galaxies like NGC~23,
NGC~7591 and NGC~7771 show a significant numbers of spaxels with ${\rm
  EW(H}\alpha)_{\rm em} <20\,$\AA \, whose line ratios (not corrected for
stellar absorption) cannot be explained by
the Dopita et al. (2006) H\,{\sc ii} region models. 
Shock models such as those of Allen et al. (2008)  could
 explain line ratios log ([N\,{\sc ii}]$\lambda$6584/H$\alpha) > -0.7$
and  log ([S\,{\sc ii}]$\lambda\lambda$6717,6731/H$\alpha) > -0.7$. 
These models are not plotted in Figure~3, but see figure~8 of 
Monreal-Ibero et al. (2010a)  and a discussion by Miller \& Veilleux (2003).  

To get a more global picture of the excitation conditions in LIRGs, in Figure~10
(left panel) we plot a  [N\,{\sc
ii}]$\lambda$6584/H$\alpha$ vs. [S\,{\sc
ii}]$\lambda\lambda$6717,6731/H$\alpha$ diagram 
for the full PMAS sample, showing a total number
of 1852 spaxels again color coded in terms of EW(H$\alpha)_{\rm em}$ 
in emission. It is clear from this figure that those spaxels
with the smallest EW(H$\alpha)_{\rm em}$  tend to show the
largest values of the line ratios, indicating that in part the
enhanced values are due to the presence of H$\alpha$ stellar
absorption.
The right panel of Figure~10 shows the same diagram 
after correcting statistically each spaxel for the presence of
H$\alpha$ in absorption, with the average values of 
${\rm H}\alpha_{\rm A}$ derived in Section~5.3. 
The effect of correcting the emission line
ratios for stellar absorption in our sample is evident, 
especially for spaxels with EW(H$\alpha)_{\rm em} <
20\,$\AA. However, as can be seen from the histograms in Figure~11, 
even after the correction for stellar absorption there
is still a significant number of spaxels with enhanced
line ratios. For example, after correction for stellar
absorption approximately 25\% of the spaxels show $\log
$ ([S\,{\sc
ii}]$\lambda\lambda$6717,6731/H$\alpha) > - 0.2$, which is the 
largest line ratio allowed by the 2Z$_\odot$ Dopita et al. (2006)
models. This suggests that there are other mechanisms apart from
 young stars ionizing the gas in our sample of LIRGs.

 We can now compare our observations  with those of
Monreal-Ibero et al. (2010a) for the VLT/VIMOS sample of LIRGs. The
VIMOS sample was observed with a mean spatial sampling of 270\,pc per
spaxel, which is  similar to that of the 
PMAS data. The Monreal-Ibero et al. (2010a) sample contains galaxies 
chosen to probe the different morphologies observed in LIRGs from relatively
isolated galaxies, to interacting galaxies to mergers. 
In that study we found that the median value of
the [N\,{\sc ii}]$\lambda$6584/H$\alpha$ ratios measured on a
spaxel-by-spaxel basis has a very weak dependence with the
morphology of the system. The distributions of the [S\,{\sc
ii}]$\lambda\lambda$6717,6731/H$\alpha$  and  the [O\,{\sc
i}]$\lambda$6300/H$\alpha$ ratios, on the other hand, tend to be more
enhanced for galaxies classified as interacting and mergers. Our PMAS
sample is smaller than the VIMOS sample, 
but it is representative of the LIRG class,
in the sense that it comprises the majority of the 
{\it  IRAS} RBGS  LIRGs at $d < 75\,$Mpc that can be
observed from Calar Alto. As such, the PMAS sample is mostly composed
of LIRGs with relatively low IR luminosities, and thus dominated by
isolated spiral galaxies or systems in weak interaction (see also Sanders \&
Ishida 2004). 
The median values  of the line ratios (on a
spaxel-by-spaxel basis) in the PMAS
sample corrected for stellar absorption are [N\,{\sc
ii}]$\lambda$6584/H$\alpha= 0.45 $ and [S\,{\sc
ii}]$\lambda\lambda$6717,6731/H$\alpha =0.44$ (see also
Figure~11). For the statistics we used only spaxels with measurements
of both line ratios. 
These ratios are 
comparable to, although slightly greater than,  those measured by Monreal-Ibero
et al. (2010a) for the VLT/VIMOS sample of LIRGs. 
The slight difference may be due to the fact that the
PMAS data probe more regions (spaxels) with low EW(H$\alpha)_{\rm
  em}$ than in the VLT/VIMOS sample (see
Rodr\'{\i}guez Zaur\'{\i}n et al. 2010b). This in turn results in the
PMAS data 
being more sensitive to diffuse emission.

\begin{figure}

\setcounter{figure}{10}
\includegraphics[width=8cm]{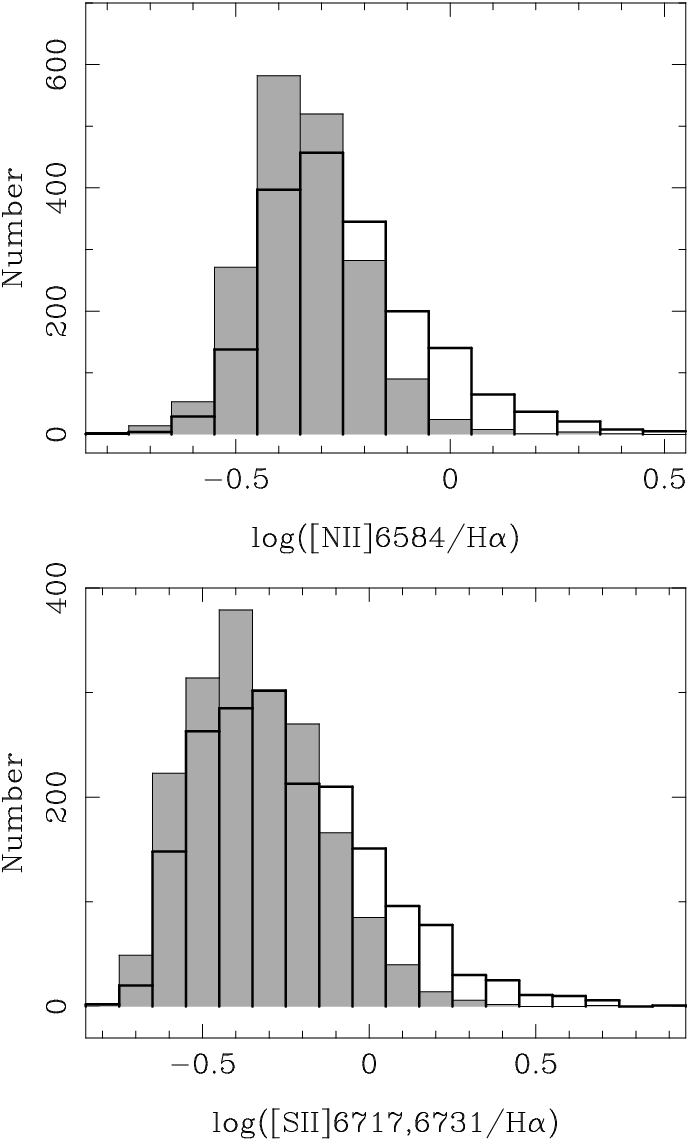}
\caption{Histograms showing the distributions of the spatially resolved  
[N\,{\sc ii}]$\lambda$6584/H$\alpha$ (upper panel) and  
[S\,{\sc ii}]$\lambda\lambda$6717,6731/H$\alpha$ (lower panel) 
line ratios for the full PMAS
sample, for a  total of 1852 spaxels. The thick line empty histograms are
the observed line ratios, while the thin line filled histograms are the line
ratios corrected for stellar absorption. } 

\end{figure}

\subsection{Diffuse emission}

Regardless of the main mechanism responsible for exciting the diffuse
gas in galaxies (e.g., massive stars, escaping photons from H\,{\sc ii}
regions, shocks), it is clear that  quantifying the fraction of diffuse
emission in galaxies is important for understanding issues such as the
early reionization of the universe. Traditionally the fraction of
diffuse emission in galaxies was obtained by 
identifying and doing photometry of H\,{\sc ii} regions, to separate the
emission coming from H\,{\sc ii} regions from that of the DIG (see
e.g., Zurita, Rozas, \& Beckman 2000; Oey et al. 2007, 
see review by Haffner et al. 2009). The main results from this method
were that the mean fraction of diffuse emission in galaxies was
$50-60\%$ with no correlation with 
 the Hubble morphological
type (Thilker et al. 2002; Oey et al. 2007). However, Oey et
al. (2007) found that starburst
galaxies showed lower
fractions of diffuse H$\alpha$ emission when compared to other
galaxies.  

\begin{table}
\caption{Abundances and fraction of diffuse ionized emission in the central
regions of LIRGs.}
\begin{tabular}{lccc}
\hline
\hline
Galaxy   & $12+ \log({\rm O/H)}$ & \multicolumn{2}{c}{DIG fraction}\\
         &                       & NICMOS & $2 \times$MW - MW   \\
\hline
NGC~23          & 8.8 & $19^{+8}_{-4}$ & $6-32$ \\
MCG~+12-02-001  & 8.7 & $8^{+3}_{-2}$  & $9-35$\\
UGC~1845        & 8.7 & $30^{+13}_{-7}$ & $1-16$ \\
NGC~2388        & 8.8 & $21^{+9}_{-5}$ & $2-17$\\
MCG~+02-20-003  & 8.7  &$50^{+13}_{-8}$ & $24-60$\\
NGC~5936        & 8.9 & $58^{+15}_{-10}$ & $1-16$\\
NGC~6701        & 8.9 & $43^{+18}_{-10}$ & $4-28$\\
NGC~7469        & -- & -- & $2-23$\\
NGC~7591        & 8.9 & $76^{+19}_{-13}$ & $47-76$\\
NGC~7771        & 8.9 & $44^{+19}_{-10}$ & $1-10$\\
\hline
\end{tabular}

Notes.--- The abundances are derived using the O3N2 index measured in
the integrated PMAS spectra (Section~6.2).
The NICMOS DIG fractions are from H\,{\sc ii} region
photometry on the {\it HST}/NICMOS Pa$\alpha$ images. The $2\times {\rm MW}$ and MW DIG
fractions are from comparing the spatially resolved  
[S\,{\sc  ii}]$\lambda$6717/H$\alpha$ ratios in our LIRGs with those
of the Milky Way
H\,{\sc ii} regions and DIG (see Section~6.3 for details).

\end{table}

In Alonso-Herrero et al. (2006) we measured the Pa$\alpha$ fluxes of
H\,{\sc ii} regions, and estimated the total Pa$\alpha$ fluxes using
{\it HST}/NICMOS narrow band images, for similar FoVs to those of the PMAS
data. Due to the relatively small FoV of the Pa$\alpha$ NICMOS images
(NIC2 camera $19\arcsec \times 19\arcsec$) the main uncertainty in measuring
Pa$\alpha$ fluxes, both in H\,{\sc ii} regions and the total 
(H\,{\sc ii} + DIG)  emission is the background removal. On
the other hand, narrow-band Pa$\alpha$ imaging  
does not suffer from problems associated with contamination by
the  [N\,{\sc ii}] doublet associated with using 
narrow-band H$\alpha$ imaging (see Blanc et al. 2009 for a discussion
on the issue). For each galaxy we measured the total Pa$\alpha$ luminosity
over the NICMOS FoV, and converted it to H$\alpha$ flux assuming case B
recombination (H$\alpha/{\rm Pa}\alpha=8.6$, Hummer \& Storey
1987). Using the total H$\alpha$ luminosity and 
the H$\alpha$ luminosity in H\,{\sc ii}
regions given in table~3 of Alonso-Herrero et al. (2006), we computed
the DIG fraction for each galaxy. The DIG
  fractions are between 8 to 76\% (Table~7). The range given for each
  galaxy 
takes into
  account the uncertainties in calculating the total Pa$\alpha$
  emission  from the {\it HST}/NICMOS narrow-band imaging (see
  Alonso-Herrero et al. 2006 for details). The DIG fractions given in 
this table for each LIRG correspond only to the central few
kiloparsecs, as neither the NICMOS nor the PMAS observations cover the
full extent of the galaxies.

Alternatively Blanc et
al. (2009) computed the DIG fraction in M51 from optical IFS by
comparing the observed [S\,{\sc  
ii}]$\lambda$6717/H$\alpha$ line ratios with the typical
values measured for
H\,{\sc ii} regions and the DIG in the Milky Way (MW) by Madsen et
al. (2006), and taking into account the difference in metallicity
between M51 and the MW. The PMAS 1\arcsec \, spaxel covers
approximately 300\,pc  
for the typical distances of our galaxies (Section~2.1). Defining the
sizes of H\,{\sc ii}  regions, even at {\it HST} resolutions, is not
straightforward, but results for LIRGs at distances similar to those of the PMAS
sample, indicate typical sizes of $\sim
150-200\,$pc (Alonso-Herrero et al. 2002). We thus expect that each
measurement of a given line ratio  
contains both H\,{\sc ii} region and DIG emission, as assumed by the Blanc
et al. (2009) method. In the MW the average 
[S\,{\sc  
ii}]$\lambda$6717/H$\alpha$ line ratios of H\,{\sc ii}
regions and the diffuse emission are $0.11\pm 0.03$ and $0.34\pm
0.13$, respectively (Madsen et al. 2006; Blanc et al. 2009). 
In Table~7 we list the DIG
fractions we obtained with the line 
[S\,{\sc  ii}]$\lambda$6717/H$\alpha$ ratios corrected for stellar
absorption and 
the Blanc et al. (2009) method. 
For each galaxy we give a range, with the smallest value
corresponding to scaling the MW H\,{\sc ii} and DIG ratios to $\sim 1.7$
times solar, and the largest ratio for solar metallicity.

Table~7 shows that the fraction of diffuse emission in
the central few kpc of our LIRGs 
varies significantly from galaxy to galaxy. Oey et al. (2007) found
that starburst galaxies, defined as galaxies having H$\alpha$ surface
brightnesses above $2.5\times 10^{39}\,{\rm erg \, s}^{-1}\,{\rm kpc}^{-2}$
  within the H$\alpha$ half radius,
  have relatively small DIG fractions ($<60\%$), when compared to
  non-starburst galaxies. According to this definition all our
  galaxies are starbursts (see Alonso-Herrero et al. 2006), 
and thus in most cases our DIG fractions
  are below 60\%. The large DIG fraction inferred for NGC~7591 may be
 explained if some of enhanced line ratios are produced in a narrow
 line region associated with the AGN in this galaxy. We also find that the
 fractions of DIG 
  emission in the central few kpc
obtained with the two methods described above mostly agree with each other,
within the uncertainties (Table~7). We note however, that the line
ratio method always tends to provide a smaller DIG fraction than the
H\,{\sc ii} region photometry method (see also Blanc et al. 2009 for
M51). The two LIRGs with the most discrepant
fractions using the two methods 
are NGC~5936 and NGC~7771. In the case of NGC~5936 we
attribute the discrepancy to the fact that, except for the very nuclear
regions, the other H\,{\sc ii} regions in the FoV of the NICMOS images
appear quite diffuse (see Alonso-Herrero et al. 2006, and Paper~I),
and thus the DIG fraction estimated from the 
NICMOS Pa$\alpha$ images is clearly just an upper limit. In the case
of NGC~7771, even if we applied the
smaller correction for the H$\alpha$ absorption feature (that is, if
the evolved stellar population had an age of a few Gyr), the DIG
fraction from the [S\,{\sc  ii}]$\lambda$6717/H$\alpha$ line ratios
would be in the range of 3 to 13\%. The most likely explanation is
that the faintest H\,{\sc ii} regions just outside the bright ring of star
formation were not identified in the NICMOS Pa$\alpha$ images, and
this resulted in an overestimate of the diffuse emission.

\section{Summary}
This is the second paper of a series presenting 
a PMAS IFS  study of 11 local (average distance of 61\,Mpc)
LIRGs covering the central few kiloparsecs of the galaxies. We 
selected the galaxies from the complete volume-limited sample
of LIRG of Alonso-Herrero et al. (2006), which has an average IR
luminosity of $ \log (L_{\rm IR}/{\rm L}_\odot) =11.32$.  As such, our
sample of LIRGs  
is mostly composed of isolated spiral galaxies, galaxies with companions,
and weakly interacting galaxies.  The PMAS observations covered a 
spectral range of $3700-7200\,$\AA, with a spectral resolution of
6.8\,\AA \, (FWHM). We used a
1\arcsec-size spaxel that provides a typical  physical sampling
of approximately 300\,pc. The PMAS observations were complemented with
our own existing {\it HST}/NICMOS observations of the near-infrared continuum
and the Pa$\alpha$ emission line. We combined the spectral information
provided by the PMAS
IFS and the spatial information
provided by the {\it HST}/NICMOS observations to study in
detail the stellar populations, the excitation
conditions of the gas, and diffuse emission in local LIRGs.

To study the stellar populations we selected the nuclear regions of
each galaxy, as well as a number of bright H\,{\sc ii} regions in NGC~23,
NGC~2388, and NGC~7771. We also extracted for each galaxy, when
possible, the average spectrum of regions with low values of EW(H$\alpha)_{\rm
  em}$. These 
regions were selected such that they did not contain high surface
brightness H\,{\sc ii} regions to  minimize the
contamination by hydrogen nebular emission in the stellar
  absorption features. 
We used BC03 and Starburst99 to fit the stellar continuum and the nebular
emission, respectively. For the modelling we assumed two stellar
populations formed in instantaneous bursts with a Salpeter IMF and
solar metallicity. 
The PMAS maps of EW(H$\alpha)_{\rm em}$ in our sample of LIRGs (see Figure~1), 
except for the nuclear region of IC~860 (Table~4),  
indicate ages of the young stellar populations
of between 5 and $\sim 10-20\,$Myr (see e.g., Leitherer et
al. 1999, and Section~3.1).  We thus chose 
ages of the ionizing stellar populations of between 1 and 20\,Myr, and
$\ge 100\,$Myr for the evolved one. Both stellar populations were
extinguished with a range of 
values of E(B-V). 

Combinations of 
evolved stellar populations with ages between 0.7 and $5-10\,$Gyr with
relatively low extinctions, and ionizing stellar populations with
higher extinctions provided reasonable fits to the optical
continua of most of our regions. The spectra of our LIRGs tend to
  show deeper MgIb and G-bands than the ULIRGs in the Rodr\'{\i}guez
  Zaur\'{\i}n et al. (2010a) sample. This suggests that the stellar
  populations in our sample are somewhat older. In fact, we found
  little evidence 
for a strong contribution to the optical light from intermediate-aged stellar
populations with ages of $\sim 100-500$\,Myr in our sample of
  LIRGs. The only exception in our 
sample is NGC~7771. Several of the regions studied in this galaxy,
including its optical nucleus, 
show clear indications for an important contribution
from an intermediate-aged stellar 
population ($\sim 100-700\,$Myr). Additionally the optical nucleus of 
NGC~7771 contains a large number of stars with ages in the range
$30-50\,$Myr, as indicated by the presence of strong He\,{\sc i} absorption
features.  

The majority of the selected
regions in our sample of LIRGs have a significant contribution 
to the 4600\,\AA \, continuum light from the ionizing stellar
populations, and for about one-third the ionizing stellar populations
have a dominant contribution.
The ages of the evolved stellar populations of our LIRGs are within
the ranges observed in massive spiral galaxies.  
The fitting of
the stellar continuum and the nebular properties yielded  
ages of the ionizing stellar populations of between 5 and 9\,Myr, and
extinctions in the range ${\rm E(B-V)}_{\rm young} =0.2-1.8$. The
extinctions from the stellar continuum fits and the Balmer decrement agree
with each other, and tend to be  higher
than those required for the evolved stellar populations.

We used the brightest emission lines to study the spatially resolved
(on a spaxel-by-spaxel basis) 
excitation conditions in the central few kiloparsecs of our sample of
LIRGs. When using traditional diagnostic diagrams involving
  bright optical emission lines, the location of the
spatially-resolved line ratios varies from LIRG
to LIRG. Some galaxies have most of their spaxels in the
H\,{\sc ii} region of the diagnostic diagrams, while others have spaxels
located both in the H\,{\sc ii} and {\it LINER} regions of the
diagrams. In particular, a significant fraction of
spaxels with EW(H$\alpha)_{\rm em} <20\,$\AA \,  have enhanced
[N\,{\sc  ii}]$\lambda$6584/H$\alpha$,  [S\,{\sc
  ii}]$\lambda\lambda$6717,6731/H$\alpha$ and [O\,{\sc
  i}]$\lambda$6300/H$\alpha$ 
line ratios, when compared to those of H\,{\sc 
  ii} regions. These spaxels tend to be associated with regions of
diffuse H$\alpha$ and Pa$\alpha$ emission in these
galaxies, that is, they do not coincide with high surface brightness
H\,{\sc ii} regions identified in the {\it HST}/NICMOS Pa$\alpha$ images. 

We also produced  spatially resolved [S\,{\sc
ii}]$\lambda\lambda$6717,6731/H$\alpha$ vs. [N\,{\sc
ii}]$\lambda$6584/H$\alpha$ diagrams for each of the galaxies, and the
full PMAS sample. These diagrams have the advantage of containing more
data points than the traditional diagnostic 
diagrams, and of having smaller corrections
for the presence of H$\alpha$ stellar absorption than diagrams using
H$\beta$.  Even so,  the corrections for stellar absorption to the
observed line ratios are
significant for spaxels with EW(H$\alpha)_{\rm em} <20\,$\AA. After
the correction for stellar absorption, there
is still a significant number of spaxels in these diagrams whose line
ratios cannot be explained as produced entirely by photoionization
by young stars (e.g., $\log $[S\,{\sc
ii}]$\lambda\lambda$6717,6731/H$\alpha > - 0.2$).
 These enhanced ratios are probably due to the combined
contributions of 
H\,{\sc ii} region emission and DIG emission. We estimated the fraction of DIG
emission in the central few kiloparsecs of our LIRG using two
different methods. The first one used H\,{\sc ii} region photometry 
from the NICMOS Pa$\alpha$ images. The second method was based on the
comparison of the spatially-resolved [S\,{\sc
  ii}]$\lambda$6717/H$\alpha$ line ratios of our LIRGs, and those of
the MW H\,{\sc ii} regions and DIG. The DIG fractions over the
central few kiloparsecs vary from
galaxy to galaxy, but are generally $<60\%$, as found for starburst
galaxies by Oey et al. (2007).

\begin{acknowledgement}

We thank an anonymous referee for useful suggestions that helped
improve the paper. We thank C. Tadhunter, 
G. Blanc and F. Rosales for enlightening
discussions. 
We are grateful to the Calar Alto staff, and in particular to S. S\'anchez, 
A. Guijarro, L. Montoya, and N. Cardiel, for their
support during the PMAS observing campaigns. 

This research has made use of the NASA/IPAC Extragalactic Database
(NED) which is operated by the Jet Propulsion Laboratory, California
Institute of Technology, under contract with the National Aeronautics
and Space Administration.

AA-H, JR-Z, LC, and SA acknowledge support
from the Spanish Plan Nacional del Espacio under grant
ESP2007-65475-C02-01. AA-H also acknowledges support for this work from the
  Spanish Ministry of Science and Innovation through Proyecto Intramural
Especial under grant number 200850I003. 
MG-M is supported by the German Federal
Department of Education and Research (BMBF) under project numbers: 
50OS0502 and 50OS0801. AM-I is supported
by the Spanish Ministry of Science and Innovation (MICINN) under program
"Specialization in International Organisms", ref. ES2006-0003.
\end{acknowledgement}

\clearpage
\begin{figure*}
\setcounter{figure}{0}

\vspace{0.2cm}

\includegraphics[width=11.cm,angle=-90]{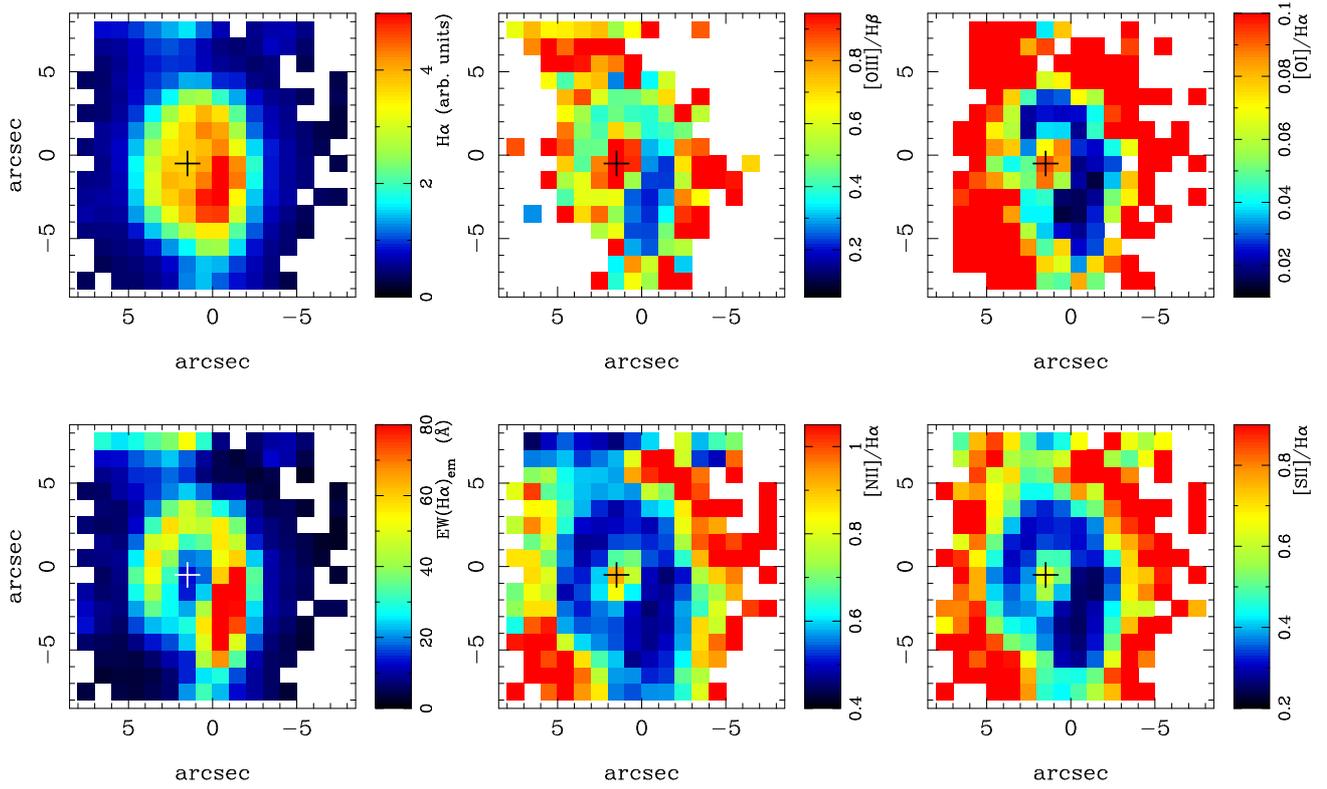}

\caption{(a) NGC~23: PMAS observed (not corrected for extinction or underlying
  stellar absoption) maps of emission line
  ratios: [O\,{\sc iii}]$\lambda$5007/H$\beta$ (top middle), [O\,{\sc
    i}]$\lambda$6300/H$\alpha$ (top right), [N\,{\sc
    ii}]$\lambda$6584/H$\alpha$ (bottom middle) and [S\,{\sc
    ii}]$\lambda \lambda$6717,6731/H$\alpha$ (bottom right). Also
    shown are the  maps of the observed H$\alpha$ flux (top left)  and
  EW(H$\alpha)_{\rm em}$ (bottom left) of the line in emission. The
FoV of the PMAS maps is $16\arcsec \times 
16\arcsec$, and the orientation is North up, East to the left. The 
crosses on the maps mark the peak of the PMAS optical 
continuum emission at $\sim 6200\,$\AA \, (see Paper I for details).
The H$\alpha$ maps are shown in a square root scale to emphasize the
low surface brightness regions.}   

\end{figure*} 

\begin{figure*}
\setcounter{figure}{0}
\vspace{0.2cm}

\includegraphics[width=11.cm,angle=-90]{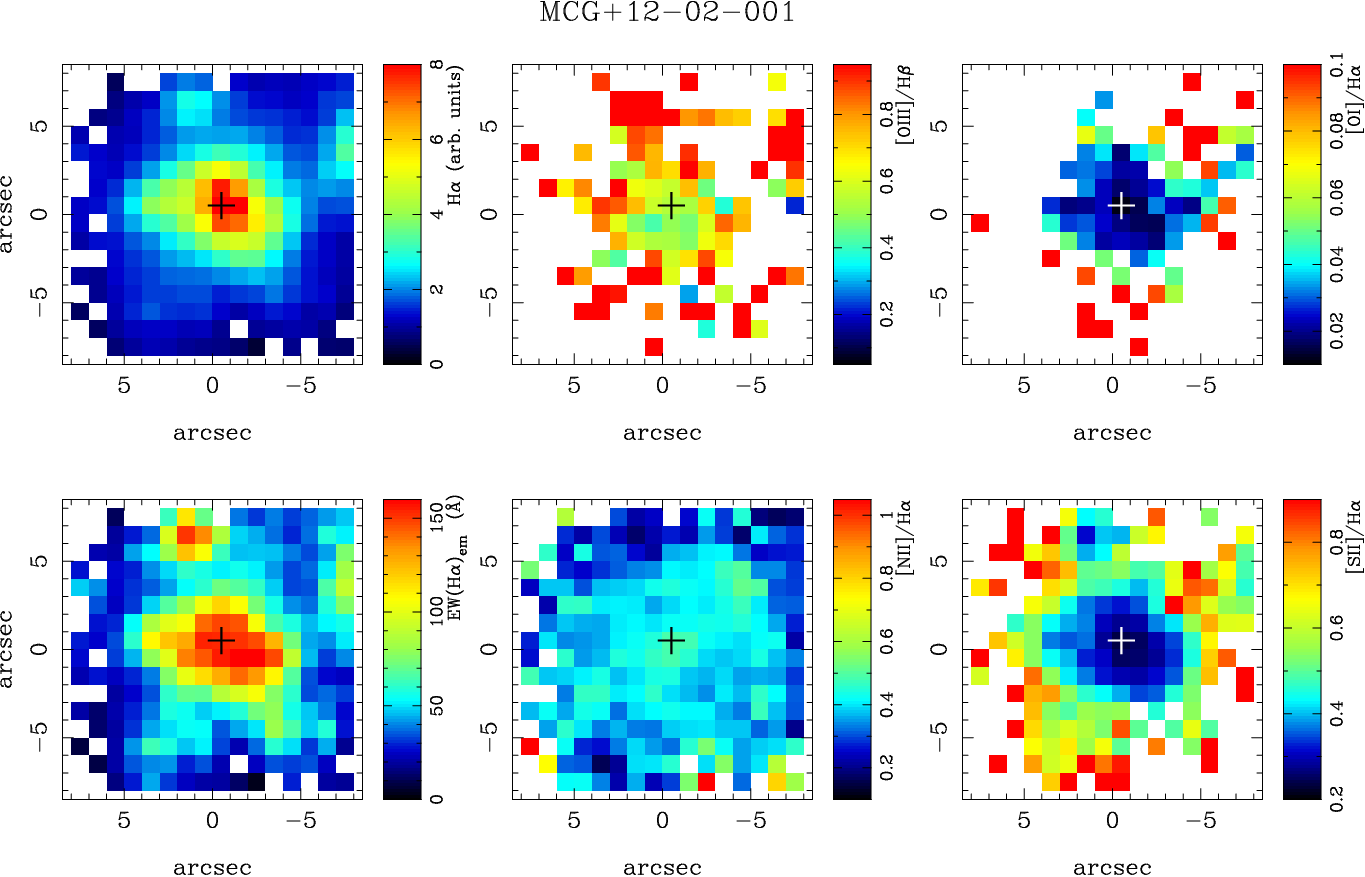}

\caption{(b) As Fig.~1a but for MCG~+12-02-001.}   

\end{figure*} 

\begin{figure*}
\setcounter{figure}{0}
\vspace{0.2cm}

\includegraphics[width=11.cm,angle=-90]{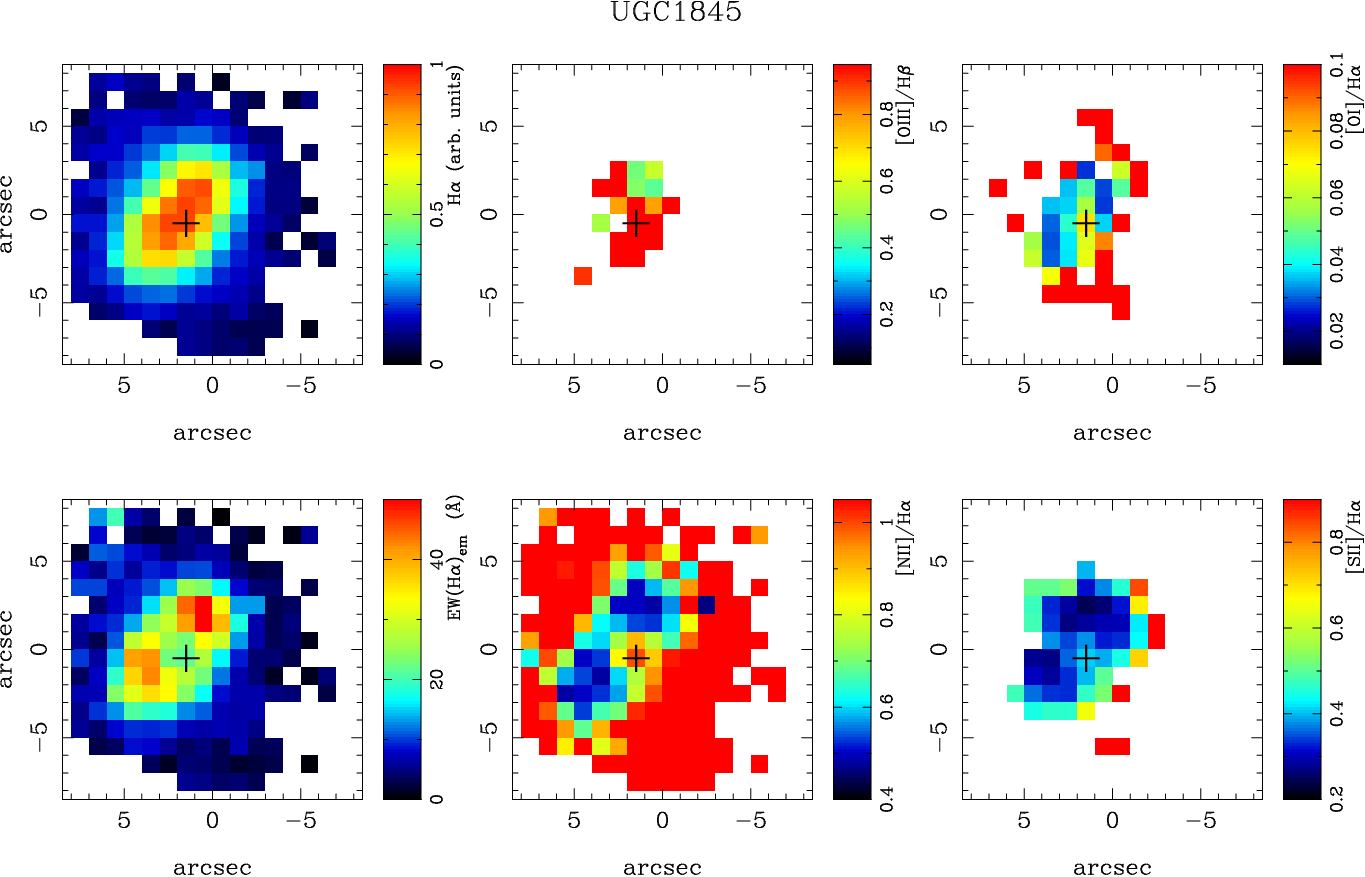}

\caption{(c) As Fig.~1a but for UGC~1845.}   

\end{figure*} 

\begin{figure*}
\setcounter{figure}{0}
\vspace{0.2cm}

\includegraphics[width=11.cm,angle=-90]{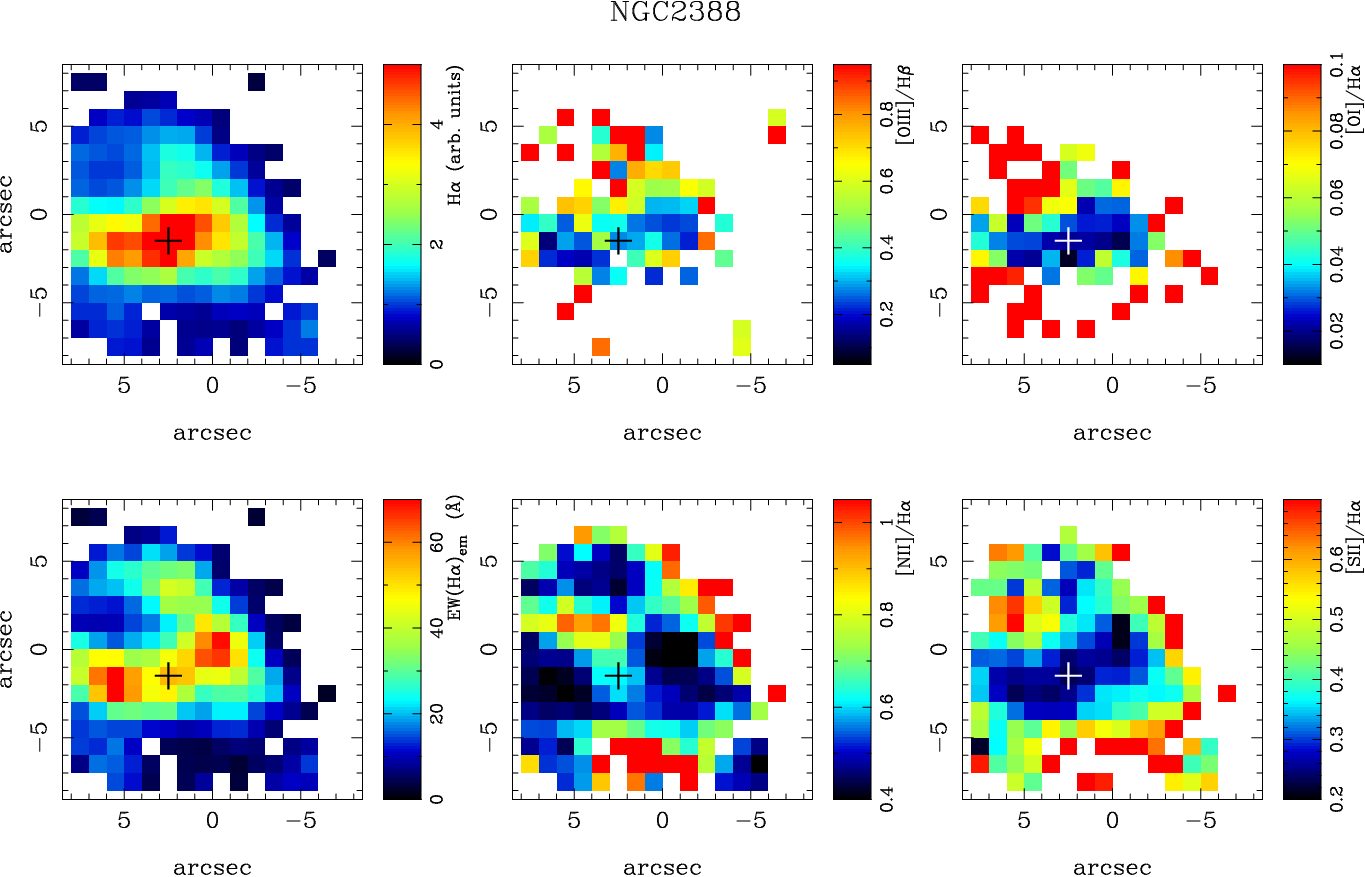}

\caption{(d) As Fig.~1a but for NGC~2388.}   

\end{figure*} 

\begin{figure*}
\setcounter{figure}{0}
\vspace{0.2cm}

\includegraphics[width=11.cm,angle=-90]{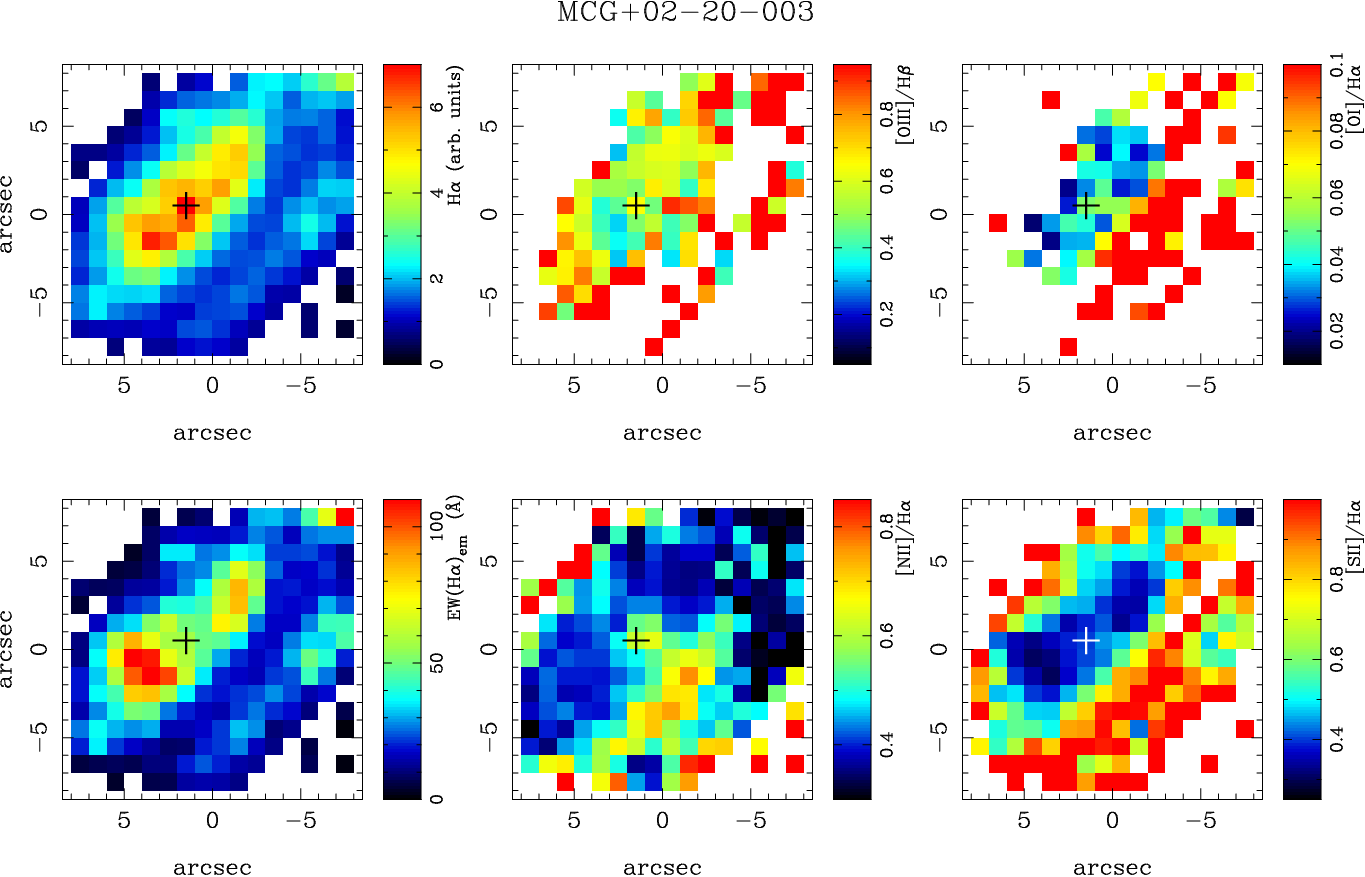}

\caption{(e) As Fig.~1a but for MCG~+02-20-003.}   

\end{figure*} 

\begin{figure*}
\setcounter{figure}{0}
\vspace{0.2cm}

\includegraphics[width=11.cm,angle=-90]{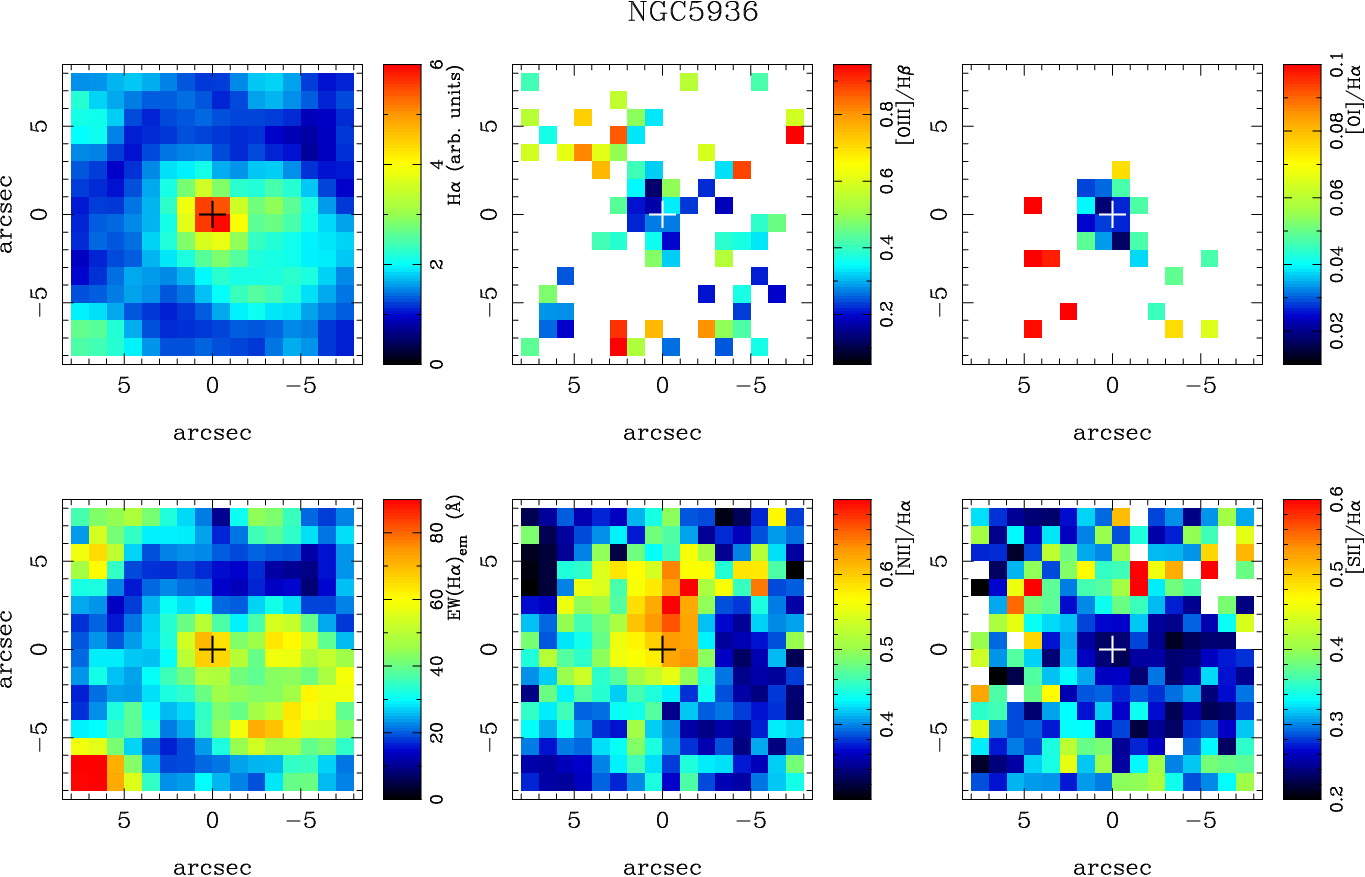}

\caption{(f) As Fig.~1a but for NGC~5936.}   

\end{figure*} 

\begin{figure*}
\setcounter{figure}{0}
\vspace{0.2cm}

\includegraphics[width=11.cm,angle=-90]{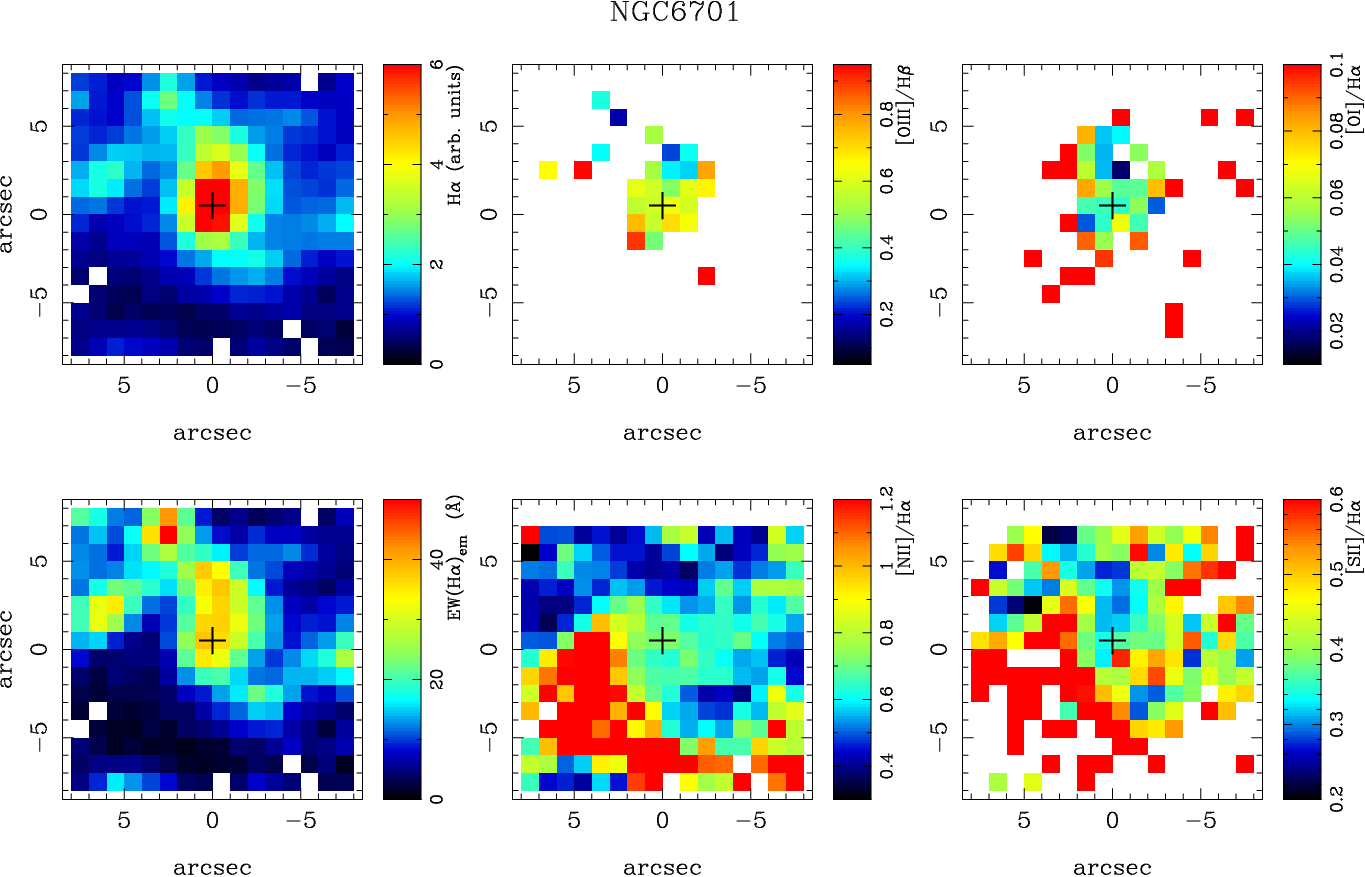}

\caption{(g) As Fig.~1a but for NGC~6701.}   

\end{figure*} 

\begin{figure*}
\setcounter{figure}{0}
\vspace{0.2cm}

\includegraphics[width=11.cm,angle=-90]{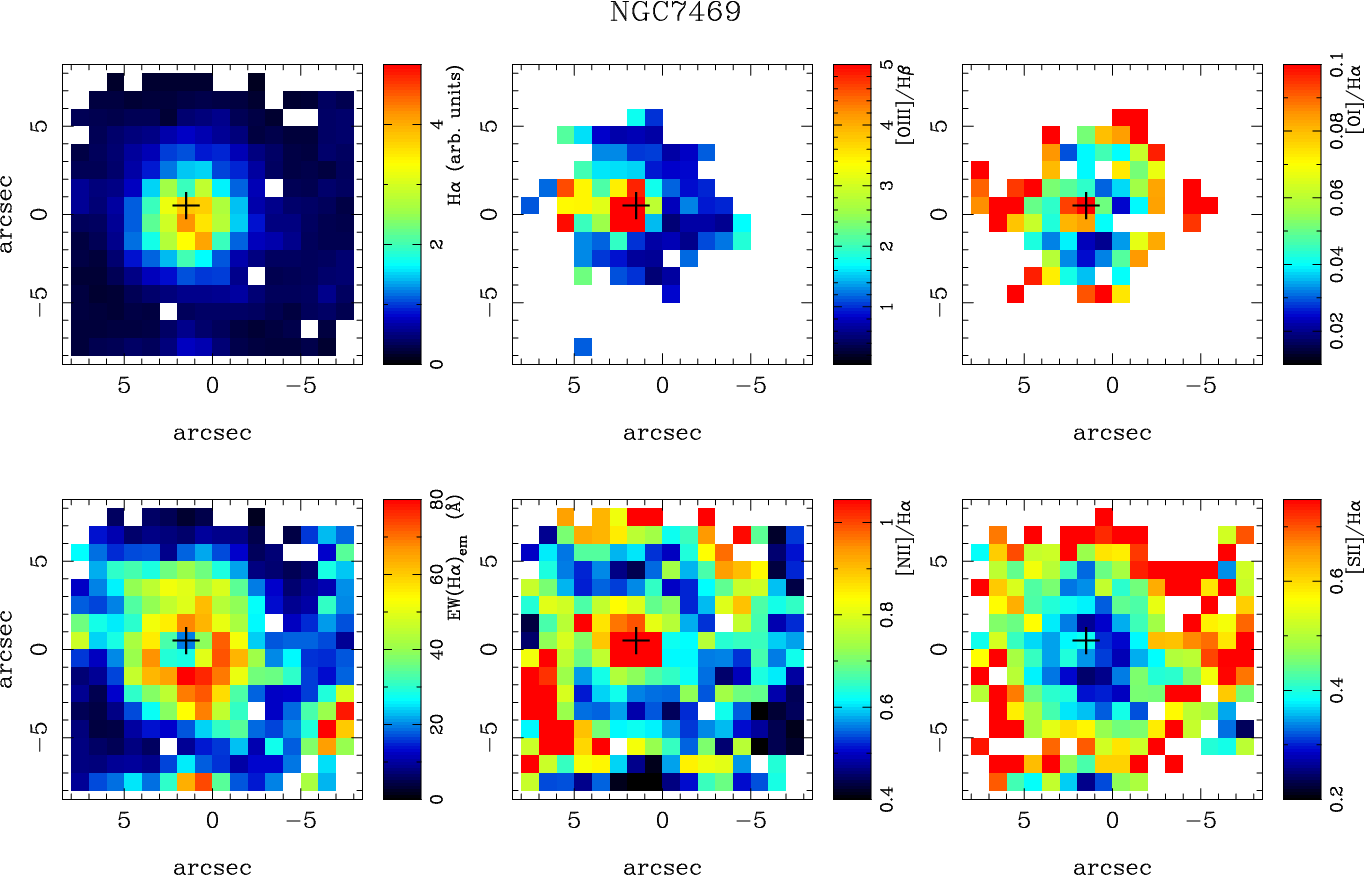}

\caption{(h) As Fig.~1a but for NGC~7469. The H$\beta$ and 
H$\alpha$  measurements
  correspond to the narrow component of the  lines (see Paper I for details).}   

\end{figure*} 

\begin{figure*}
\setcounter{figure}{0}
\vspace{0.2cm}

\includegraphics[width=11.cm,angle=-90]{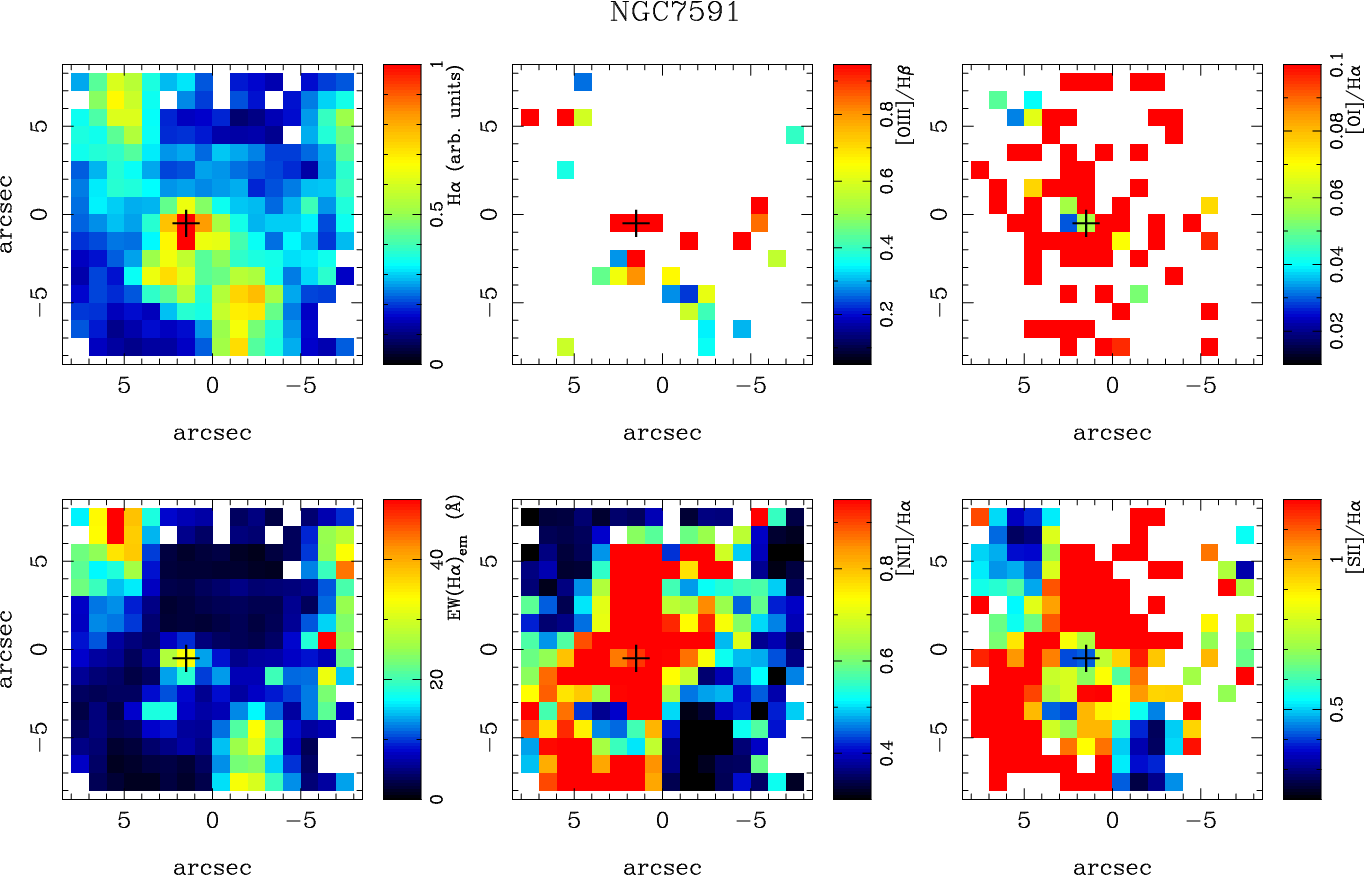}

\caption{(i) As Fig.~1a but for NGC~7591.}   

\end{figure*}

\begin{figure*}
\setcounter{figure}{0}
\vspace{0.2cm}

\includegraphics[width=15.cm]{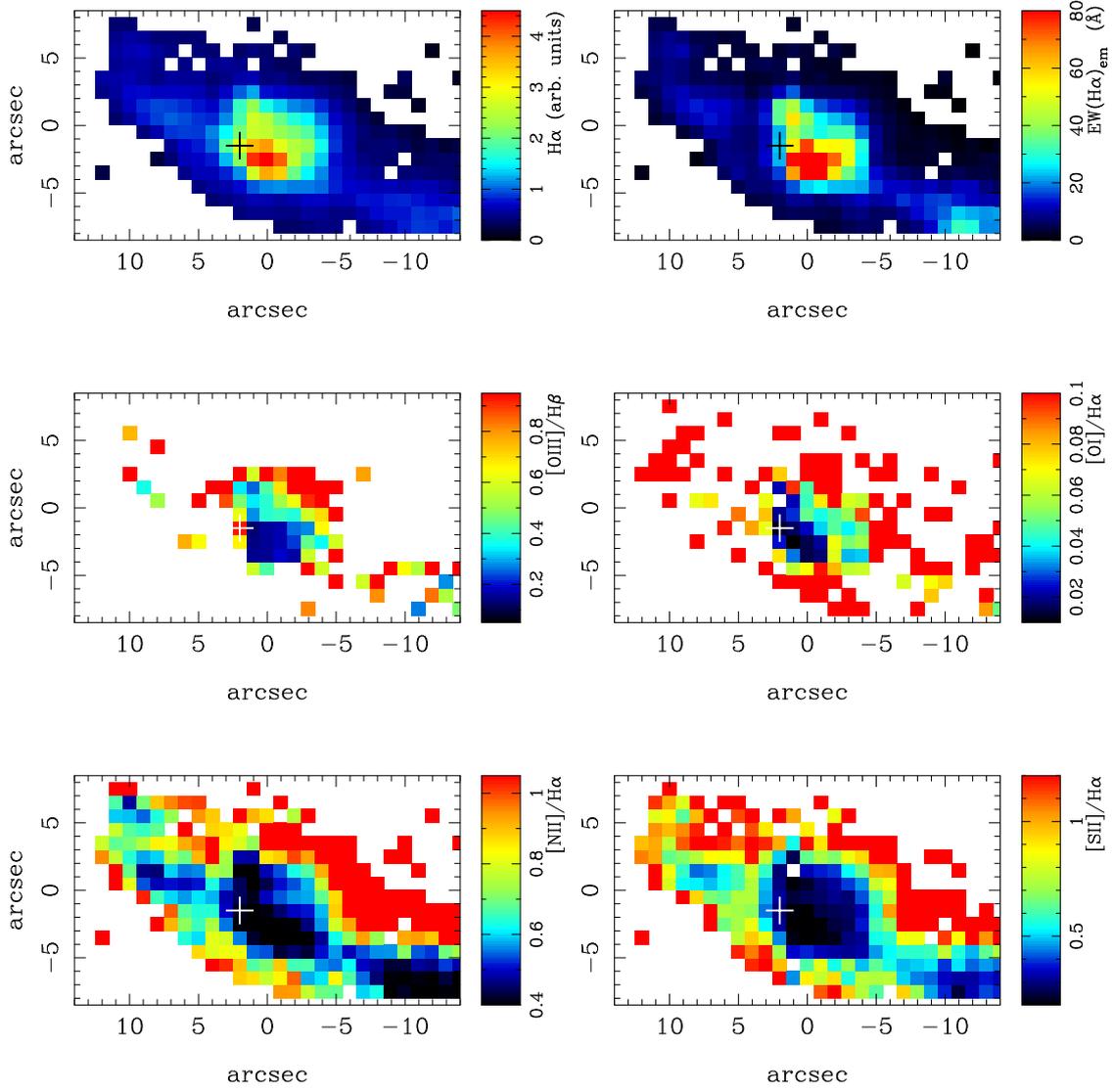}

\caption{(j) As Fig.~1a but for NGC~7771, except that the FoV for this
galaxy covers the central $28\arcsec \times
16\arcsec$ (Section~2.1 for details). The maps are shown: 
H$\alpha$ flux (top left), 
  EW(H$\alpha)_{\rm em}$ (top right),  
[O\,{\sc iii}]$\lambda$5007/H$\beta$ (middle left), [O\,{\sc
    i}]$\lambda$6300/H$\alpha$ (middle right), [N\,{\sc
    ii}]$\lambda$6584/H$\alpha$ (bottom left) and [S\,{\sc
    ii}]$\lambda \lambda$6717,6731/H$\alpha$ (bottom right). }   

\end{figure*} 

\clearpage

\begin{figure*}
\setcounter{figure}{1}
\includegraphics[width=7.cm,angle=-90]{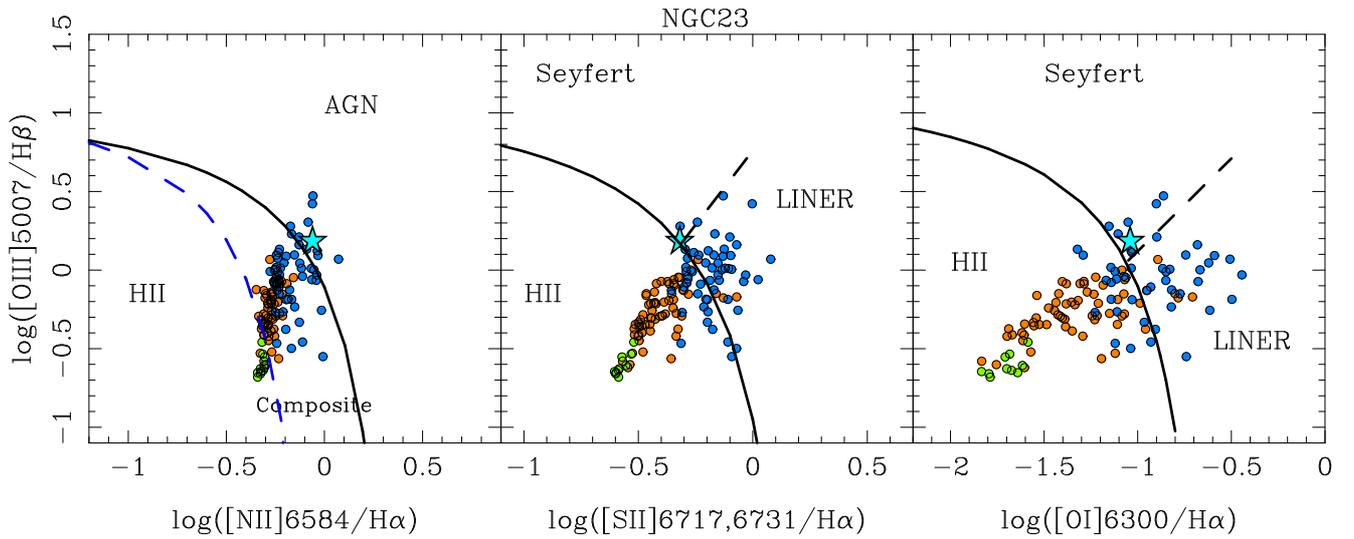}
\caption{(a) Observed 
diagnostic diagrams for the spatially resolved measurements of 
NGC~23. The line ratios are not corrected for extinction or underlying stellar
absorption.
The star symbols are  the nuclear
  region measurements given in Paper~I. 
The line ratios for individual spaxels are 
  color coded according to the value of the 
EW(H$\alpha)_{\rm em}$  of the line
    in emission: green
  dots EW(H$\alpha)_{\rm em} > 60\,$\AA, orange  dots  20\,\AA $< {\rm
    EW(H}\alpha)_{\rm em} \le 
  60\,$\AA,  blue dots 5\,\AA $< {\rm EW(H}\alpha)_{\rm em} \le
  20\,$\AA,  and red dots ${\rm EW(H}\alpha)_{\rm em} \le 5\,$\AA. The
  solid lines are the ``maximum starburst lines''  defined by Kewley
  et al. (2001) from theoretical modeling, and the thick dashed lines
  are the empirical separation between AGN and H\,{\sc ii} regions of
  Kauffmann et al. (2003), and between Seyfert and LINERs of Kewley et
  al. (2006). 
}
\end{figure*}

\begin{figure*}
\setcounter{figure}{1}
\includegraphics[width=7.cm,angle=-90]{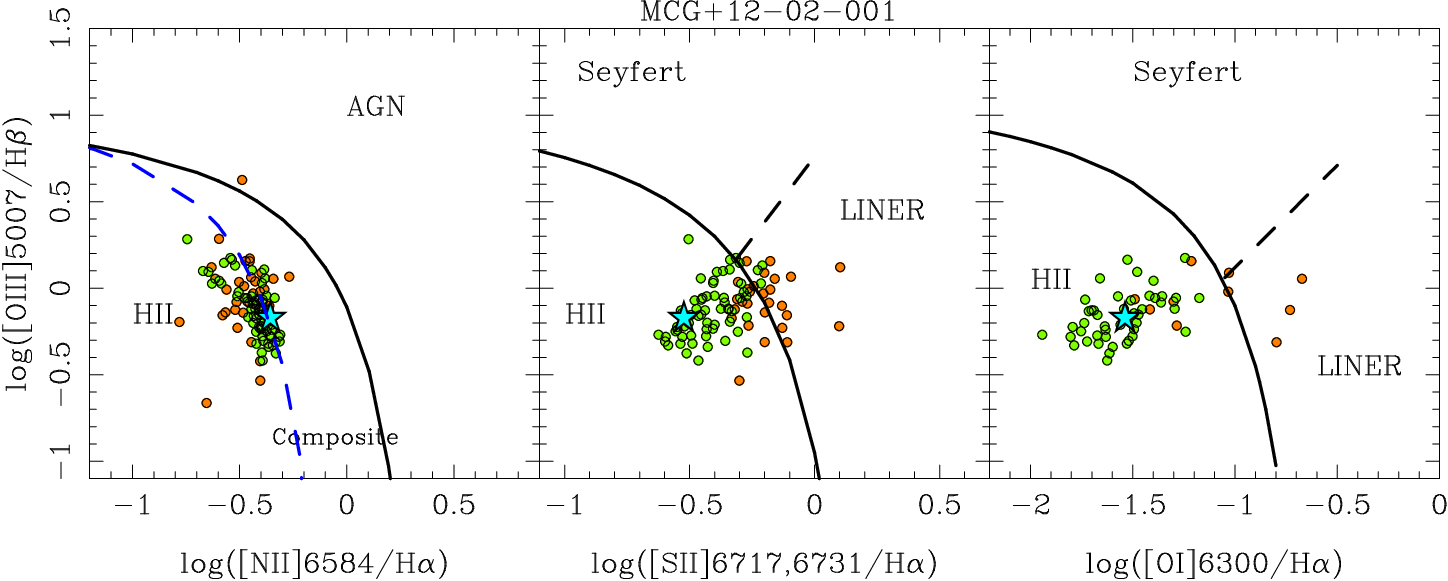}
\caption{(b) As Fig.~2a but for MCG~+12-02-001.}  
\end{figure*}

\begin{figure*}
\setcounter{figure}{1}
\includegraphics[width=7.cm,angle=-90]{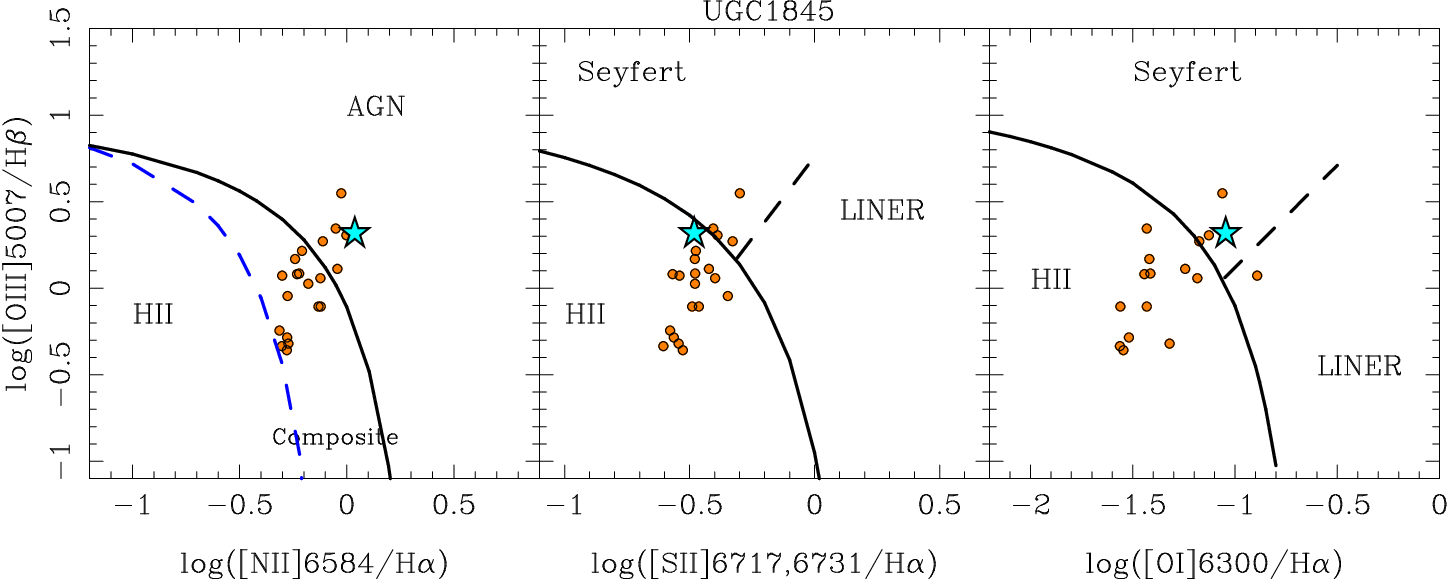}
\caption{(c) As Fig.~2a but for UGC~1845.}  
\end{figure*}

\begin{figure*}
\setcounter{figure}{1}
\includegraphics[width=7.cm,angle=-90]{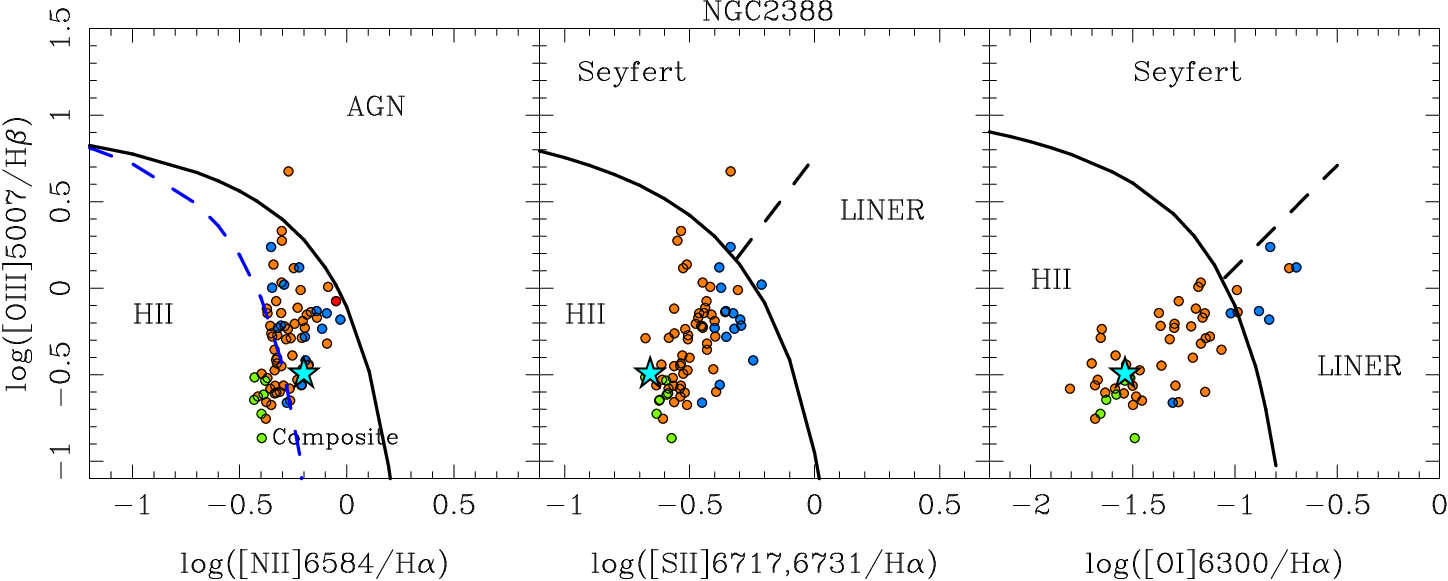}
\caption{(d) As Fig.~2a but for NGC~2388.}  
\end{figure*}

\begin{figure*}
\setcounter{figure}{1}
\includegraphics[width=7.cm,angle=-90]{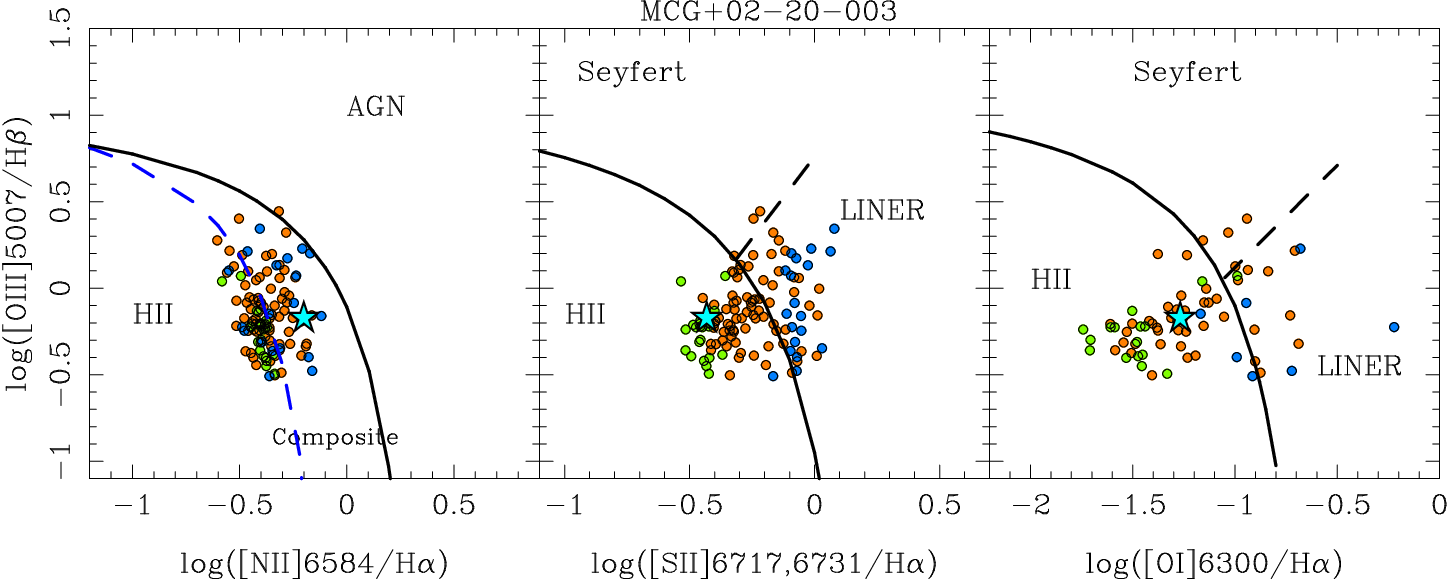}
\caption{(e) As Fig.~2a but for MCG~+02-20-003.}  
\end{figure*}

\begin{figure*}
\setcounter{figure}{1}
\includegraphics[width=7.cm,angle=-90]{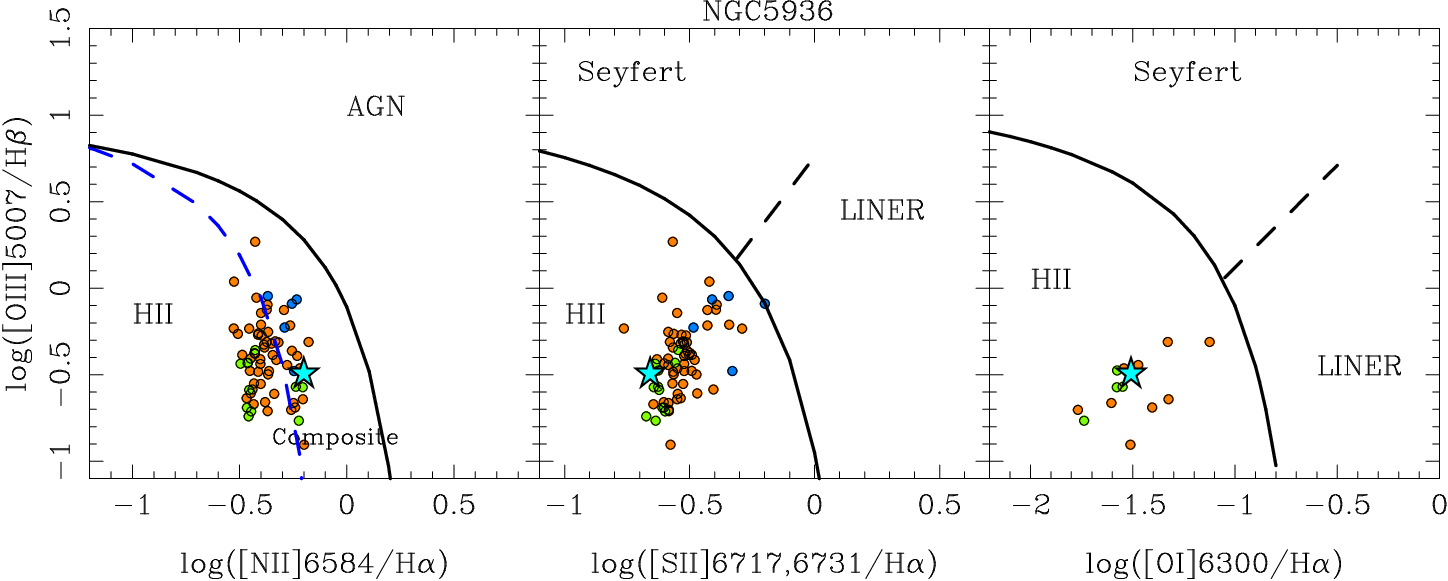}
\caption{(f) As Fig.~2a but for NGC~5936.}  
\end{figure*}

\begin{figure*}
\setcounter{figure}{1}
\includegraphics[width=7.cm,angle=-90]{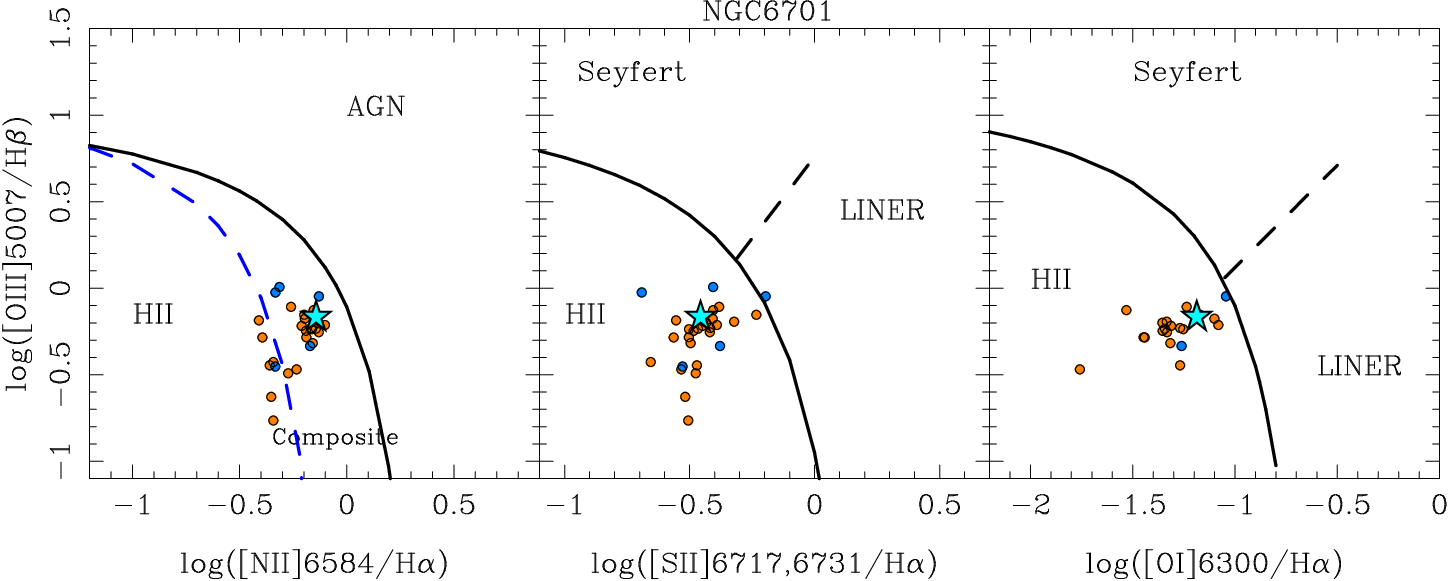}
\caption{(g) As Fig.~2a but for NGC~6701.}  
\end{figure*}

\begin{figure*}
\setcounter{figure}{1}
\includegraphics[width=7.cm,angle=-90]{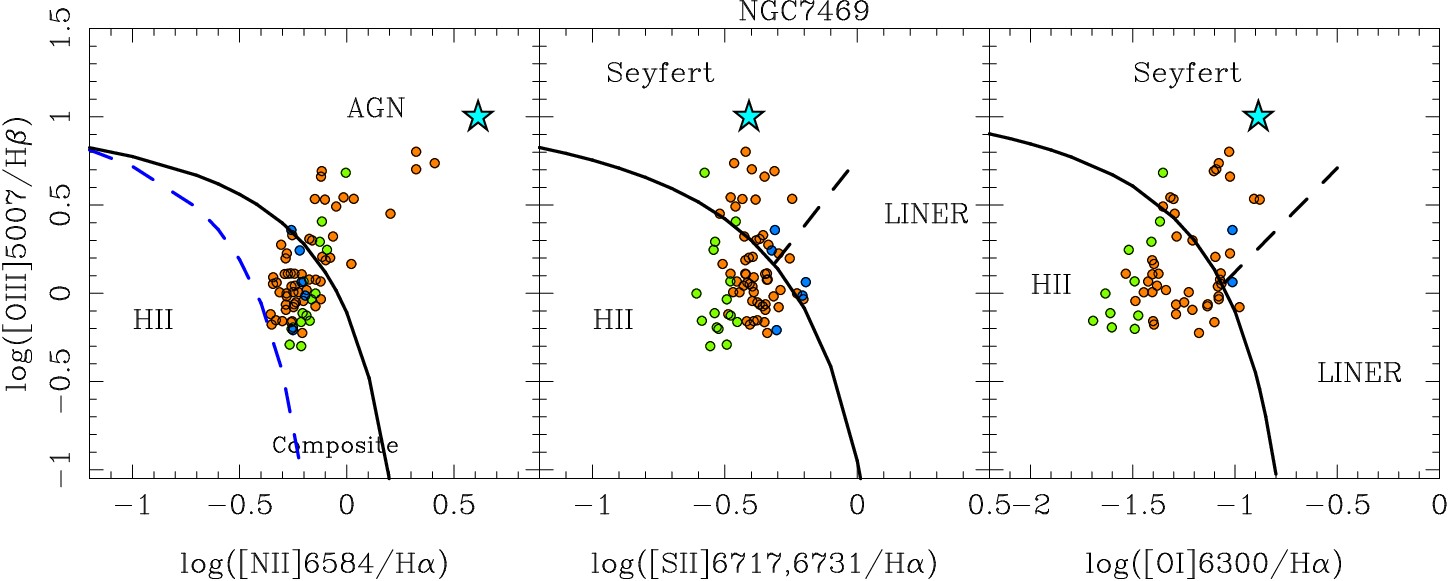}
\caption{(h) As Fig.~2a but for NGC~7469. The H$\beta$ and H$\alpha$
  measurements correspond to the narrow component of the lines (see
  Paper I for details).}  
\end{figure*}

\begin{figure*}
\setcounter{figure}{1}
\includegraphics[width=7.cm,angle=-90]{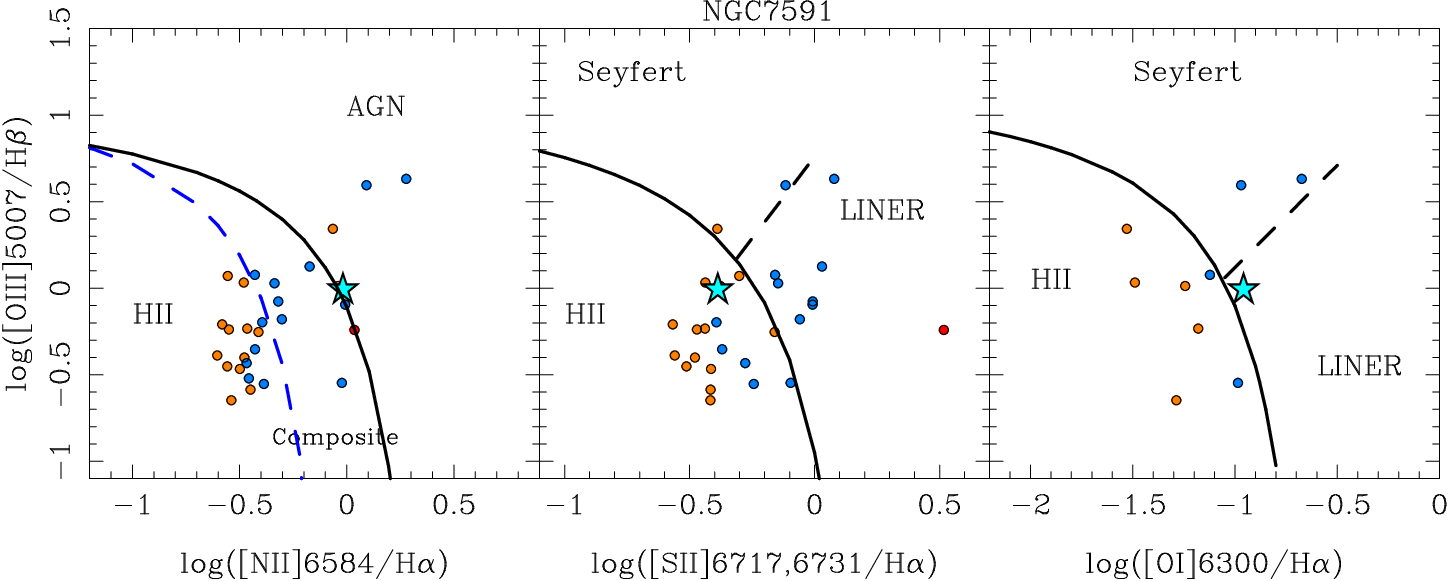}
\caption{(i) As Fig.~2a but for NGC~7591.}  
\end{figure*}

\begin{figure*}
\setcounter{figure}{1}
\includegraphics[width=7.cm,angle=-90]{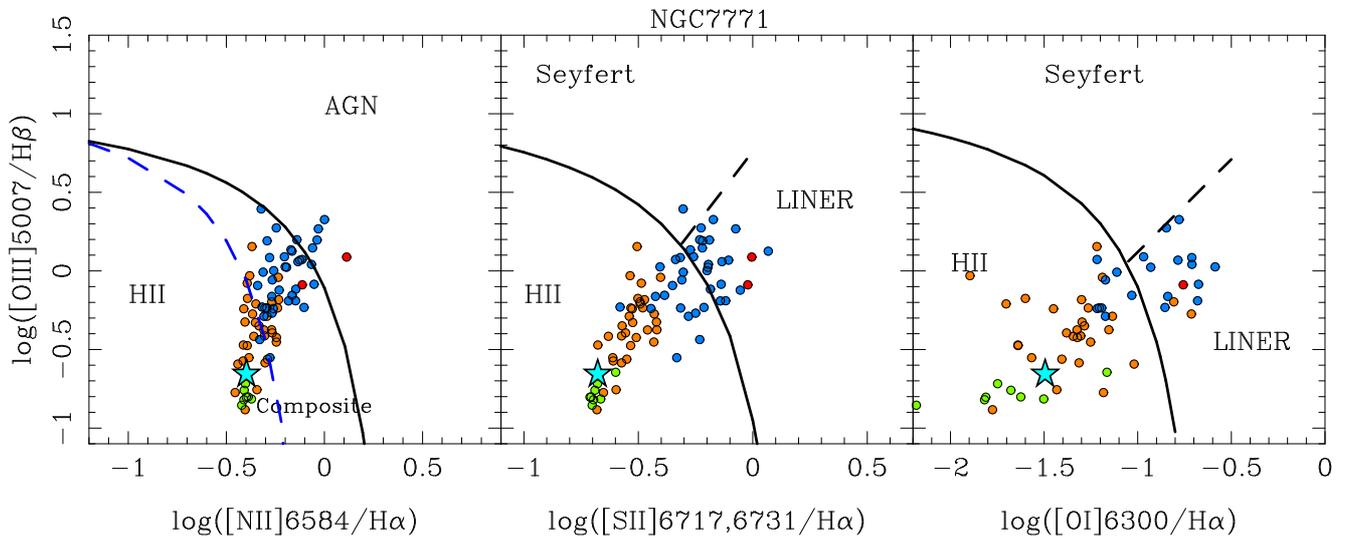}
\caption{(j) As Fig.~2a but for NGC~7771.}  
\end{figure*}

\begin{figure*}[h]
\setcounter{figure}{2}
\includegraphics[width=7cm,angle=-90]{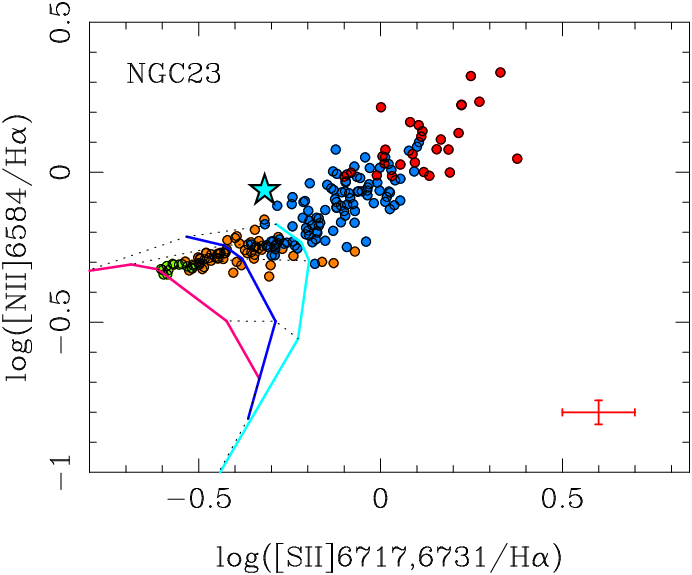}
\hspace{1cm}
\includegraphics[width=7cm,angle=-90]{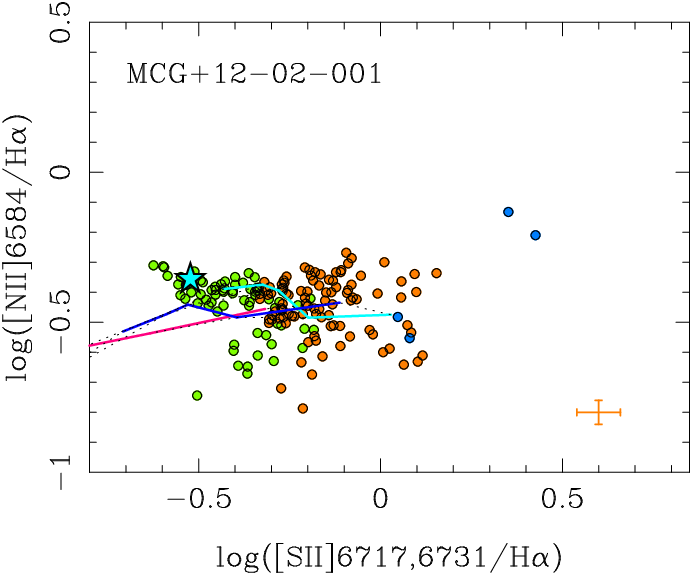}

\includegraphics[width=7cm,angle=-90]{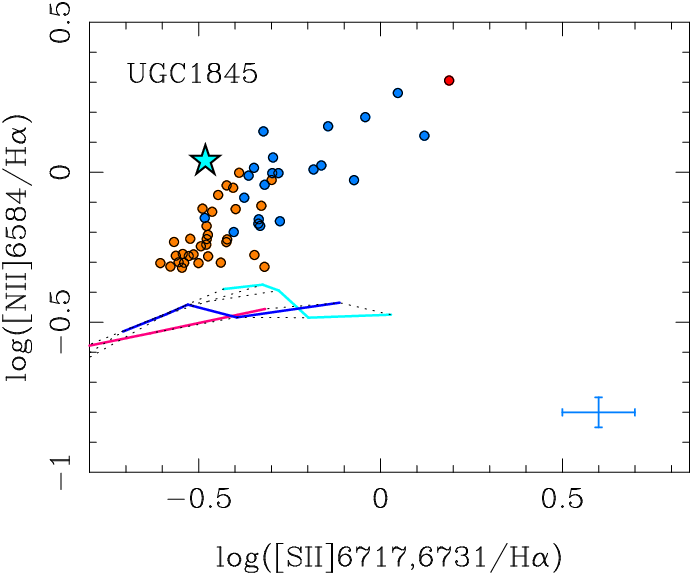}
\hspace{1cm}
\includegraphics[width=7cm,angle=-90]{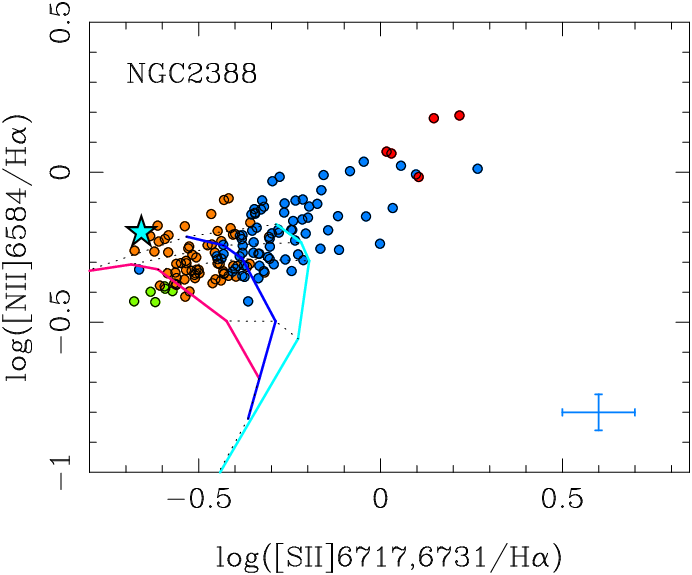}

\includegraphics[width=7cm,angle=-90]{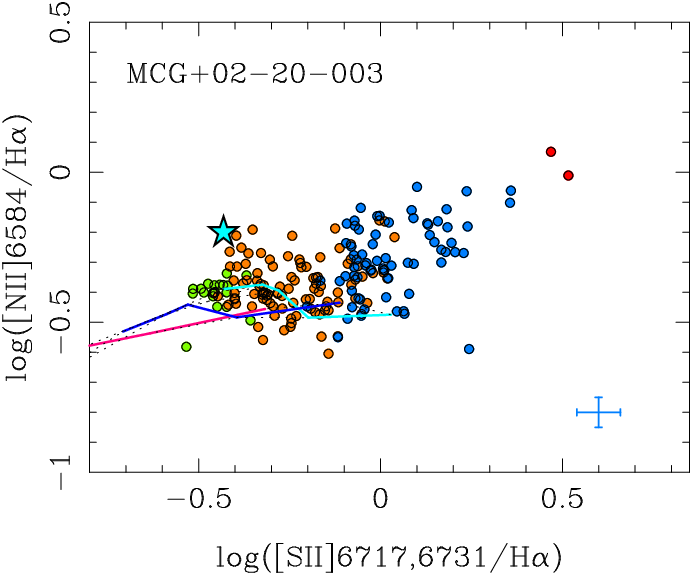}
\hspace{1cm}
\includegraphics[width=7cm,angle=-90]{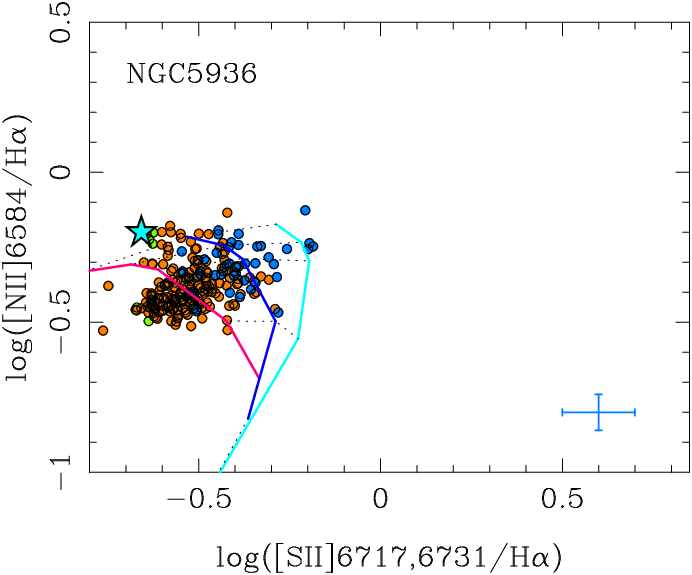}
\caption{ Spatially resolved observed
[N\,{\sc ii}]$\lambda$6584/H$\alpha$ vs. 
[S\,{\sc ii}]$\lambda\lambda$6717,6731/H$\alpha$ diagrams for each of
the galaxies in our sample. The individual spaxel
  measurements are color coded according to EW(H$\alpha)_{\rm em}$ as
  in Figure~2. The spaxels corresponding to the nuclear 
  region are shown as a star-like symbol. The line ratios are not
  corrected for underlying 
  stellar absorption or extinction. The error bars are the  
typical uncertainties in the measured line ratios for the smallest range
of EW(H$\alpha)_{\rm em}$, color-coded as the data points, measured
for each galaxy. The uncertainties in the line ratios of spaxels with 
  EW(H$\alpha)_{\rm
  em}$ greater than the plotted range are always smaller. 
The solid  lines are the Dopita et al. (2006)
  models for evolving 
 H\,{\sc ii} regions for instantaneous star formation and 
2Z$_\odot$ metallicity in all cases
except for the MCG~+12-02-001, UGC~1845 and MCG~+02-20-003 plots where
we show the 
Z$_\odot$ models (see Table~7 for the estimated abundances). The pink, dark blue and light blue lines correspond to 
$\log \mathcal R$ parameters of $-2$, $-4$, and $-6$. The
dashed lines are isochrones for ages 0.1, 0.5, 1, 2 and 3\,Myr from
top to bottom. The $\mathcal R$ parameter 
is defined as the ratio of the mass of the ionizing cluster to 
the pressure of the interstellar medium.} 
\end{figure*} 

\begin{figure*}
\setcounter{figure}{2}
\includegraphics[width=7cm,angle=-90]{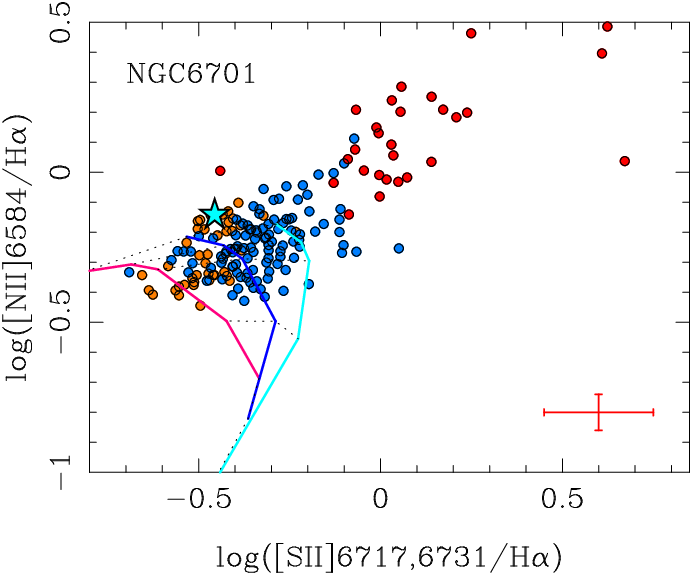}
\hspace{1cm}
\includegraphics[width=7cm,angle=-90]{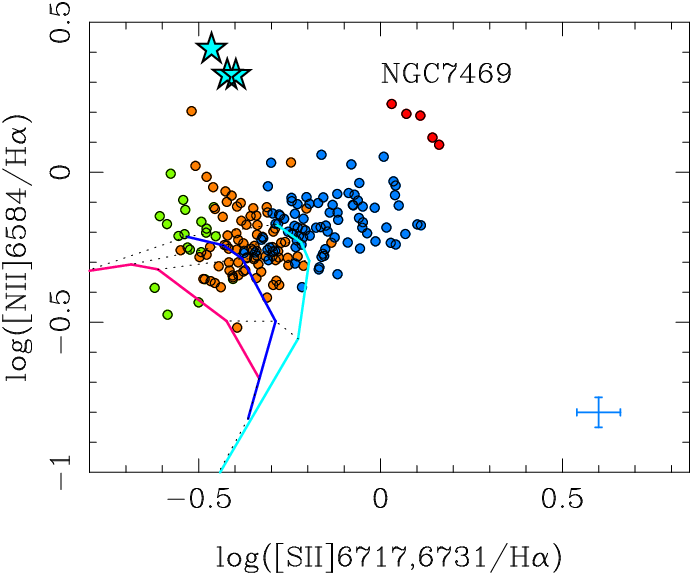}

\includegraphics[width=7cm,angle=-90]{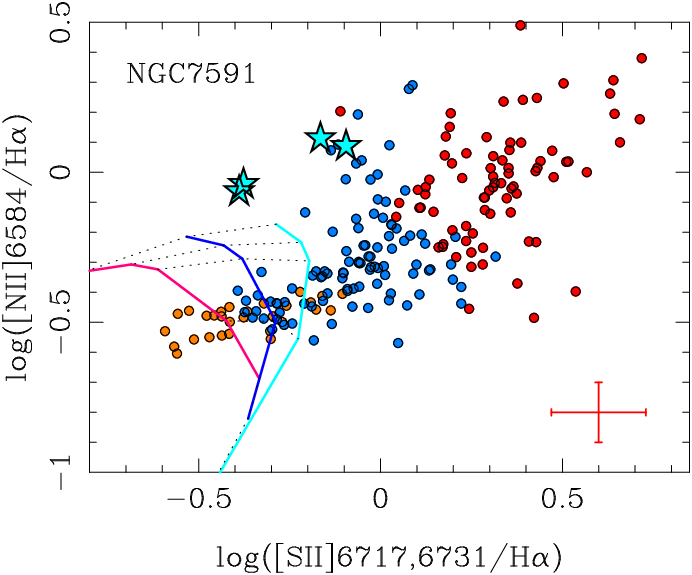}
\hspace{1cm}
\includegraphics[width=7cm,angle=-90]{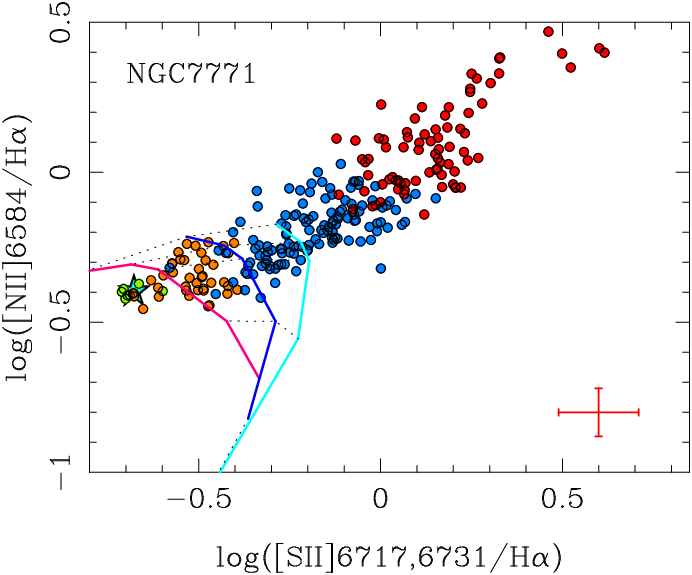}

\caption{ Continued. For NGC~7469 and NGC~7591 we identified the
  nuclear regions as 
  the 4 central brightest spaxels at 6200\,\AA, as these two galaxies
  contain an active nucleus and the seeing (FWHM) was larger than the
  spaxel size. }
\end{figure*}

\end{document}